\pdfoutput=1
\documentclass[cernpreprint,PAPER,USenglish,texlive=2016,texmf]{atlasdoc}
\usepackage{atlaspackage}
\usepackage{atlasbiblatex}
\usepackage{atlascontribute}
\usepackage{atlasphysics}
\usepackage{atlasheavyion}
\usepackage{physics}
\usepackage{dcolumn}
\usepackage{subfig}
\usepackage{placeins}
\usepackage{pPbHBTPaperDraft-defs}

\AtlasTitle{Femtoscopy with identified charged pions in proton-lead collisions at $\sqrt{s_{\mathrm{NN}}}=5.02$ \TeV{} with ATLAS}
\PreprintIdNumber{CERN-EP-2017-004}
\AtlasJournalRef{Phys. Rev. C 96 (2017) 064908}
\AtlasDOI{10.1103/PhysRevC.96.064908}
\AtlasAbstract{Bose-Einstein correlations between identified charged pions are measured for \mbox{$p$+Pb} collisions at $\sqrt{s_{\mathrm{NN}}}=5.02~\text{Te\kern -0.1em V}$ using data recorded by the ATLAS detector at the LHC corresponding to a total integrated luminosity of $28~\mathrm{nb}^{-1}$. Pions are identified using ionization energy loss measured in the pixel detector. Two-particle correlation functions and the extracted source radii are presented as a function of collision centrality as well as the average transverse momentum ($k_{\mathrm{T}}$) and rapidity ($y^{\star}_{\pi\pi}$) of the pair. Pairs are selected with a rapidity $-2 < y^{\star}_{\pi\pi} < 1$ and with an average transverse momentum $0.1 < k_{\mathrm{T}} < 0.8~\text{Ge\kern -0.1em V}$. The effect of jet fragmentation on the two-particle correlation function is studied, and a method using opposite-charge pair data to constrain its contributions to the measured correlations is described. The measured source sizes are substantially larger in more central collisions and are observed to decrease with increasing pair $k_{\mathrm{T}}$. A correlation of the radii with the local charged-particle density is demonstrated. The scaling of the extracted radii with the mean number of participating nucleons is also used to compare a selection of initial-geometry models. The cross-term $R_\mathrm{ol}$ is measured as a function of rapidity, and a nonzero value is observed with $5.1 \sigma$ combined significance for $-1 < y^{\star}_{\pi\pi} < 1$ in the most central events.}

\hypersetup{pdftitle={ATLAS draft},pdfauthor={The ATLAS Collaboration}}

\begin{document}

\maketitle

\tableofcontents

\section{Introduction}
\label{sec:intro}
Studies of multi-particle correlations in proton-lead (\pPb) 
\cite{Abelev:2012ola,HION-2012-13,Chatrchyan:2013nka,HION-2013-04, 
Khachatryan:2015waa} 
and proton-proton (\pp) \cite{Aad:2015gqa,} collisions at the LHC and in deuteron-gold (\dAu) \cite{Adare:2013piz,Adare:2014keg,Adamczyk:2015xjc}
and helium-3--gold (\HeAu)~\cite{Adare:2015ctn} collisions at RHIC have shown that these
correlation functions exhibit features similar to those observed in
nucleus-nucleus collisions \cite{Ackermann:2000tr, Adare:2008ae, Alver:2009id, ALICE:2011ab, Chatrchyan:2011eka, ATLAS:2012at} that are attributed to collective dynamics
of the strongly coupled quark-gluon plasma. In particular,
two-particle angular correlations studied in high multiplicity \pPb\cite{HION-2013-04, CMS:2012qk, Abelev:2012ola} and \pp\cite{Aad:2015gqa, Khachatryan:2010gv} collisions
at the LHC show a ``ridge'' -- an enhancement in the correlation
function at small relative azimuthal angle (\dphi) that extends over a
range of relative pseudorapidity (\deta). The ridge in both
systems is generally understood to result from a combination of sinusoidal
modulations of the two-particle correlation function of different
harmonics~\cite{Abelev:2012ola,HION-2012-13,Chatrchyan:2013nka,HION-2013-04, 
Khachatryan:2015waa}. Hydrodynamic calculations show that such a modulation can
arise from initial-state spatial anisotropies that, through the
collective expansion of the medium, are imprinted on the
azimuthal-angle distributions of the final-state particles (see
e.g.~Refs.~\cite{Huovinen:2001cy,Romatschke:2003ms,Heinz:2013th} and references therein).
Such hydrodynamic models can also reproduce the modulation observed in \pPb, \dAu, and \HeAu collisions \cite{PhysRevC.85.014911,PhysRevC.88.044915,PhysRevC.88.014903,PhysRevC.89.044902}, but the suitability of hydrodynamics in ``small'' systems remains a topic of active debate.
Alternatively, the observed modulation in these collisions has been explained by so-called ``glasma'' models that invoke saturation of the
nuclear parton distributions \cite{PhysRevD.87.051502,PhysRevD.87.054014,Kovchegov:2014yza,McLerran:2013una,PhysRevD.87.094034}. To disentangle these competing explanations and, more specifically, to test whether collective phenomena
are present in \pPb\ collisions at the LHC, additional measurements are required to constrain the source geometry.

Hanbury Brown and Twiss (HBT) correlations, which probe the space-time extent of a
particle-emitting source (see \Ref{\cite{Lisa:2005dd}} and references
therein), may provide valuable insight into the problems described above.
The HBT method
originated in astronomy \cite{HanburyBrown:1956, HanburyBrown:1954},
where space-time correlations of photons due to wave function
symmetrization are used to measure the size of distant stars.  The
procedure can be adapted to the extremely small sources encountered in hadronic
collisions if identical-particle Bose-Einstein correlations are
instead studied in relative momentum space \cite{PhysRev.120.300}.
The two-particle correlation function $C(\mathbf{q})$, parameterized as a function of relative momentum, is 
sensitive to the two-particle source density function $S(\mathbf{r})$ through the
two-particle final-state wavefunction \cite{Lisa:2005dd}:
\begin{equation} \label{eq:wavefunction_kernel}
C_\mathbf{k}(\mathbf{q}) - 1 = \int d^3 r \, S_{\mathbf{k}}(\mathbf{r}) \left(\left|\braket{\mathbf{q}}{\mathbf{r}}\right|^2 - 1 \right) \, ,
\end{equation}
where $\mathbf{q}$ and $\mathbf{k}$ are, respectively, the relative and average momentum of a pair of particles, $\mathbf{r}$ is the distance between the origin points of the two particles, and the two-particle source function $S_{\mathbf{k}}$ is normalized so that $\int d^3 r \, S_{\mathbf{k}}(\mathbf{r}) = 1$.
In the case of a non-interacting identical boson wave-function, the term within the parentheses of \Eqn{\ref{eq:wavefunction_kernel}} is a cosine and the correlation function is enhanced by the Fourier transform of the source function.
Thus, the Bose-Einstein modification of the relative momentum
distributions produces an enhancement at small $\mathbf{q}$ whose range in $\mathbf{q}$ is inversely related to the
size of the source.

In a typical HBT analysis, the correlation functions are fit to a
function of relative momentum that is often a Gaussian or exponential function, or a stretched exponential function that can interpolate between these two.
The parameters of the fits that relate to the space-time extent of the source function are referred to as the ``HBT radii''.
Measurements of Bose-Einstein correlations in \pp collisions at center-of-mass energies
$\sqrt{s} = 0.9~\TeV$ and $\sqrt{s} = 7~\TeV$ have been made by the
ATLAS \cite{Aad:2015sja}, CMS \cite{Khachatryan:2011hi}, and
ALICE \cite{Aamodt:2011kd} experiments.  At both energies the source radii are
observed to decrease with rising transverse momentum.
It is also observed that the extracted radii increase with particle multiplicity but
saturate at the highest multiplicities.

Although Bose-Einstein correlations are the most straightforward to
measure experimentally, any final-state interaction can in principle
be used to image the source density.  The term ``femtoscopy'' is
often used to refer to any measurement that provides spatio-temporal
information about a hadronic source \cite{Lednicky:2002fq}. The measured
source radii are interpreted as the dimensions of the region of homogeneity of the source at
freeze-out, after all interactions between final-state particles and
the bulk have ceased; thus, they are sensitive to the space-time
evolution of the event. In particular, an increase in radii at low average transverse momentum \kt indicates radial expansion since higher-momentum particles
are more likely to be produced earlier in the event \cite{Kolb:2003dz}.
The \kt-scaling of HBT radii in \pPb systems is of significant interest when studied as a function of centrality, an experimental proxy for the impact parameter. Thus, these measurements can provide insight into the conditions necessary for hydrodynamic behavior in small systems.

In many HBT measurements, the correlation functions are evaluated in
one dimension using the invariant relative momentum
$\qinv \equiv \sqrt{-q_\mu q^\mu}$, where $q = p^a - p^b$ for a pair of particles $a$ and $b$ with four-momenta $p^a$ and $p^b$. In three dimensions, HBT correlations are studied using the ``out-side-long''
convention \cite{Pratt:1986cc, PhysRevC.37.1896, Bertsch:1989vn,Csorgo:1989kq}.
In this system, \qout, the outwards component, is the projection along $\mathbf{k}_\mathrm{T}$; \qside, the sideways component, is the projection along $\hat{\mathbf{z}} \times \mathbf{k}_\mathrm{T}$ (with the $z$-axis along the beamline); and \qlong is
the longitudinal component. The relative momentum of the pair is evaluated in the
longitudinally comoving frame (LCMF), i.e., the frame boosted such that $k_{z} = 0$.
This formulation of the HBT analysis has the advantage that it decomposes the correlation
function into components that emphasize distinct physical effects. In
particular, the spatial extent of the source in the longitudinal and
transverse directions is likely to be different. The out and side
radii are also expected to differ due to the effects of the
Lorentz boost in the out direction and, if the system exhibits
collectivity, due to space-momentum correlations. In a fully boost-invariant system,
observables evaluated in the LCMF should be independent of $k_{z}$ (or rapidity). The inherent
asymmetry of \pPb\ collisions seen, for example, in the charged-particle pseudorapidity distributions \cite{Adam:2014qja,Aad:2015zza}, provides a
unique opportunity to study the correlations between 
source sizes and the pair's rapidity, collision centrality, or the local (in rapidity)
charged-particle density. The results of such a study may provide
insight into or constrain theoretical models of the underlying dynamics
responsible for producing the final-state particles.

To address the topics and questions discussed above, this paper presents
measurements of correlations between identified charged pions in
5.02~\TeV\ \pPb\ collisions which were performed by the ATLAS
experiment at the LHC.  While femtoscopic methods have already been
applied to \pPb systems at the LHC \cite{PhysRevC.91.034906,
Abelev:2014pja}, this paper presents a new data-driven technique to
constrain the significant background contribution from jet
fragmentation, referred to in this paper as the ``hard process''
background. It also provides new measurements of the dependence of the source radii
on the pair's rapidity \kys, calculated assuming both particles have the mass of the pion, over the range $-2 < \kys < 1$. Results are presented for one- and three-dimensional
source radii as a function of the pair's average transverse momentum, \kt, over
the range $0.1 < \kt < 0.8$~\GeV\ and for several \pPb\ centrality
intervals with the most central case being 0--1\%.
The \pPb\ collision centrality is characterized using \sumETPb, 
the total transverse energy measured in the Pb-going forward calorimeter (FCal) \cite{Aad:2015zza}.
It is defined such that central events, with large \sumETPb, have a low centrality percentage, and peripheral events, with a small \sumETPb, have a high centrality percentage.
  Using the measured centrality dependence of the
source radii, the scaling of the system size with the number of nucleon
participants \Npart is also investigated, using a generalization of the Glauber model \cite{Miller:2007ri}.

\FloatBarrier

\section{ATLAS detector}
\label{sec:atlas}
The ATLAS detector is described in detail in Ref.~\cite{Aad:2008zzm}.
The measurements presented in this paper have been performed using the inner detector, minimum-bias trigger scintillators~(MBTS), FCal, zero-degree calorimeter (ZDC), and the trigger and data acquisition systems.
The inner detector \cite{Aad:2010bx}, which is immersed in a 2 T axial magnetic field, is used to reconstruct
charged particles within $|\eta| < 2.5$.\footnote{ATLAS uses a
right-handed coordinate system with its origin at the nominal
interaction point (IP) in the center of the detector and the $z$-axis
along the beam pipe.  The $x$-axis points from the IP to the center of
the LHC ring, and the $y$-axis points upward.  Cylindrical coordinates
$(r,\phi)$ are used in the transverse plane, $\phi$ being the
azimuthal angle around the beam pipe.  The pseudorapidity is defined
in terms of the polar angle $\theta$ as $\eta=-\ln\tan(\theta/2)$.}
It consists of a silicon pixel detector, a semiconductor tracker (SCT) made of double-sided silicon microstrips, and a transition radiation tracker made of straw tubes.
All three detectors consist of a barrel and two symmetrically placed endcap sections. A particle traveling from the interaction point (IP) with $|\eta|<2$ crosses at least 3 pixel layers, 4 double-sided microstrip layers and typically 36 straw tubes.
In addition to hit information, the pixel detector provides time-over-threshold for each hit pixel which is proportional to the deposited energy and which is used to provide measurements of specific energy loss (\dEdx) for particle identification.

The FCal covers a pseudorapidity region of
$3.1 < |\eta| < 4.9$ and is used to estimate the centrality of each
collision. The FCal uses liquid argon as the active medium with
tungsten and copper absorbers.
The MBTS, consisting of two
arrays of scintillation counters, are positioned at $z = \pm 3.6$ m
and cover $2.1 < |\eta| < 3.9$.
The ZDCs, situated approximately $140$~m from the nominal IP, detect neutral particles, mostly neutrons and photons, that have $|\eta|>8.3$.
They are used to distinguish pileup events (bunch crossings involving more than one collision) from central collisions by detecting spectator nucleons that did not participate in the interaction.
The calorimeters use tungsten plates as absorbers and quartz rods sandwiched between the tungsten plates as the active medium.

Events used for the analysis presented in this paper were primarily obtained from a combination of minimum-bias (MinBias) triggers that required either at least two hit scintillators in the MBTS or at least one hit on each side of the MBTS. An additional requirement on the number of hits in the SCT was imposed on both of these minimum-bias triggers to remove false triggers. To increase the number of events available in the highest \sumETPb\ interval, the analysis includes a separate sample of events selected by a trigger (HighET) that required a total transverse energy in both sides of the FCal of at least 65~\GeV.

\FloatBarrier

\section{Data sets}
\label{sec:data}
\subsection{LHC data}
\label{subsec:data}

This analysis uses data from the LHC 2013 \pPb{} run at \pPbenergy{} with an integrated luminosity of 28 \inb.
The Pb ions had an energy per nucleon of $1.57~\TeV$ and collided with the $4~\TeV$ proton beam to yield a center-of-mass energy \pPbenergy\ with a longitudinal boost of $\ycm = 0.465$ in the proton direction relative to the ATLAS laboratory frame.
The \pPb\ run was divided into two periods between which the directions of the proton and lead beams were
reversed. 
The data in this paper are presented using the convention that the proton beam travels in the forward ($+z$) direction and the lead beam travels in the backward ($-z$) direction.
When the data from these two periods are combined, the MinBias triggers sampled a total luminosity, after prescale, of $24.5~\mu\text{b}^{-1}$ and yielded a total of 44 million events; the HighET trigger sampled a
total luminosity of $41.4~\mu\text{b}^{-1}$ after prescale and yielded 700 thousand events.

\subsection{Monte Carlo event generators}
\label{subsec:generators}

The effects of charged-particle reconstruction and selection are studied in a \pPb sample generated using \Hijing \cite{Gyulassy:1994ew} and simulated with the GEANT4 package \cite{Agostinelli:2002hh}.
Five million events are generated at a center-of-mass energy per nucleon-nucleon pair of \pPbenergy with a longitudinal boost of $\ycm = 0.465$ in the proton direction.
The sample is fully reconstructed with the same conditions as the data \cite{SOFT-2010-01}.

Four additional Monte Carlo generator samples are used to study the background from hard processes, as described in \Sect{\ref{subsec:hard_process}}.
No detector simulation is performed on these samples, as the net effects of the simulation and reconstruction were studied using the fully reconstructed \pPb simulation events and found to be negligible. The two-particle reconstruction effects occur only at very low $\mathbf{q}$ (as discussed in \Sect{\ref{subsec:syst_description}}), but these generated samples are used only to study correlations from jet fragmentation which span a much broader range of $\mathbf{q}$.
In each of the following samples, 50 million (250 million for \PYEight) minimum-bias events are generated at a center-of-mass energy per nucleon-nucleon pair of \pPbenergy.

\begin{enumerate}
\item \Hijing~\pPb\\
The energy and boost settings are the same as in the nominal \pPb reconstructed simulation, except that the minimum hard-scattering transverse momentum is adjusted as described in \Sect{\ref{subsec:hard_process}}. This boost is applied only in the \pPb sample.

\item \Hijing~\pp\\
The generator is run with all settings the same as in the \pPb sample, except that both incoming particles are protons.

\item \Pythia~8 \pp \cite{Sjostrand:2007gs}\\
The set of generator parameters from ATLAS ``UE AU2-CTEQ6L1'' \cite{ATL-PHYS-PUB-2012-003} is used with \Pythia 8.209, which utilizes the CTEQ 6L1 \cite{Pumplin:2002vw} parton distribution function (PDF) from LHAPDF6 \cite{Buckley:2014ana}.

\item \Herwig++ \pp \cite{Bahr:2008pv}\\
The NNLO MRST PDF \cite{Martin:2002dr} is used with \Herwig++ 2.7.1.
\end{enumerate}

\begin{figure}[b]
\begin{minipage}[t]{1.0\textwidth}
\centering
\includegraphics[width=0.55\textwidth]{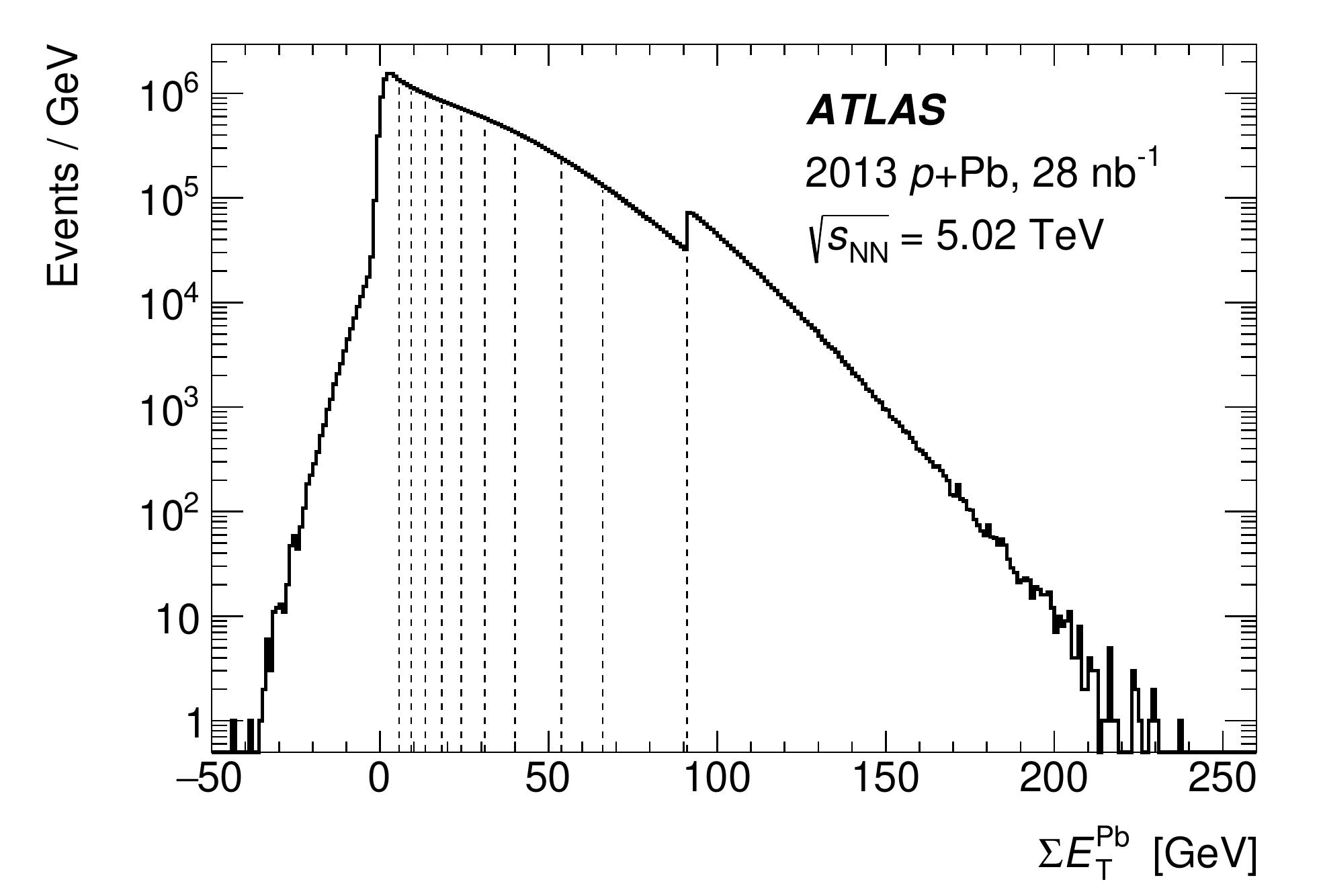}
\end{minipage}
\caption{The distribution of the total transverse energy in the
  forward calorimeter in the Pb-going direction (\sumETPb) for the events used in this analysis.
  Dashed lines are shown at the boundaries of the centrality intervals, and the discontinuity at $\sumETPb = 91.08~\GeV$ corresponds to the lower \sumETPb boundary of the 0--1\% centrality interval.}
\label{fig:fcal_et}
\end{figure}

\subsection{Event selection and centrality}
\label{subsec:event_track_selection}
In the offline analysis, charged-particle tracks and collision vertices are reconstructed using the same algorithms and methods applied in previous minimum-bias \pp\ and \pPb\ measurements \cite{Aad:2010ac,Aad:2015zza}.
Events included in this analysis are required to pass either of the two MinBias triggers or the HighET trigger, to have a hit on each side of the MBTS with a difference in average particle arrival times measured on the two sides of the MBTS which is less than $10~\textrm{ns}$, a reconstructed primary vertex (PV), and at least two tracks satisfying the selection criteria listed in \Sect{\ref{subsec:tracks}}.
Events that have more than one reconstructed vertex (including secondary vertices) with either more than ten tracks or a sum of track transverse momentum (\pt) greater than 6~\GeV\ are rejected.
An upper limit is placed on the activity measured in the Pb-going ZDC to further reject pileup events.

The centralities of the \pPb\ events are characterized following the
procedures described in \Ref{\cite{Aad:2015zza}}, using \sumETPb, the
total transverse energy in the Pb-going side of the FCal.  The use of
the FCal for measuring centrality has the advantage that it is not
sensitive to multiplicity fluctuations in the kinematic region covered
by the inner detector, where the measurements are
performed. Measurements are presented in this paper for the
centrality intervals listed in Table~\ref{table:npart}. The events
selected using the HighET trigger are used only 
in the 0--1\% centrality interval. Figure~\ref{fig:fcal_et} shows the
distribution of \sumETPb\ values obtained from events included in this
measurement. The discontinuity in the spectrum occurs at the low edge
of the 0--1\% centrality interval, above which the HighET events are included.

For each centrality interval, the average multiplicity of charged particles with $\pt > 100~\MeV$ and $|\eta| < 1.5$, \avgdNdeta, and the corresponding average number of participating nucleons, \avgNpart, are obtained from a previous publication \cite{Aad:2015zza}.
Since this analysis uses finer centrality intervals (no wider than 10\% of the total centrality range) than
those used in \Ref{\cite{Aad:2015zza}}, a linear interpolation over the Glauber \avgNpart is used to construct additional values for \avgdNdeta based on the published results.
This interpolation is justified by the result in \Ref{\cite{Aad:2015zza}} that charged-particle multiplicity is proportional to \avgNpart in the peripheral region.
The values and uncertainties from this procedure are listed in Table~\ref{table:npart}.

\renewcommand{\arraystretch}{1.1}
\begin{table}
\begin{center}
\begin{tabular}{r || c | c | c || c}
\hline
 & \multicolumn{3}{c||}{\avgNpart} & \\ \cline{2-4}
Centrality & Glauber & GGCF $\omega_{\sigma} = 0.11$ & GGCF $\omega_{\sigma} = 0.2$ & $\avgdNdeta$\\
\hline \hline
0--1\% & $ 18.2^{+2.6}_{-1.0} $ & $ 24.2^{+1.5}_{-2.1} $ & $ 27.4^{+1.6}_{-4.5} $ & $58.1\ \pm 0.1\ \ \pm 1.9\ $ \\[2pt]
\hline
1--5\% & $ 16.10^{+1.66}_{-0.91} $ & $ 19.5^{+1.2}_{-1.3} $ & $ 21.4^{+1.5}_{-2.0} $ & $45.8\ \pm 0.1\ \ \pm 1.3\ $ \\[2pt]
\hline
5--10\% & $ 14.61^{+1.21}_{-0.82} $ & $ 16.5^{+1.0}_{-1.0} $ & $ 17.5^{+1.1}_{-1.1} $ & $38.5\ \pm 0.1\ \ \pm 1.1\ $ \\[2pt]
\hline
10--20\% & $ 13.05^{+0.82}_{-0.73} $ & $ 13.77^{+0.79}_{-0.81} $ & $ 14.11^{+0.86}_{-0.79} $ & $32.34 \pm 0.05 \pm 0.97$ \\[2pt]
\hline
20--30\% & $ 11.37^{+0.65}_{-0.63} $ & $ 11.23^{+0.62}_{-0.67} $ & $ 11.17^{+0.68}_{-0.62} $ & $26.74 \pm 0.04 \pm 0.80$ \\[2pt]
\hline
30--40\% & $ 9.81^{+0.56}_{-0.57} $ & $ 9.22^{+0.50}_{-0.54} $ & $ 8.97^{+0.60}_{-0.49} $ & $22.48 \pm 0.03 \pm 0.75$ \\[2pt]
\hline
40--50\% & $ 8.23^{+0.48}_{-0.55} $ & $ 7.46^{+0.41}_{-0.43} $ & $ 7.15^{+0.54}_{-0.39} $ & $18.79 \pm 0.02 \pm 0.69$ \\[2pt]
\hline
50--60\% & $ 6.64^{+0.41}_{-0.52} $ & $ 5.90^{+0.36}_{-0.34} $ & $ 5.60^{+0.47}_{-0.30} $ & $15.02 \pm 0.02 \pm 0.62$ \\[2pt]
\hline
60--70\% & $ 5.14^{+0.35}_{-0.43} $ & $ 4.56^{+0.32}_{-0.26} $ & $ 4.32^{+0.41}_{-0.23} $ & $11.45 \pm 0.01 \pm 0.56$ \\[2pt]
\hline
70--80\% & $ 3.90^{+0.24}_{-0.30} $ & $ 3.50^{+0.22}_{-0.18} $ & $ 3.34^{+0.29}_{-0.16} $ & $ \ \  8.49 \pm 0.02 \pm 0.51$ \\[2pt]
\hline
\end{tabular}
\caption{The average number of nucleon participants \avgNpart \cite{Aad:2015zza} for each centrality interval in the Glauber model as well as the two choices for the Glauber-Gribov model with color fluctuations (GGCF) \cite{Alvioli:2013vk} (and references therein), along with the average multiplicity with $\pt > 100 \MeV$ and $|\eta| < 1.5$ also obtained from Ref.~\cite{Aad:2015zza}. The parameter $\omega_{\sigma}$ represents the size of fluctuations in the nucleon-nucleon cross section. Asymmetric systematic uncertainties are shown for \avgNpart. The uncertainties in \avgdNdeta are given in the order of statistical followed by systematic.}
\label{table:npart}
\end{center}
\end{table}

\subsection{Charged-particle selection and pion identification}
\label{subsec:tracks}
Reconstructed tracks used in the HBT analysis are required to have $|\eta| < 2.5$ and $\pt > 0.1~\GeV$ and to satisfy a standard set of selection criteria \cite{Aad:2010ac}: a minimum of one pixel hit is required, and if the track crosses an active module in the innermost layer, a hit in that layer is required; for a track with \pt greater than 0.1, 0.2, or 0.3~\GeV\ there must be at least two, four, or six hits respectively in the SCT; the transverse impact parameter with respect to the primary vertex, $d_{0}^{\textrm{PV}}$, must be such that $|d_0^{\textrm{PV}}| < 1.5~\textrm{mm}$; and the corresponding longitudinal impact parameter must satisfy $|z_{0}^{\textrm{PV}}\sin\theta| < 1.5 \textrm{ mm}$.
To reduce contributions from secondary decays, a stronger constraint on the pointing of the track to the primary vertex is applied.
Namely, neither $|d_{0}^{\textrm{PV}}|$ nor $|z_{0}^{\textrm{PV}}\sin \theta|$ can be larger than three times its uncertainty as derived from the covariance matrix of the track fit. 

\begin{figure}[t]
\centering
\subfloat[]{
\includegraphics[width=.49\linewidth]{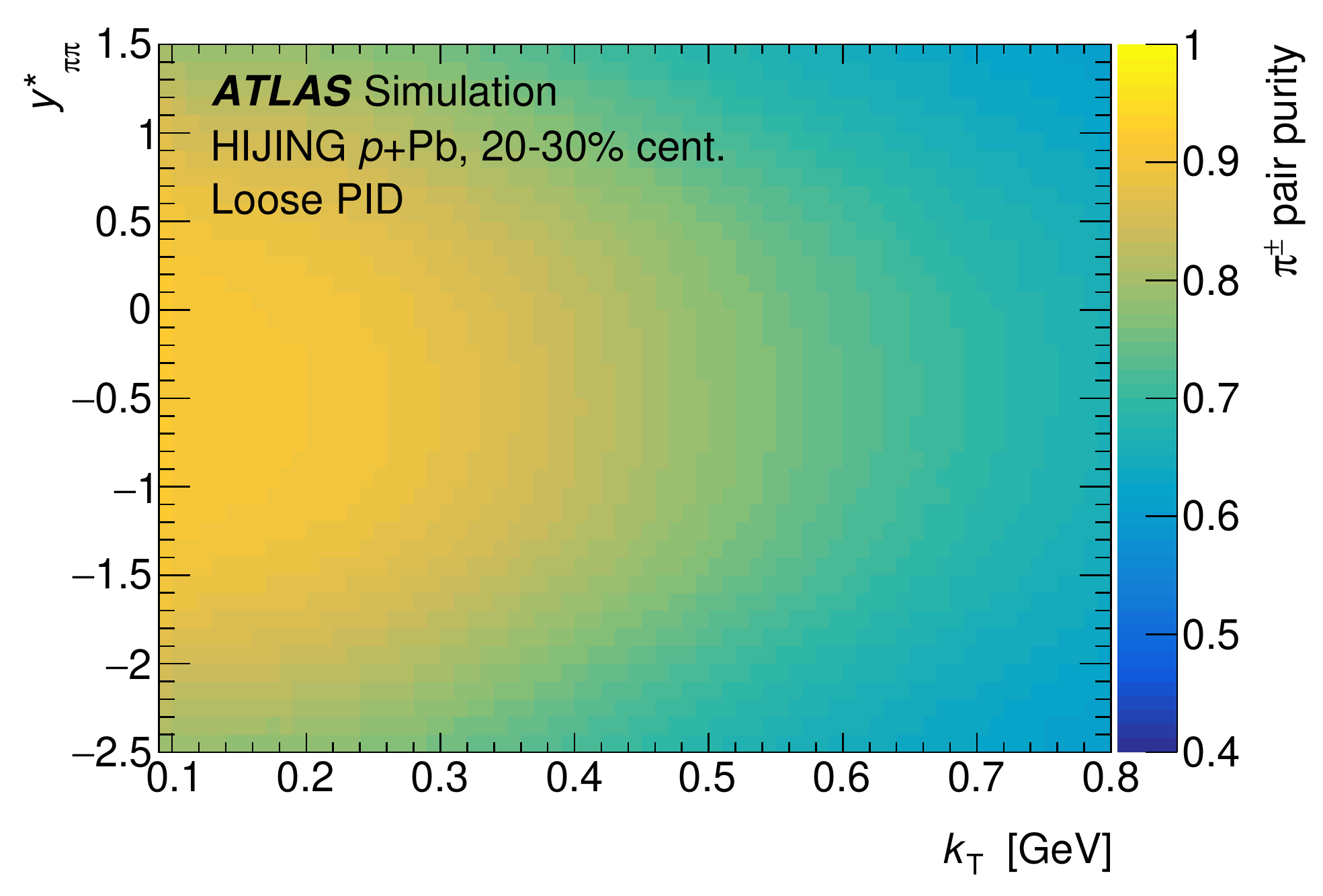}
}
\subfloat[]{
\includegraphics[width=.49\linewidth]{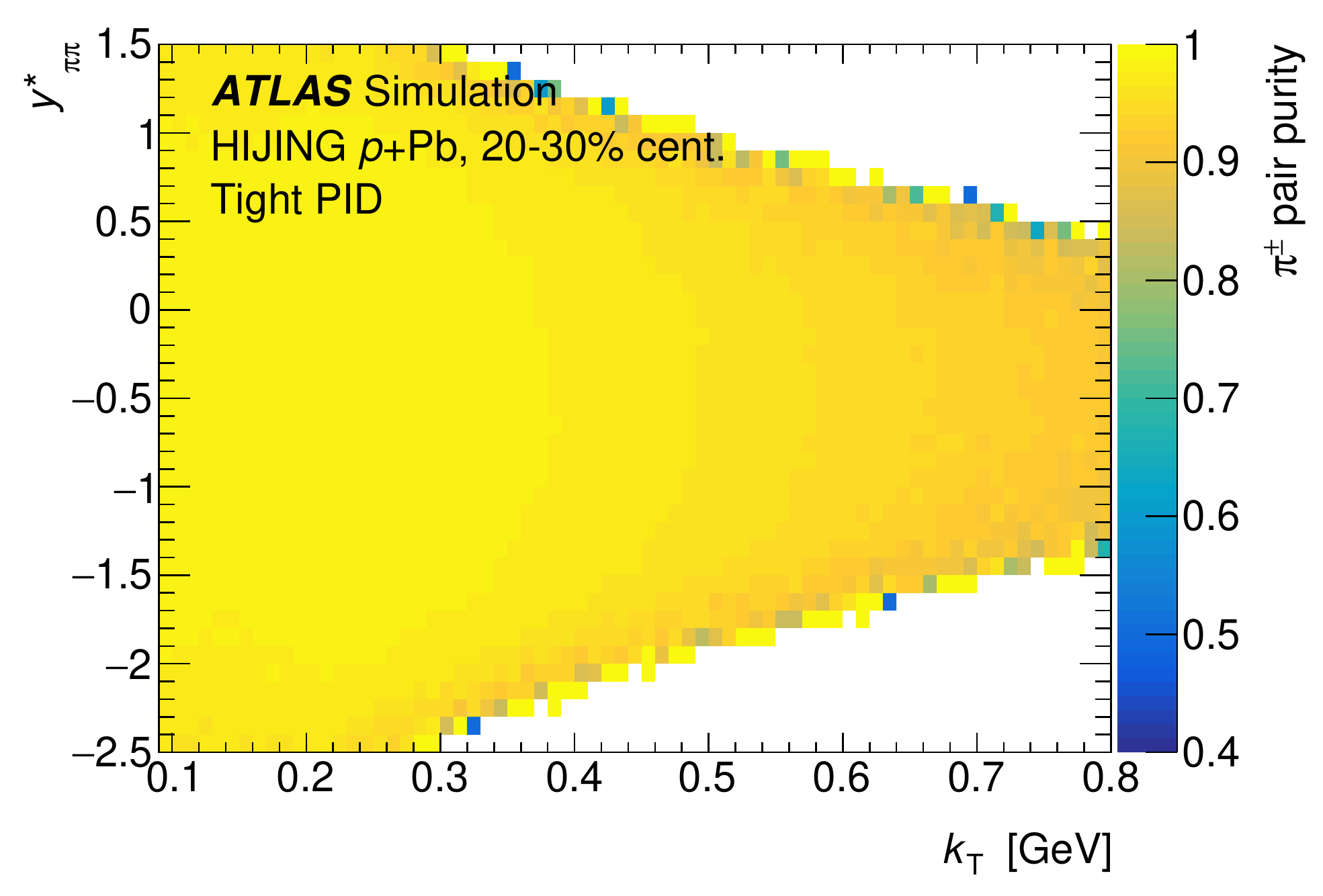}
}\\
\subfloat[]{
\includegraphics[width=.6\linewidth]{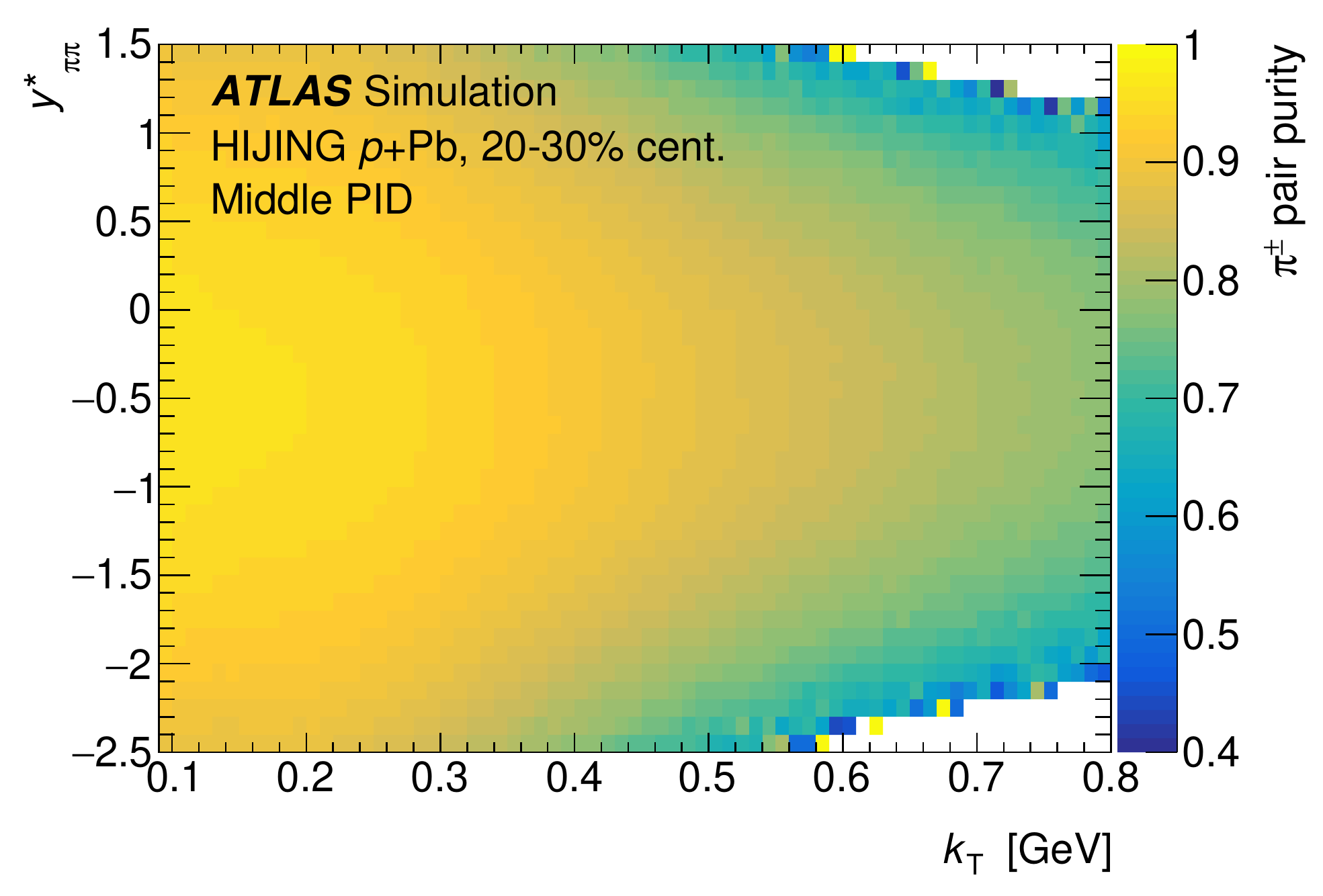}
}
\caption{Purity of identified pion pairs with the loose (a), tight (b), and middle (c) PID selections. The purities are estimated using fully simulated \Hijing \pPb events, as a function of the pair's average transverse momentum \kt and rapidity \kys. The rapidity is calculated using the pion mass for both reconstructed charged particles.}
\label{fig:pid_purity}
\end{figure}

Particle identification (PID) is performed through measurements of the specific energy loss \dEdx derived from the ionization charge deposited in the pixel clusters associated with a track.
The \dEdx of a track is calculated as a truncated mean of the \dEdx in individual pixel clusters as described in Ref. \cite{Aad:2014rca}, since the truncated mean gives a better resolution than the mean.
Relative likelihoods that the track is a $\pi$, $K$, and $p$ are formed by fitting the \dEdx distributions to $\sqrt{s} = 7~\TeV$ \pp data in several momentum intervals as explained in \Ref{\cite{ATLAS-CONF-2011-016}}.
Three PID selection levels are defined: one designed to have a high efficiency for pions, one designed to result in high pion purity, and one in between that was chosen as the nominal selection level and is used throughout the analysis if other PID selections are not explicitly mentioned.
The efficiency and purity of these selections are studied in the fully reconstructed simulated sample.
The resulting purity of track pairs in the nominal selection is shown in \Fig{\ref{fig:pid_purity}} as a function of pair's \kt and \kys.
The results are also evaluated at the looser and tighter PID definitions (also in \Fig{\ref{fig:pid_purity}}), and the differences are incorporated into the systematic uncertainty (see \Sect{\ref{sec:systematics}}).

\subsection{Pair selection}
\label{subsec:pair}

Track pairs are required to have $|\dphi| < \pi/2$ to avoid an enhancement in the correlation function arising primarily from dijets.
This enhancement does not directly affect the signal region but can influence the results by affecting the overall normalization factor in the fits.
The pair's rapidity \kys, measured with respect to the nucleon-nucleon
center of mass, must lie in the range $-2 < \kys < 1$. This requirement is
more stringent than the single-track requirement $|\eta| < 2.5$.
When analyzing track pairs of opposite charge, common particle resonances are removed via requirements on the invariant mass so that $\left|m_{\pi\pi} - m_{\rho^0}\right| > 150$~\MeV, $\left|m_{\pi\pi} - m_{\KzeroS}\right| > 20$~\MeV\ and $\left|m_{KK} - m_{\phi (1020)}\right| > 20~\MeV$, where $m_{ab}$ is the pair's invariant mass calculated with particle masses $m_a$ and $m_b$.
The values are chosen according to the width of the resonance (for the $\rho^0$) or the scale of the detector's momentum resolution [$\KzeroS$ and $\phi (1020)$].
These requirements are applied when forming both the same- and mixed-event distributions (defined in \Sect{\ref{sec:correlation_function}}).

\begin{figure}[tb]
\centering
\subfloat[]{
\includegraphics[width=.49\linewidth]{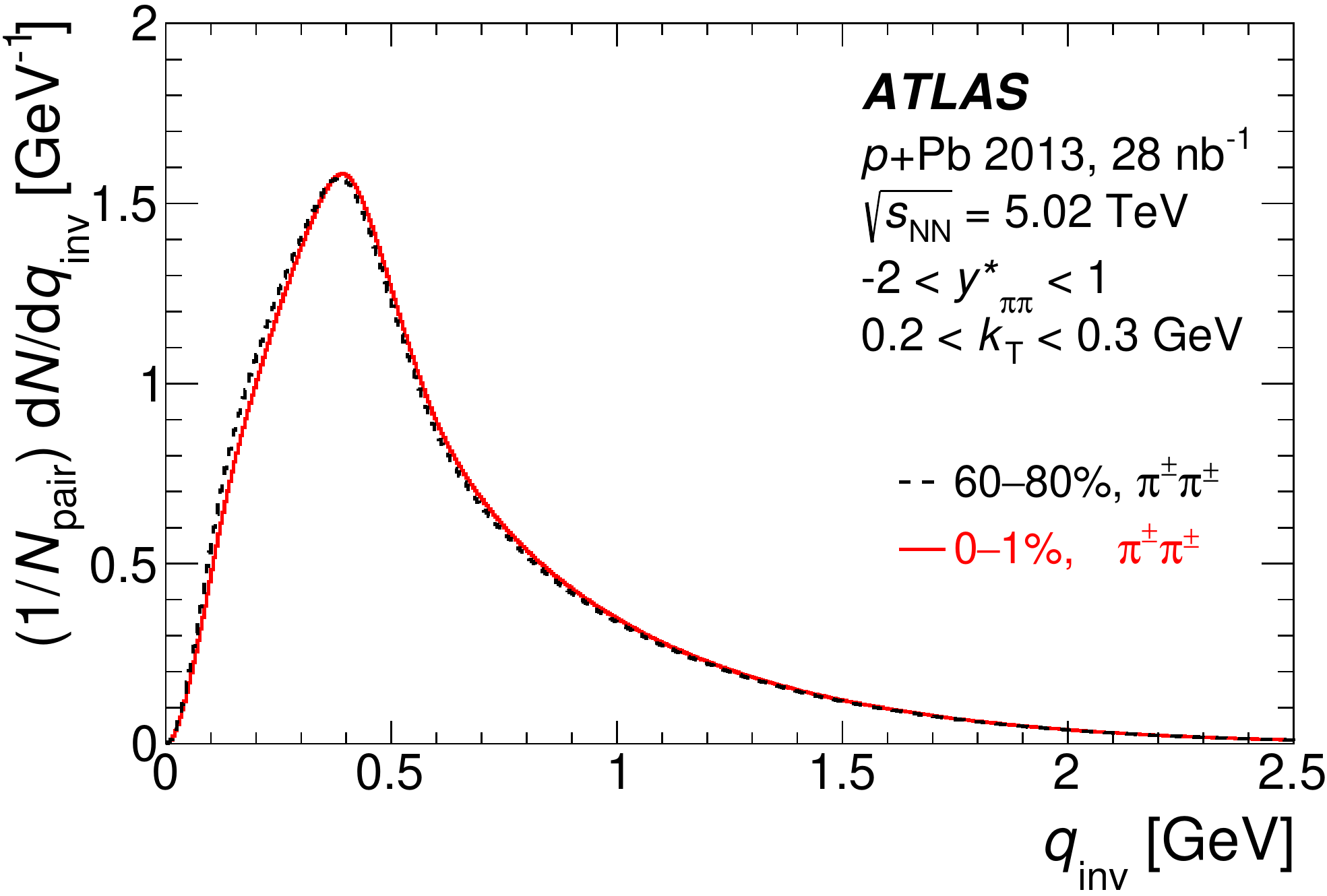}
}
\subfloat[]{
\includegraphics[width=.49\linewidth]{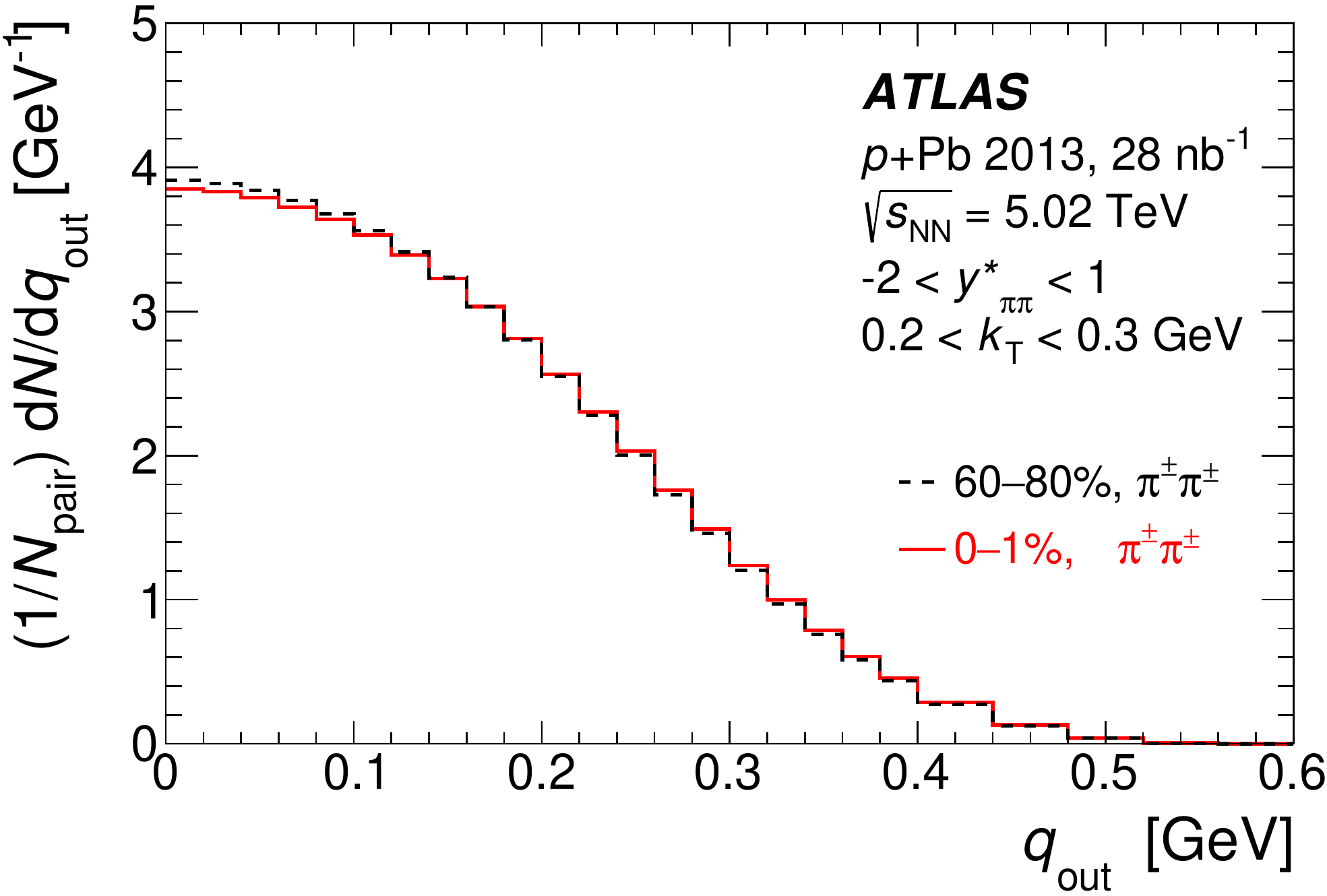}
}\\
\subfloat[]{
\includegraphics[width=.49\linewidth]{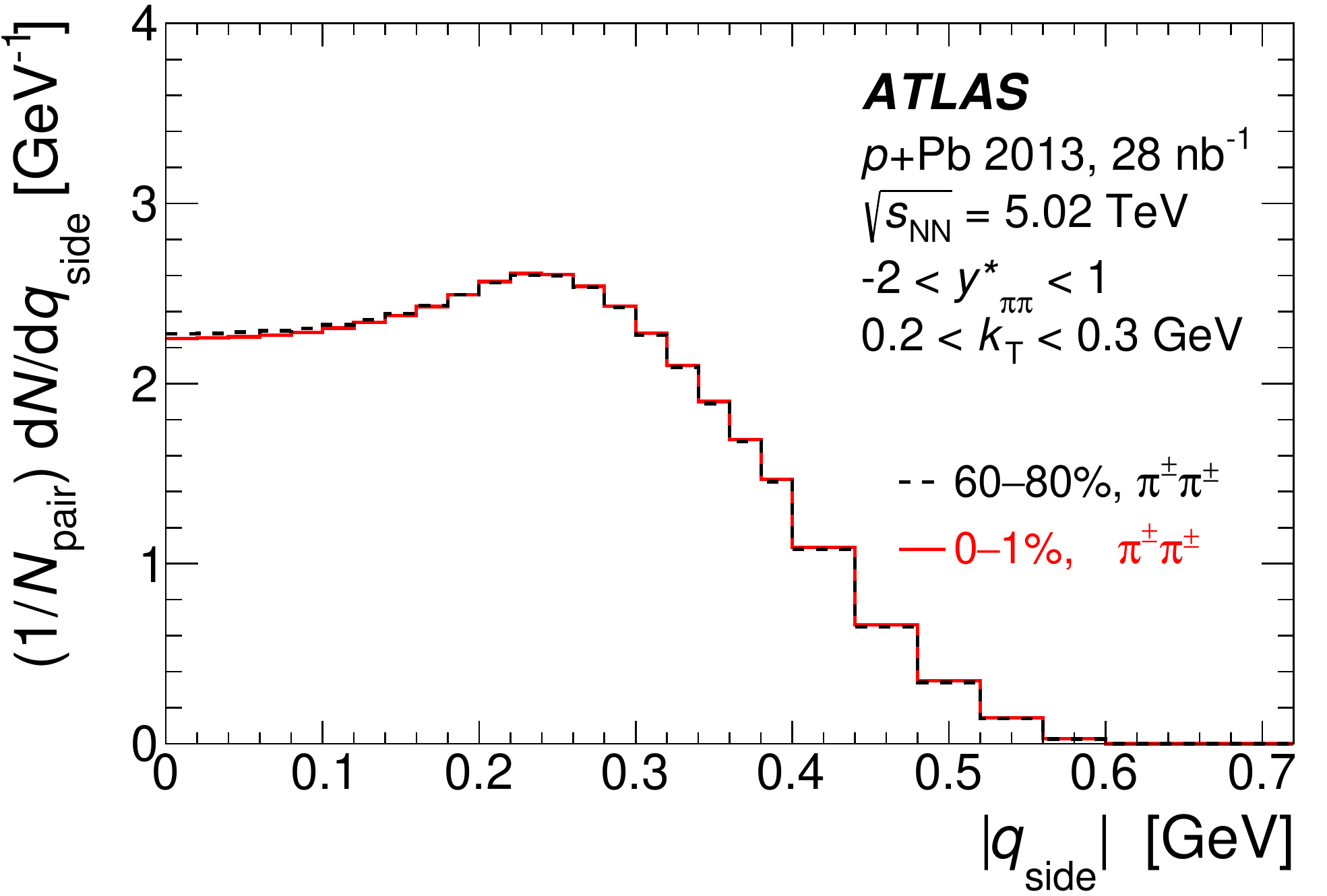}
}
\subfloat[]{
\includegraphics[width=.49\linewidth]{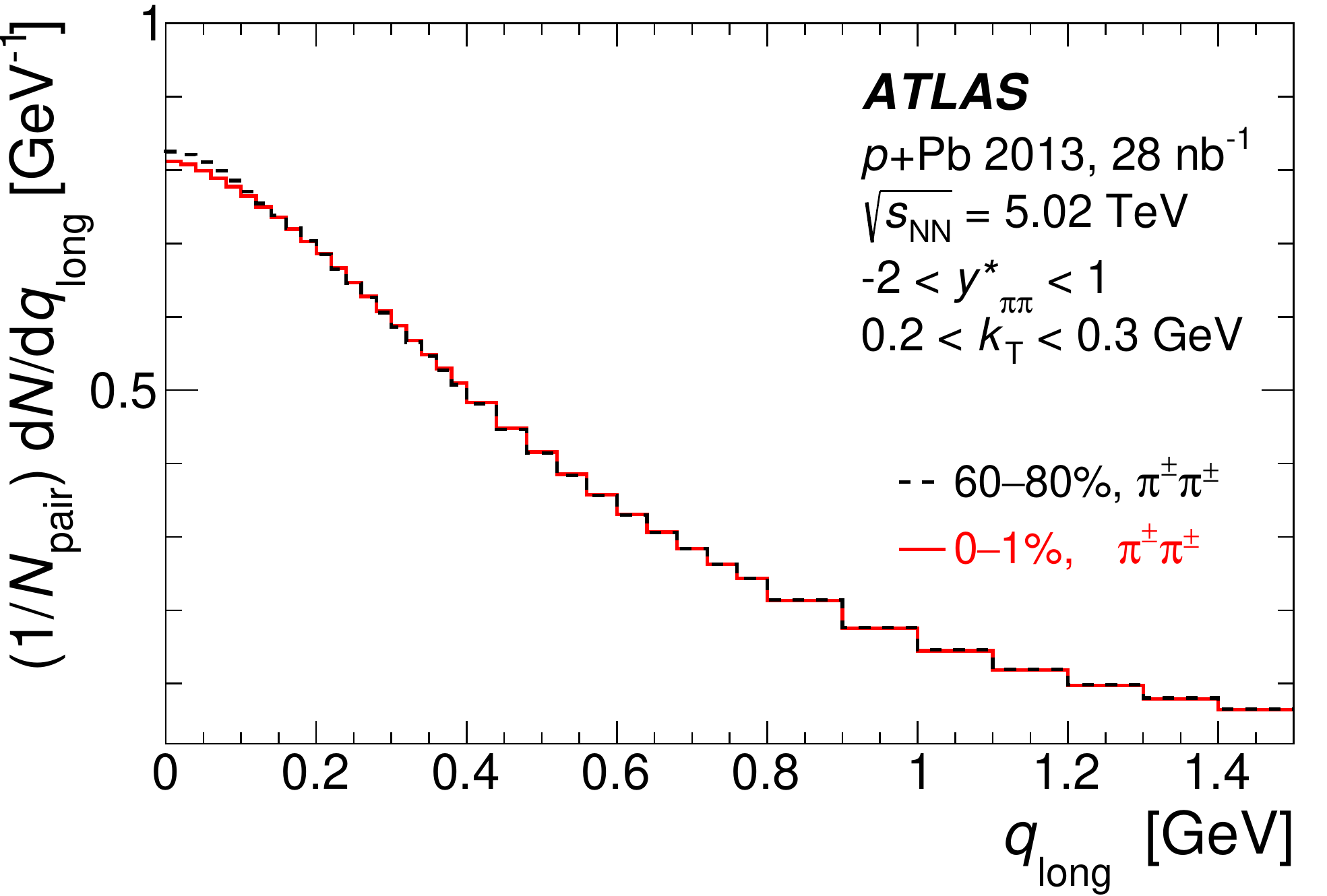}
}
\caption{Pair-normalized distributions of the invariant relative momentum \qinv\ (a) and the three-dimensional relative momentum components \qlong\ (b), \qside\ (c) and \qout\ (d) for identified same-charge pion pairs with $0.2 < \kt < 0.3$~\GeV\ obtained from two centrality intervals: 60--80\% and 0--1\%.}
\label{fig:pairqdists}
\end{figure}

The \qinv, \qlong, $|\qside|$, and \qout\ distributions of the pairs obtained through these procedures are shown in \Fig{\ref{fig:pairqdists}} for the 0--1\% and 60--80\% centrality intervals.
The one-dimensional \qinv\ distribution necessarily decreases to zero at $\qinv = 0$ due to the scaling of the phase-space volume element $d^3 q \propto q^2 dq$.
In contrast the three-dimensional quantities remain finite at zero
relative momentum. The distributions are nearly identical for the two
centrality intervals, although differences can be seen at small relative
momentum in all four distributions.

\FloatBarrier

\section{Correlation function analysis}
\label{sec:correlation_function}
The two-particle correlation function is defined as the ratio of two-particle to single-particle momentum spectra:
\[ C\left(p^a, p^b\right) \equiv \dfrac{\left(\dfrac{dN^{ab}}{d^3p^a d^3p^b}\right)}{\left(\dfrac{dN^{a}}{d^3p^a}\right)\left(\dfrac{dN^{b}}{d^3p^b}\right)} \,,\]
for pairs of particles with four-momenta $p^a$ and $p^b$.
This definition has the useful feature that most single-particle efficiency, acceptance, and resolution effects cancel in the ratio.
The correlation function is expressed as a function of the relative momentum\footnote{While $q$ here refers to the relative four-momentum, it is also used generically to refer to either the Lorentz-invariant \qinv or three-vector $\mathbf{q}$.
The correlation function is studied in terms of both these variables but the description of the analysis is nearly identical for both cases.} $q \equiv p^a - p^b$ in intervals of average momentum $k \equiv \left(p^a + p^b\right)/2$.

The relative momentum distribution $A(q) \equiv \left. dN/dq \right|_{\textrm{same}}$ (\Fig{\ref{fig:pairqdists}}) is formed by selecting like-charge (or unlike-charge) pairs of particles from each event in an event class, which is defined by the collision centrality and $z$ position of the primary vertex ($z_\textrm{PV}$).
The combinatorial background $B(q) \equiv \left. dN/dq \right|_{\textrm{mix}}$ is constructed by event mixing, that is, by selecting one particle from each of two events in the same event class as $A(q)$.
Each particle in the background fulfills the same selection requirements as those used in the same-event distribution.
Event classes are categorized by centrality so that events are only compared to others with similar multiplicities and momentum distributions.
Events are sorted by $z_\textrm{PV}$ so that the background distribution is constructed with pairs of tracks originating from nearby space points, which is necessary for $B(q)$ to accurately represent the as-installed detector.
The $A(q)$ and $B(q)$ distributions are combined over $z_\textrm{PV}$ intervals in such a way that each of them samples the same $z_\textrm{PV}$ distribution.
The ratio of the distributions defines the correlation function:
\begin{equation}
C_{\mathbf{k}}(q) \equiv \frac{A_{\mathbf{k}}(q)}{B_{\mathbf{k}}(q)} \label{eq:same_over_mixed} \,.
\end{equation}

\subsection{Parameterization of the correlation function}
\label{subsec:parameterization}

Assuming that all particles are identical pions created in a fully chaotic source and that they have no final-state interaction, the enhancement in the correlation function is the Fourier transform of the source density.

The Bowler-Sinyukov formalism \cite{Bowler:1991vx,Sinyukov:1998fc} is used to account for final-state corrections:
\[ \label{eq:correlation_function}
C_\mathbf{k}(q) = (1-\lambda) + \lambda K(q)C_{\textrm{BE}}(q) \,,
\]
where $K$ is a correction factor for final-state Coulomb interactions, and $C_{\textrm{BE}}(q) = 1 + \mathcal{F} \left[ S_\mathbf{k} \right](q)$ with $\mathcal{F}\left[ S_\mathbf{k} \right](q)$ denoting the Fourier transform of the two-particle source density function $S_\mathbf{k}(r)$.
Several factors influence the value of the parameter $\lambda$.
Including non-identical particles decreases this parameter, as does coherent particle emission.
Products of weak decays or long-lived resonances also lead to a decrease in $\lambda$, as they are emitted at a length scale greater than can be resolved by femtoscopic methods given the momentum resolution of the detector.
These additional contributions to the source density are not Coulomb-corrected within the Bowler-Sinyukov formalism.
When describing pion pairs of opposite charge, there is no Bose-Einstein enhancement and $C_{\textrm{BE}} \rightarrow 1$.

Coulomb interactions suppress the correlation at small relative momentum for identical charged particles.
The particular choice of correction factor $K(q)$ is determined using the formalism in \Ref{\cite{Maj:2009ue}}.
This uses the approximation that the Coulomb correction is effectively applied not from a point source, but over a Gaussian source density of radius \Reff:
\begin{equation}
  K(\qinv) = G(\qinv) \left[ 1 + \frac{8\Reff}{\sqrt{\pi}a} {}_2 F_2 \left( \frac{1}{2}, 1; \frac{3}{2}, \frac{3}{2}; -\Reff^2 \qinv^2 \right) \right] \,,\label{eqn:kdef}
\end{equation}
where $a = 388~\textrm{fm}$ is the Bohr radius \cite{Bohr:1913} of a two-pion state, ${}_2 F_2$ is a generalized hypergeometric function, and $G(\qinv)$ is the Gamow factor \cite{PhysRevC.20.2267, Gyulassy:1980xb}
\[G(\qinv) = \frac{4\pi}{a \qinv} \frac{1}{\mathrm{e}^{4\pi/a \qinv} - 1} \; .\]

For opposite-charge pairs, $a$ is taken to be negative, since its definition includes a product of the two charges.

The Bose-Einstein enhancement in the invariant correlation functions is fit to an exponential form:
\begin{equation} C_{\textrm{BE}}(\qinv) = 1 + \mathrm{e}^{- \Rinv \qinv } \,, \label{eqn:cbe_qinv}\end{equation}
where \Rinv is the Lorentz-invariant HBT radius. This function corresponds to an underlying Breit-Wigner source density.

The Bose-Einstein component of the three-dimensional correlation functions is fit to a function of the form
\begin{equation}
C_{\textrm{BE}}(\mathbf{q}) = 1 + \mathrm{e}^{- \left\| R \mathbf{q} \right\|} \;, \label{eqn:cbe_qosl}
\end{equation}
where $R$ is a symmetric matrix of the form
\begin{align}
  R = \left( \begin{array}{ccc}
  \Rout & 0 & \Rol \\
  0 & \Rside & 0 \\
  \Rol & 0 & \Rlong
  \end{array} \right) \; . \label{eq:r_matrix}
\end{align}
The off-diagonal entries other than \Rol can be argued to vanish by the average azimuthal symmetry of the source.
In hydrodynamic models the out-long term \Rol is sensitive to spatio-temporal correlations, and therefore, to the lifetime of the source \cite{Chapman:1994yv,Bozek:2014fxa}. It couples radial and transverse expansion, and is expected to vanish in the absence of either. If the source is fully boost invariant then this term vanishes, so an observation of a nonzero value demonstrates that the homogeneity region is not boost invariant.

In order to reduce computational demands, a few symmetry arguments are considered. The order of the pairs is chosen such that \qout is always positive, which can be done so long as $C(-\mathbf{q}) = C(\mathbf{q})$. The average azimuthal symmetry of the source is invoked in order to allow only the absolute value of \qside to be considered. The sign of \qlong cannot be similarly discarded if a nonzero \Rol is allowed.

A Gaussian form for the Bose-Einstein enhancement is often used in the three-dimensional correlation function.
However, this form was found to give a poor description of ATLAS data, relative to an exponential form.
This was also observed in the ATLAS \pp results in \Ref{\cite{Aad:2015sja}}.
The chosen form of $\mathcal{F}[S_\mathbf{k}](\mathbf{q})$ must be taken into account when interpreting source radii, and there is no simple correspondence between parameters estimated using one form and those from another. An ad hoc factor of $\sqrt{\pi}$ is often invoked to relate Gaussian radii to exponential radii by assuming that the first $q$-moment of the invariant correlation function should be preserved, but this assumption is not rigorously justified and the argument fails in general for three-dimensional correlation functions.

\subsection{Hard-process contribution}
\label{subsec:hard_process}

Additional non-femtoscopic enhancements to the correlation functions at $\qinv \lesssim 0.5 - 1~\GeV$ are observed in both the opposite-charge ($+-$) and the same-charge ($\pm\pm$) pairs.
As discussed later in this section, the enhancement is more prominent in $+-$ pairs than it is in $\pm\pm$ pairs.
Monte Carlo (MC) generators do not simulate the final-state two-particle interactions used for femtoscopy, but they do describe the background correlations.
The MC generators used in this section to constrain the background description are described in \Sect{\ref{subsec:generators}}. 

\begin{figure}[b]
\centering
\subfloat[]{
\includegraphics[width=.49\linewidth]{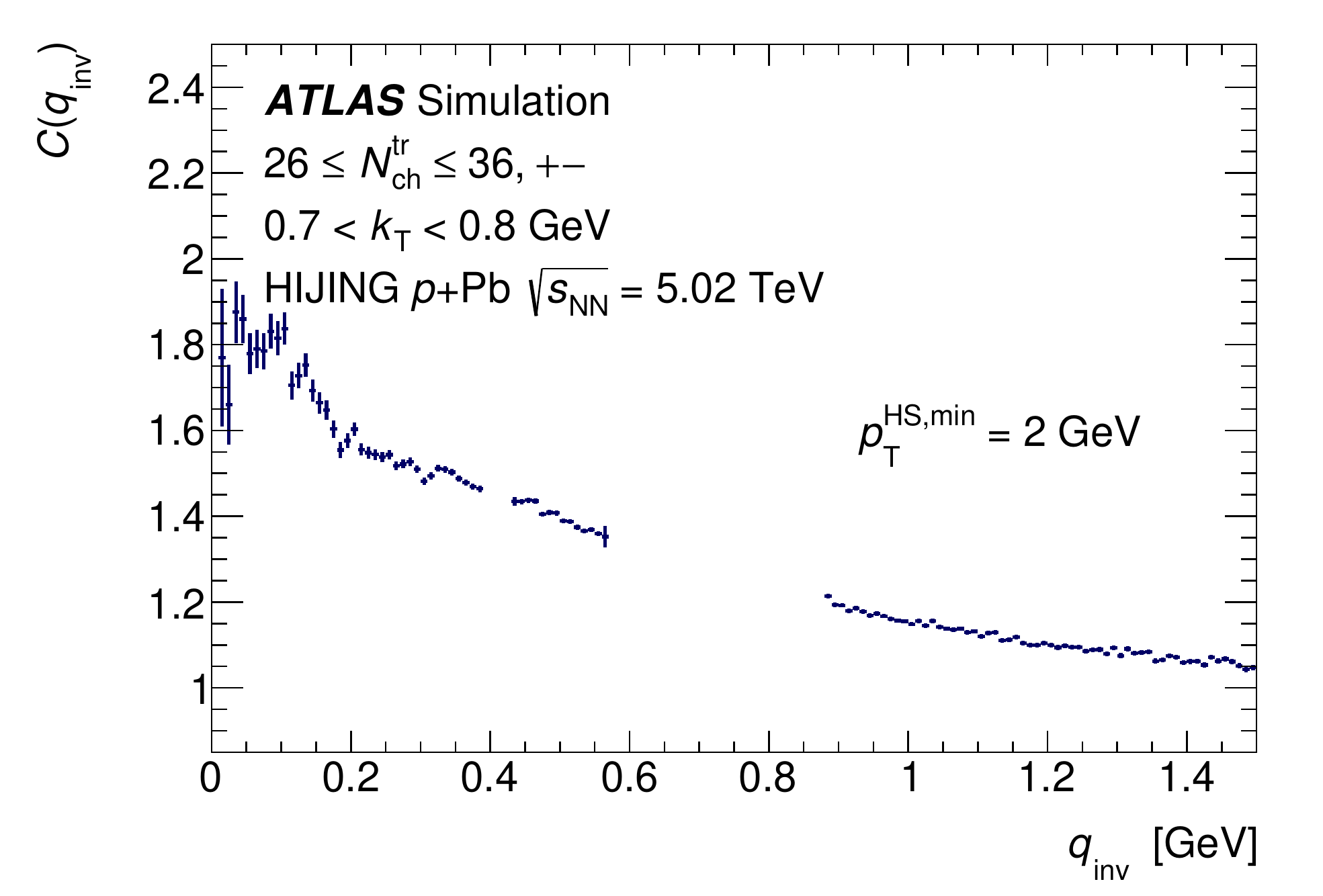}
}
\subfloat[]{
\includegraphics[width=.49\linewidth]{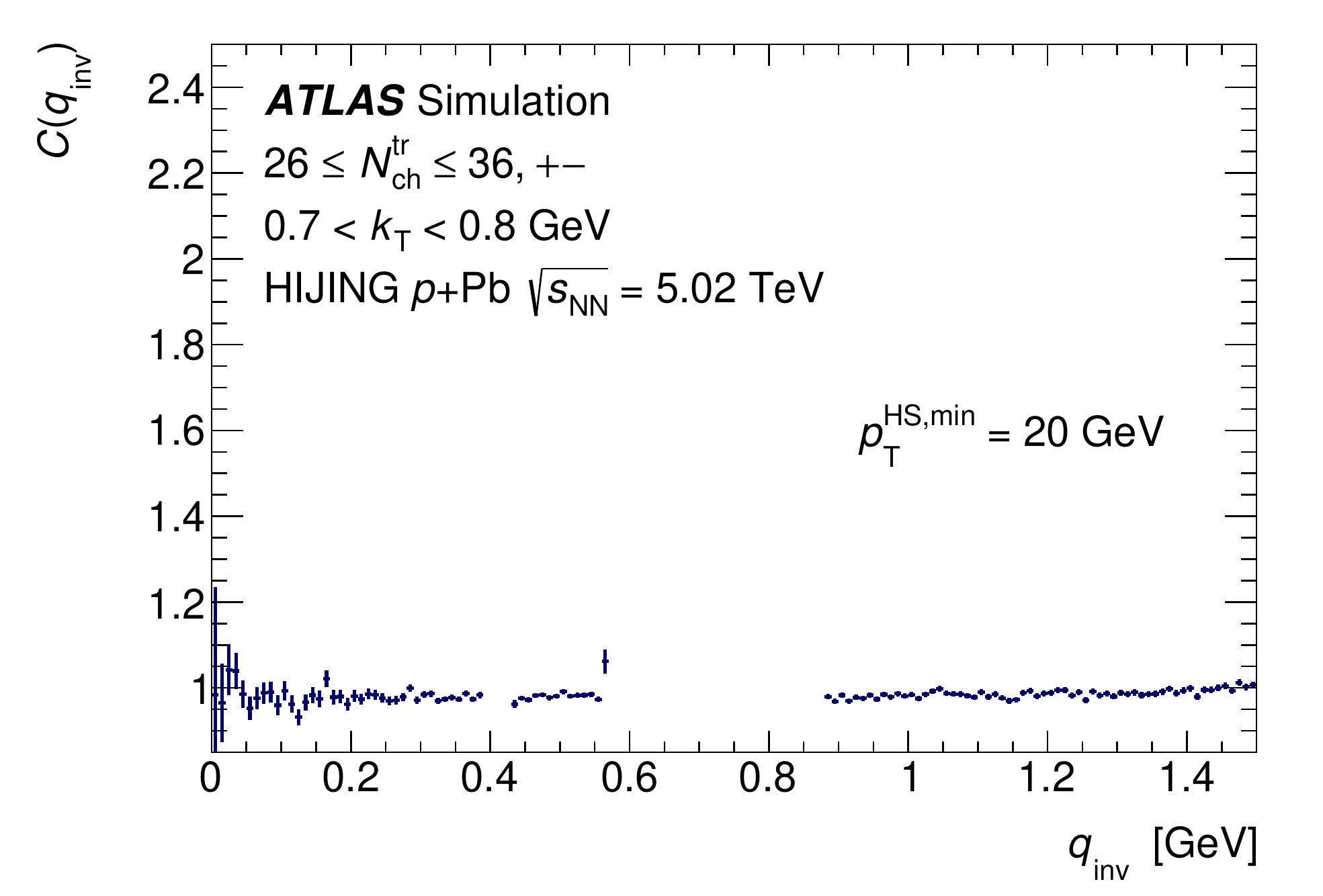}
}\\
\subfloat[]{
\includegraphics[width=.49\linewidth]{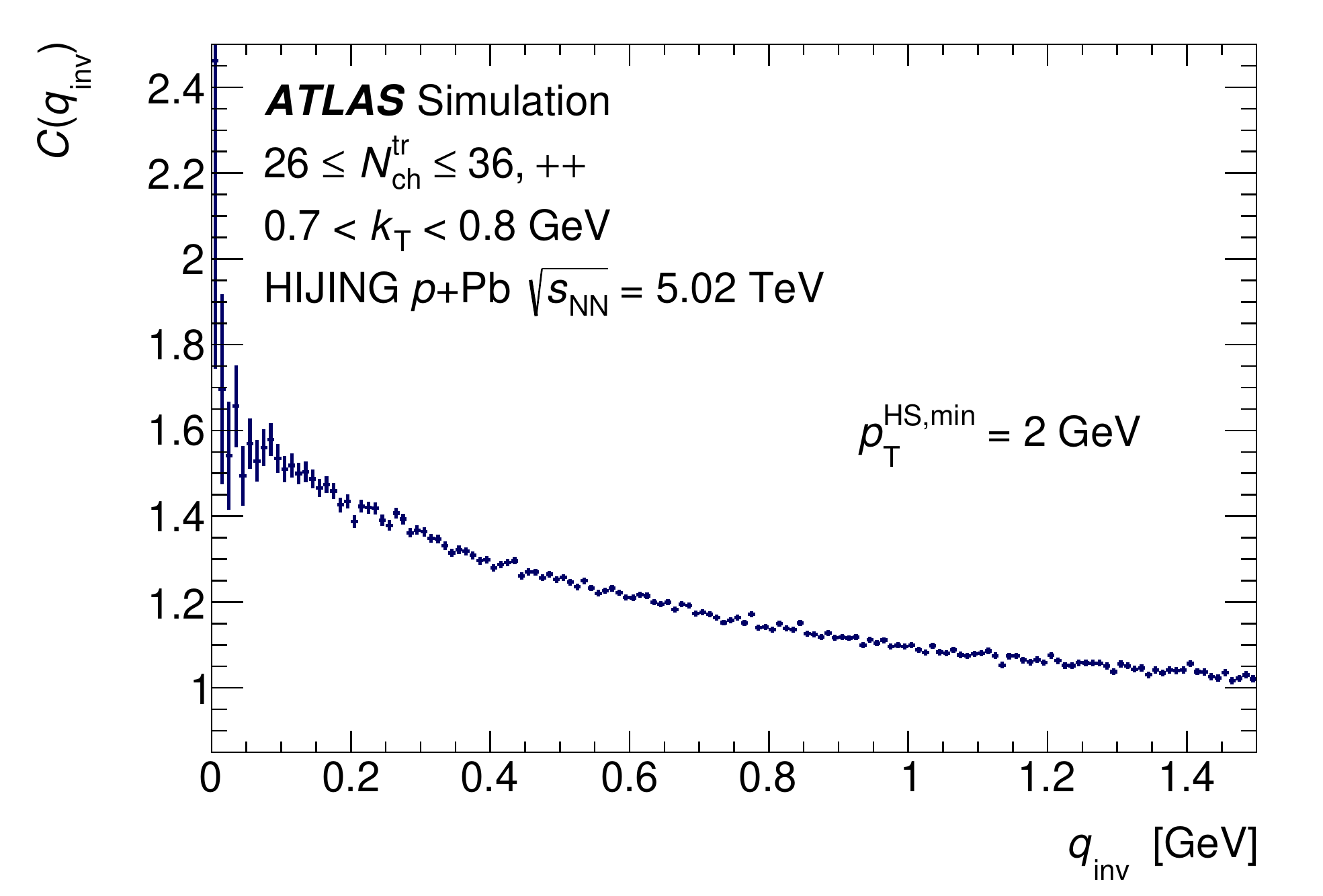}
}
\subfloat[]{
\includegraphics[width=.49\linewidth]{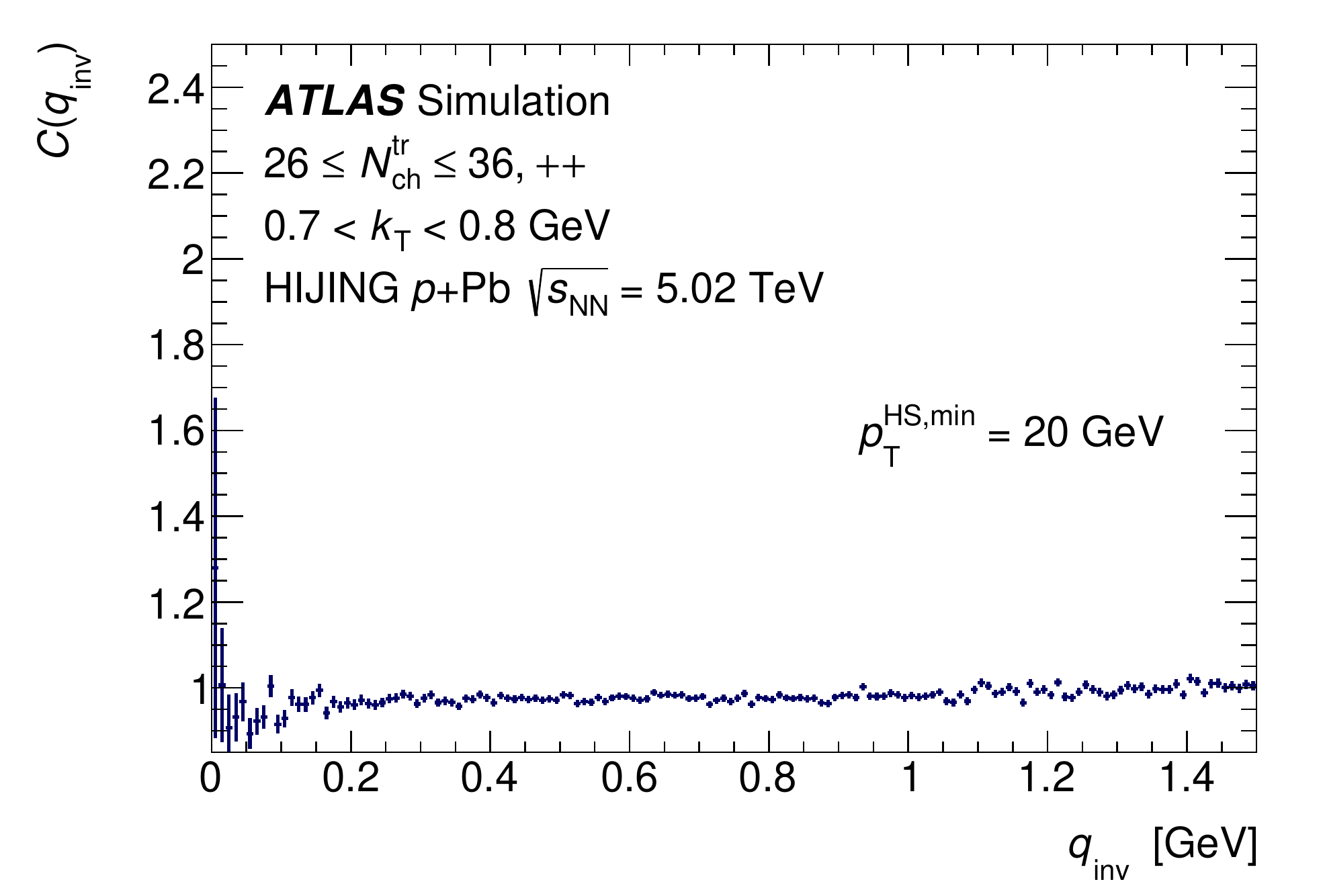}
}
\caption{Correlation functions of charged particles from \Hijing~for opposite- (a,b) and same-charge (c,d) pairs with transverse momentum $0.7 < \kt < 0.8$~\GeV, using events with a generated multiplicity $26 \leq N_\mathrm{ch}^\mathrm{tr} \leq 36$. The generator is run with the minimum hard-scattering $\pt^{\textrm{HS, min}}$ at the default setting of $2~\GeV$ (a,c) and increased to $20~\GeV$ (b,d) to remove the contribution from hard processes. The gaps in the opposite-charge correlation functions are a result of the requirements described in \Sect{\ref{subsec:pair}}, which remove the largest resonance contributions.}
\label{fig:hijing_min_hs_pt}
\end{figure}

The non-femtoscopic enhancement is more prominent for higher \kt and lower multiplicities.
This suggests that the correlation is primarily due to jet fragmentation.
This hypothesis is verified by studying correlation functions in \Hijing, by increasing the minimum hard-scattering \pt ($\pt^{\textrm{HS,min}}$) from 2 to 20~\GeV.
Increasing $\pt^{\textrm{HS,min}}$ has the effect of suppressing most hard processes in typical events.
Without the resulting jet fragmentation, the non-femtoscopic enhancement is removed from the correlation function, as demonstrated in \Fig{\ref{fig:hijing_min_hs_pt}} by comparing the panels on the left and right of each row.

The amplitude of the hard-process contribution tends to be larger in the Monte Carlo events than it is in the data.
Thus, attempting to account for it by studying the double-ratio $C^{\textrm{data}}(q) / C^{\textrm{MC}}(q)$ leads to a depletion that is apparent in the region where the Bose-Einstein enhancement disappears \cite{Aad:2015sja}.
Another commonly used method is to parameterize the mini-jet contribution using simulation and to allow one or more parameters of the description to vary in the fit \cite{Abelev:2014pja, PhysRevC.91.034906}.

To avoid too much reliance on either a full MC description or arbitrary additional free parameters, a data-driven method is derived here to constrain the correlations from jet fragmentation.
Opposite-charge correlation functions are used to predict the jet contribution in the same-charge correlation function.
This poses two challenges.
Firstly, resonance decays appear prominently in the opposite-charge correlations.
The most prominent of these are removed by requirements on the invariant mass of the opposite-charge pairs (as described in \Sect{\ref{subsec:pair}}), and the fits to the opposite-charge correlation functions are restricted to $\qinv > 0.1~\GeV$.
The lower bound on the domain of the fit reduces sensitivity to effects such as three-body decays that are unrelated to jet fragmentation, which is significant over a broader range of $q$.
Secondly, jet fragmentation does not affect opposite-charge and same-charge correlations in an identical manner.
This is in part because opposite-charge pairs are more likely to have a closer common ancestor in a jet's fragmentation into hadrons.

To account for the remaining differences between $+-$ and $\pm\pm$ pairs, a study of both classes of correlation functions is performed in \PYEight.
In order to isolate the effect of jet fragmentation, decays from the relatively longer-lived particles $\eta$, $\eta'$, and $\omega$ are excluded.
Pairs of particles from two-body resonance decays are also neglected, in order to remove mass peaks in the correlation function.
The same-pair mass cut around the $\rho$ resonance that is used in the data is also applied in \PYEight events, since the removal of the corresponding region of phase space has a significant effect on the shape of the correlation function.

\subsubsection{Jet fragmentation in \qinv}
\label{subsubsec:invariant_bkgd}

To describe the jet fragmentation in the invariant correlation functions, fits are performed in \PYEight \pp to a stretched exponential function of the form
\begin{equation}
\Omega\left(\qinv\right) = \mathcal{N} \left(1 + \lbkgdinv \mathrm{e}^{-|\Rbkgdinv \qinv|^{\abkgdinv}} \right) \,,\label{eq:bkgd_form}
\end{equation}
where $\mathcal{N}$ is a normalization factor and the other parameters depend on the charge combination and on \kt.
The $\Omega(\qinv)$ function above is applied as a multiplicative factor to the femtoscopic correlation function. The strategy employed is to estimate these parameters for same-charge correlation functions based on values determined using opposite-charge correlations.
First, the shape parameter \abkgdinv is determined with fits to same-charge correlation functions, with all parameters allowed to be free.
It is only weakly dependent on multiplicity, so a function is fit to parameterize \abkgdinv in \PYEight as a function of \kt (with \kt in \GeV):
\[
\abkgd^{\mathrm{inv}} \left(\kt\right) = 2 - 0.050 \ln\left(1 + \mathrm{e}^{50.9(\kt - 0.49)}\right) \, .
\]
The fits are well described by a Gaussian form ($\abkgdinv = 2$) at $\kt \lesssim 0.4~\GeV$, and \abkgdinv decreases to a value around $1.3$ in the highest \kt interval.

\begin{figure}[t]
\centering
\subfloat[]{
\includegraphics[width=0.49\linewidth]{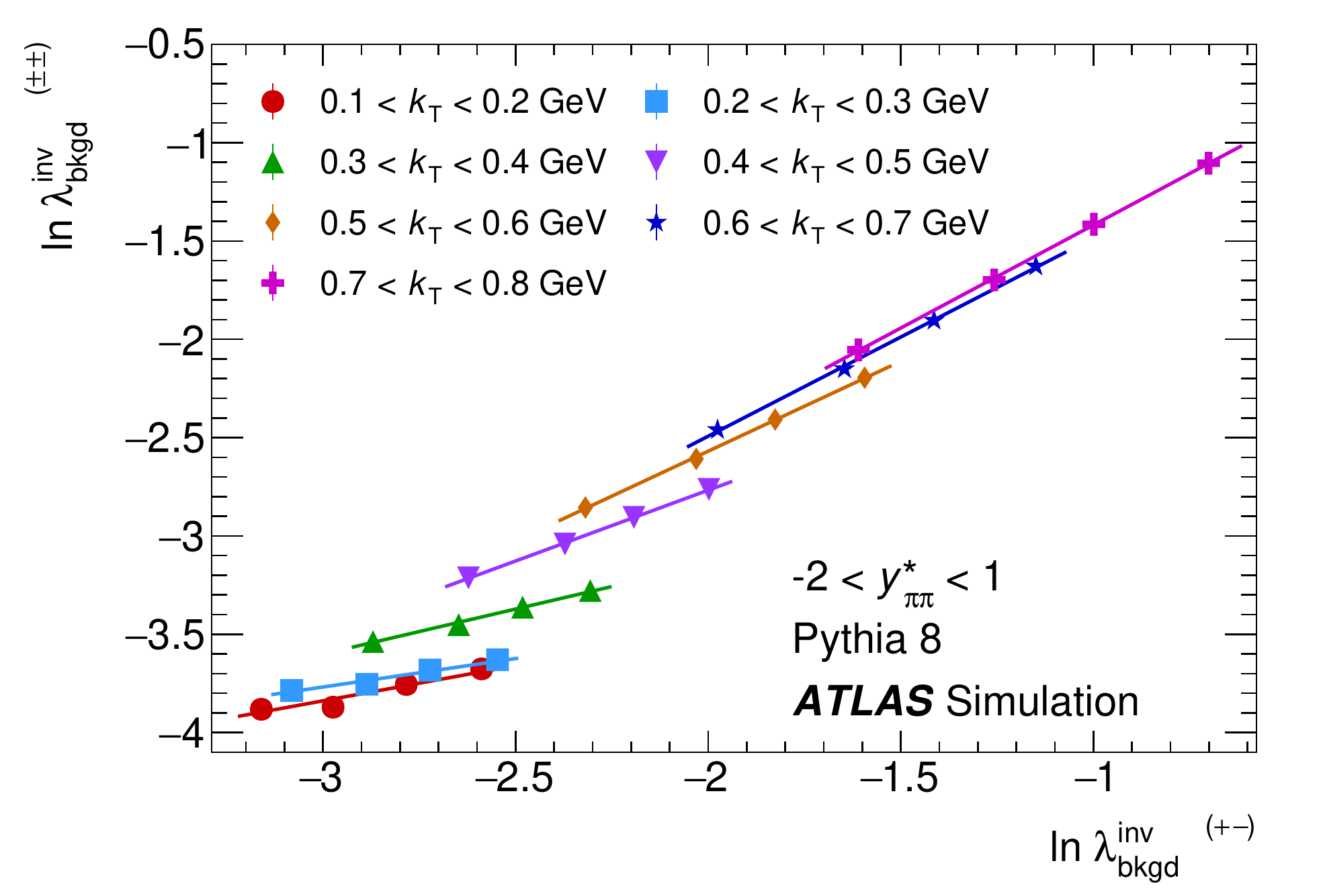}
}
\subfloat[]{
\includegraphics[width=0.49\linewidth]{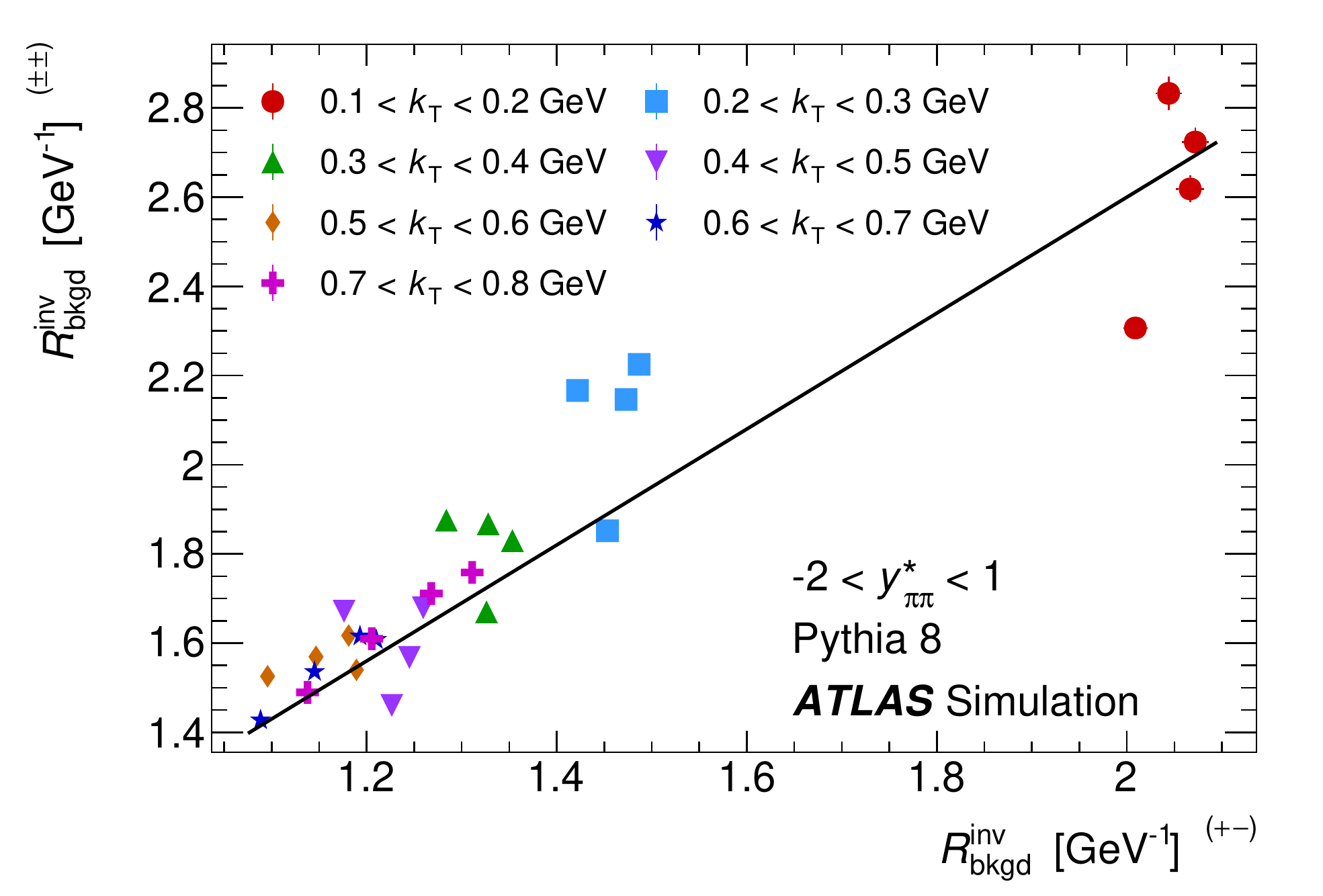}
}
\caption{Comparison of jet fragmentation parameters between opposite- and same-charge correlation functions. The amplitude is shown in (a), and the width is shown in (b). The lines are fits of the data to \Eqnrange{\ref{eqn:rBackPar}}{\ref{eq:lambdaBackPar}}. For each \kt interval, four multiplicity intervals are shown ($26 \leq \Nch \leq36$, $37 \leq \Nch \leq48$, $49 \leq \Nch \leq64$, and $65 \leq \Nch$).}
\label{fig:inv_bkgd_charge_comp}
\end{figure}

The fits are performed again to the \PYEight correlation functions, with \abkgdinv now fixed to the same value in same- and opposite-charged pairs, and a comparison is made between the width parameters $\Rbkgdinv{}^{+-}$ and $\Rbkgdinv{}^{\pm\pm}$.
The width of the jet fragmentation correlation for same-charge pairs is found to be correlated to that for opposite-charge pairs, as shown in the right plot of \Fig{\ref{fig:inv_bkgd_charge_comp}}. Four intervals of charged particle multiplicity, \Nch, calculated for particles with $\pt > 100~\MeV$ and $|\eta| < 2.5$ are shown: $26 \leq \Nch \leq36$, $37 \leq \Nch \leq48$, $49 \leq \Nch \leq64$, and $65 \leq \Nch$. The relationship between the invariant background widths is modeled as a direct proportionality,
\begin{equation}
\Rbkgdinv{}^{\pm\pm} = \rho \Rbkgdinv{}^{+-} \,,\label{eqn:rBackPar}
\end{equation}
with a value of $\rho = 1.3$ extracted from \PYEight.
This proportionality begins to break down at low \kt, but the model becomes increasingly accurate at larger \kt, where hard processes give a larger contribution to the correlation function.

Next, $\Rbkgdinv{}^{\pm\pm}$ is fixed from $\Rbkgdinv{}^{+-}$ using the value of $\rho$, and the fits are performed again to parameterize the relationship between the amplitudes,
\begin{equation}
\lbkgdinv{}^{\pm\pm} = \mu(\kt) \left(\lbkgdinv{}^{+-}\right)^{\nu(\kt)} \,.\label{eq:lambdaBackPar}
\end{equation}

As shown in the left-hand plot of \Fig{\ref{fig:inv_bkgd_charge_comp}}, $\mu$ and $\nu$ are fit in each \kt interval to describe four multiplicity intervals.
The power-law scaling of \Eqn{\ref{eq:lambdaBackPar}} is found to provide a good description of the relation between the same- and opposite-charge amplitudes across all four multiplicity intervals studied.
The multiplicity-independence of $\mu$ and $\nu$ is important in justifying the use of these parameters in \pPb.

The correspondence between opposite- and same-charge pairs in both \pp and \pPb systems is studied in \Hijing, since the study described in this section is performed with \PYEight in a \pp system.
While the mapping is mostly consistent between the two systems, it is found that $\mu$ is larger in \pPb than in \pp by $8.5 \%$ on average.
When the mapping is applied to the data, this attenuation factor (along with a corresponding systematic uncertainty described in \Sect{\ref{sec:systematics}}) is also taken into account.

\begin{figure}[!ht]
\centering
\includegraphics[width=0.5\linewidth]{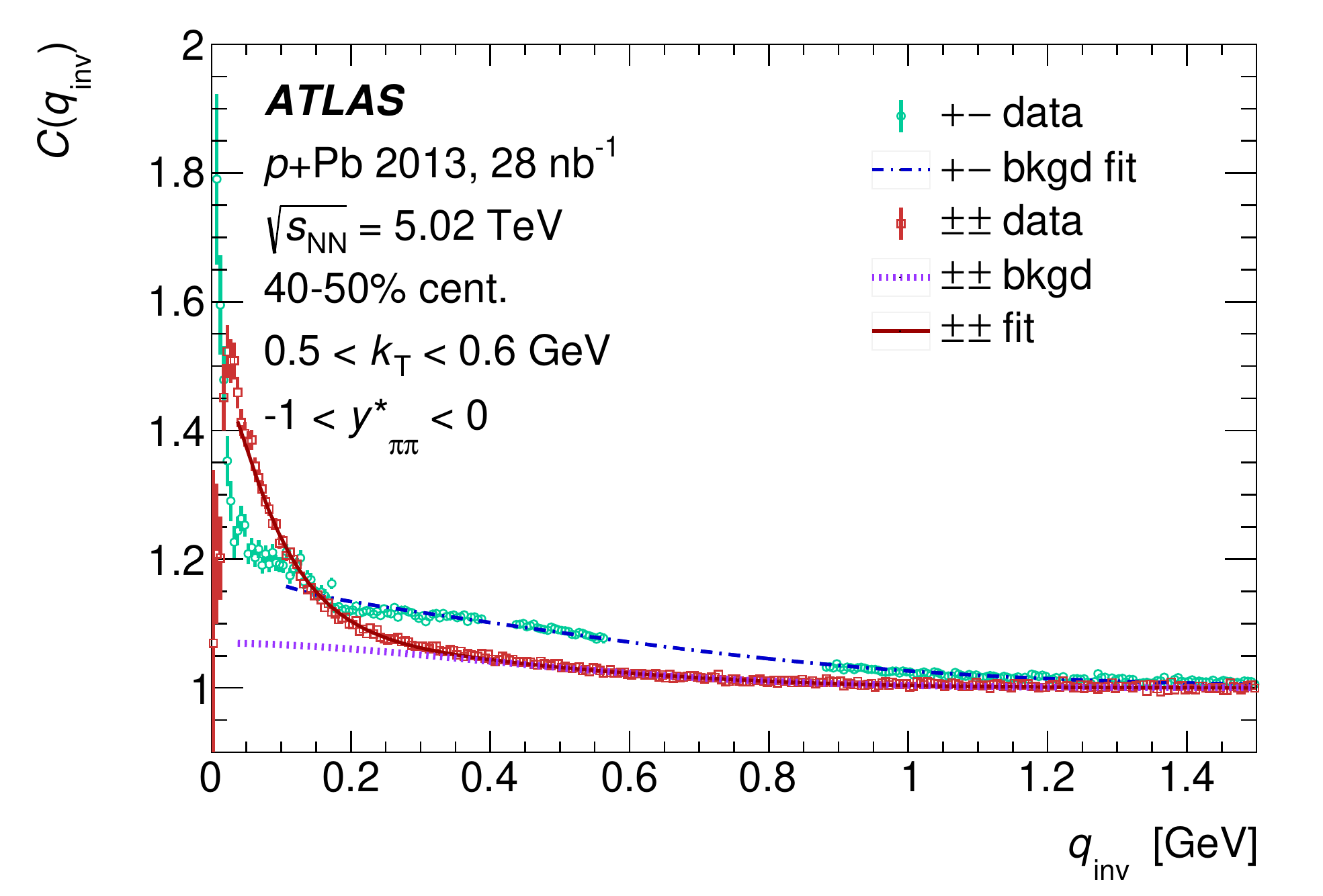}
\caption{Correlation functions in \pPb data for opposite-charge (teal circles) and same-charge (red squares) pairs. The opposite-charge correlation function, with the most prominent resonances removed, is fit to a function of the form in \Eqn{\ref{eq:bkgd_form}} (blue dashed line). The violet dotted line is the estimated jet contribution in the same-charge correlation function, also of the form of \Eqn{\ref{eq:bkgd_form}}, and the dark red line is the full fit of \Eqn{\ref{eq:correlation_function_full}} to the same-charge data.}
\label{fig:hp_example}
\end{figure}

With $\abkgdinv(\kt)$, $\mu(\kt)$, $\nu(\kt)$, and $\rho$ determined from Monte Carlo generator samples, the mapping can be applied to the \pPb data.
As illustrated in \Fig{\ref{fig:hp_example}}, the $+-$ correlation function is fit to \Eqn{\ref{eq:bkgd_form}} for $\qinv > 0.1~\GeV$, with \abkgd fixed from \PYEight and $\lbkgdinv{}^{+-}$ and $\Rbkgdinv{}^{+-}$ as free parameters.
The $\mu$, $\nu$, and $\rho$ parameters are used to infer $\lbkgdinv{}^{\pm\pm}$ and $\Rbkgdinv{}^{\pm\pm}$, which are fixed before the femtoscopic part of the correlation function is fit to $\pm\pm$ data.

\subsubsection{Jet fragmentation in three dimensions}
\label{subsubsec:osl_bkgd}

In the longitudinally comoving frame of a particle pair produced in a jet, the axis of the jet is aligned on average with the ``out'' direction and the plane transverse to the jet's momentum is spanned by the ``side'' and ``long'' directions. In three dimensions the correlation from jet fragmentation is factorized into components which separately describe the ``out'' direction and both the ``side'' and ``long'' directions:
\begin{align}
\Omega(\mathbf{q}) = 1 + \lbkgdosl \exp\left(-\left| \Rbkgdout \qout \right|^{\abkgdout} - \left| \Rbkgdsl \qsl \right|^{\abkgdsl}\right) \label{eqn:jet_frag_form_3d}
\end{align}
where $\qsl = \sqrt{\qside^2 + \qlong^2}$, \lbkgdosl is the background amplitude, and \abkgdout and \abkgdsl parameterize the shape of the fragmentation contribution along and transverse to the jet axis, respectively.
The shape parameters \abkgdout and \abkgdsl are taken from \PYEight and fixed to 1.5 and 1.7 respectively.
Fits of these parameters to \PYEight correlation functions are not fully consistent with these chosen numerical constants at all \kt and multiplicities.
However, the impact of the somewhat arbitrary choice of fixing these parameters is tested by varying them both by 0.1, and the changes in the results are less than 1\%.

\begin{figure}[!ht]
\centering
\subfloat[]{
\includegraphics[width=0.49\linewidth]{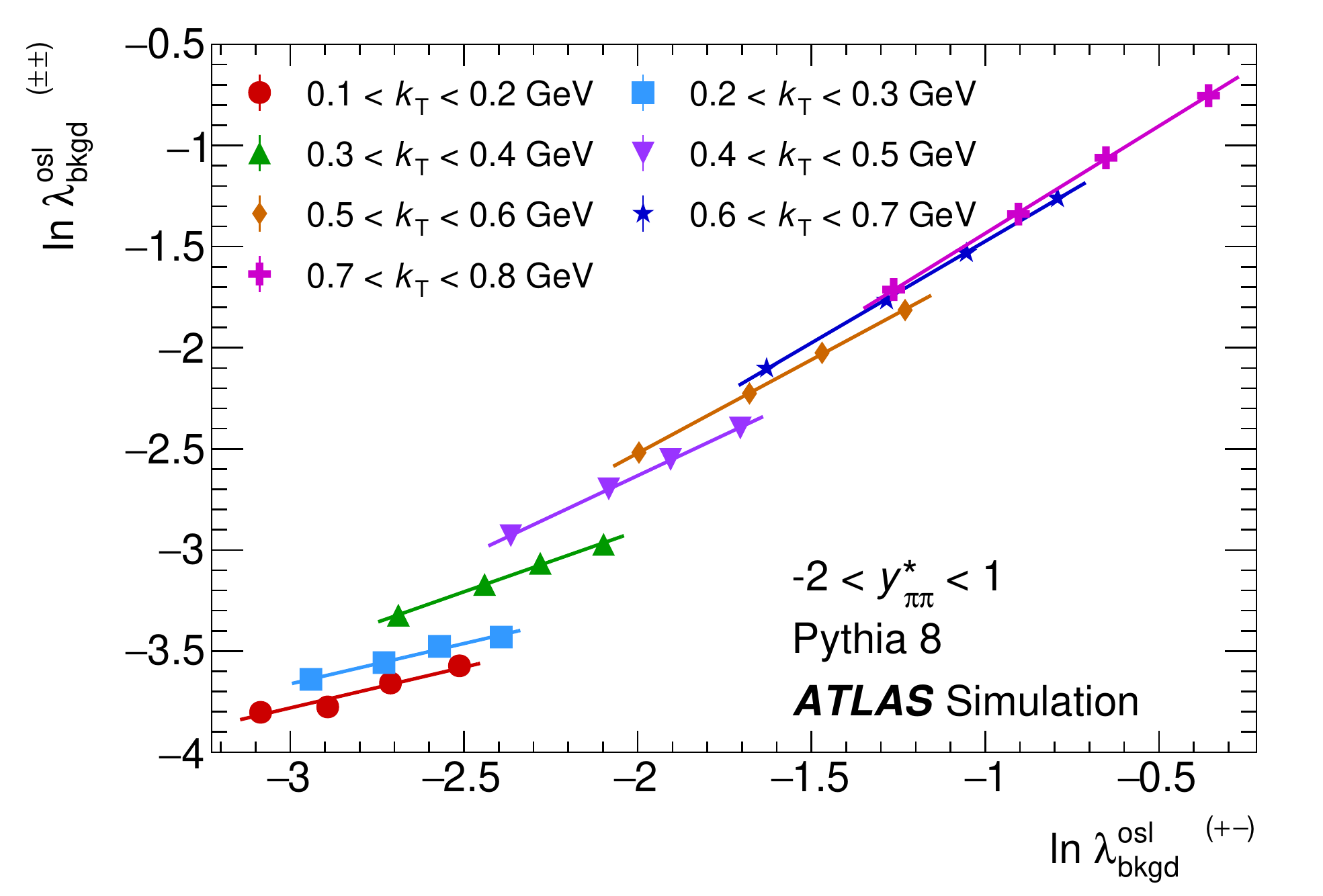}
}\\
\subfloat[]{
\includegraphics[width=0.49\linewidth]{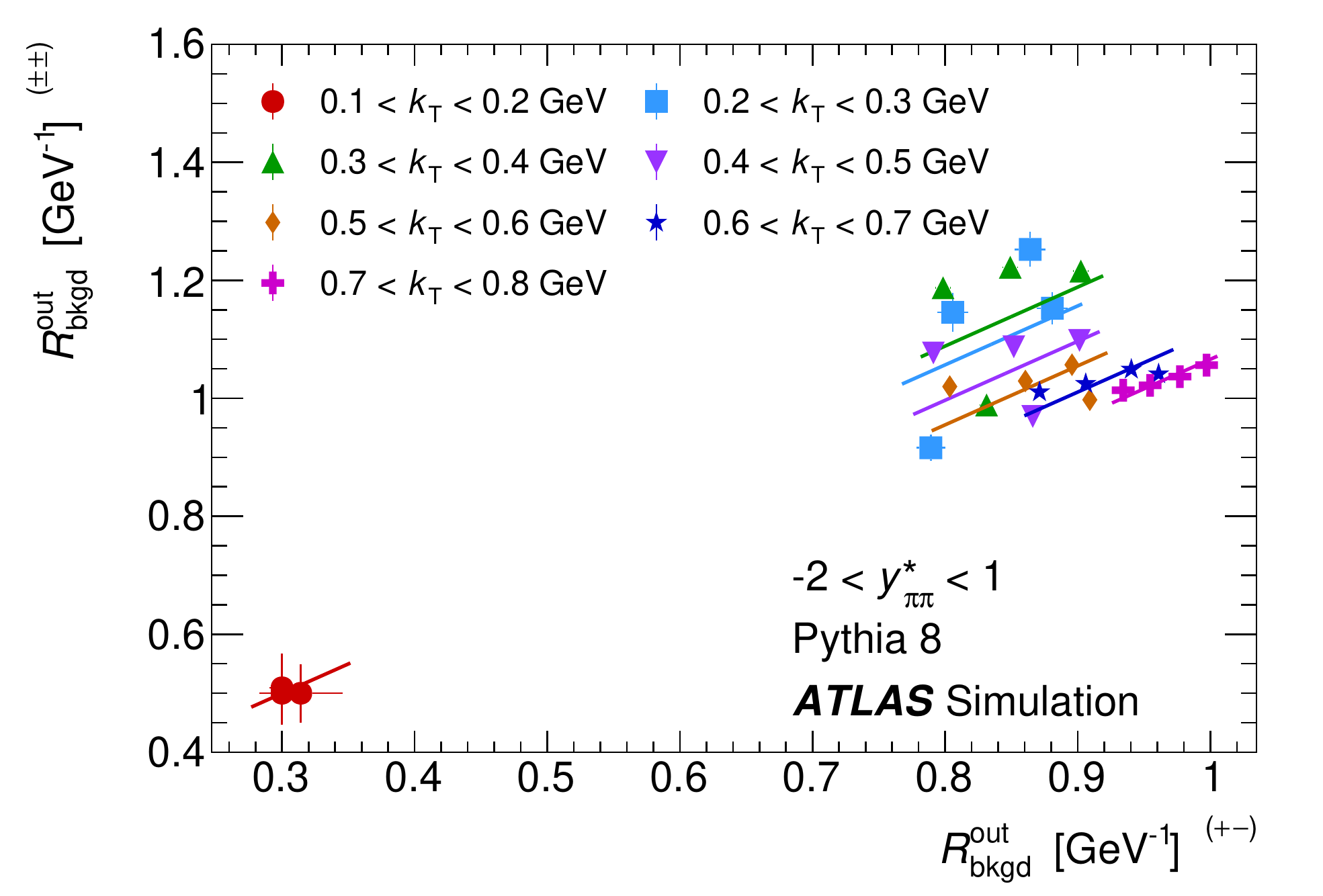}
}
\subfloat[]{
\includegraphics[width=0.49\linewidth]{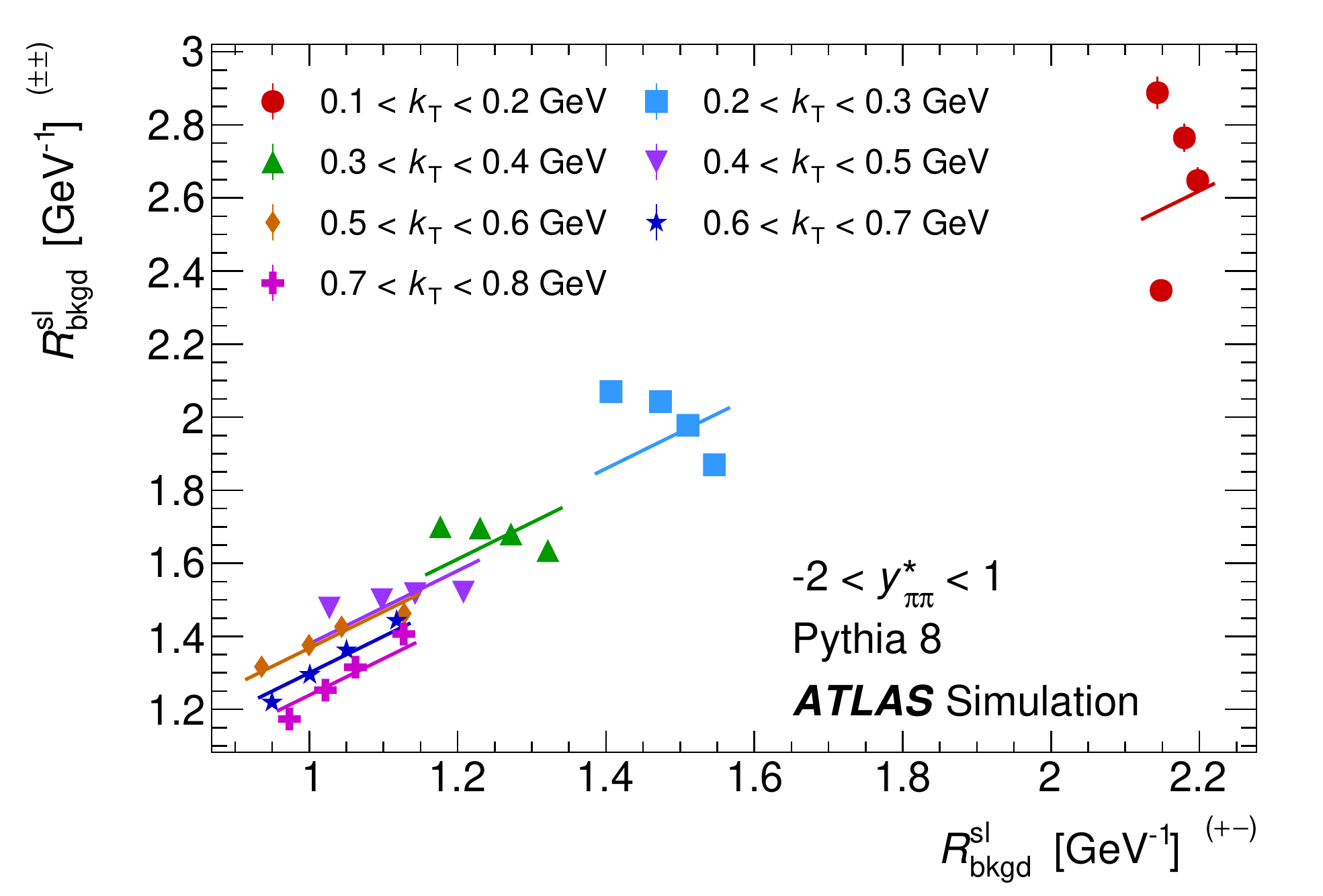}
}
\caption{Comparison of jet fragmentation parameters between opposite- and same-charge correlation functions. The amplitude is shown in (a), and the two widths in (b,c). The lines are fits of the data to \Eqnrange{\ref{eqn:lambdaBackParOSL}}{\ref{eqn:rBackParSL}}. For each \kt interval, four multiplicity intervals are shown ($26 \leq \Nch \leq36$, $37 \leq \Nch \leq48$, $49 \leq \Nch \leq64$, and $65 \leq \Nch$).}
\label{fig:osl_bkgd_charge_comp}
\end{figure}

Similarly to the procedure used for the \qinv correlation functions, the width parameters are compared between opposite- and same-charge correlation functions (bottom panels of \Fig{\ref{fig:osl_bkgd_charge_comp}}); however, in three dimensions the relationships are parameterized as a function of \kt. Next, as for \qinv, the amplitudes for three-dimensional jet correlations are compared between opposite- and same-charge pairs (top panel of \Fig{\ref{fig:osl_bkgd_charge_comp}}).
While the relationships between opposite- and same-charge correlations are not well described everywhere by the fitted lines, the model becomes increasingly accurate at larger \kt, where hard processes give a larger contribution to the correlation function.
The functional forms of the mappings from opposite- to same-charge three-dimensional parameters are:
\begin{align}
\lbkgdosl{}^{\pm\pm} &= \mu\left(\kt\right) \left(\lbkgdosl{}^{+-}\right)^{\nu(\kt)} \label{eqn:lambdaBackParOSL}\\
\Rbkgdout{}^{\pm\pm} &= \Rbkgdout{}^{+-} + \Delta\Rbkgdout\left(\kt\right) \label{eqn:rBackParOut}\\
\Rbkgdsl{}^{\pm\pm} &= \Rbkgdsl{}^{+-} + \Delta\Rbkgdsl\left(\kt\right) \,. \label{eqn:rBackParSL}
\end{align}
The widths used in the three-dimensional background description are related by a \kt-dependent additive factor [ \Eqns{\ref{eqn:rBackParOut}}{\ref{eqn:rBackParSL}}] because a simple proportionality [ \Eqn{\ref{eqn:rBackPar}}] is not as successful in describing the behavior.

The invariant and three-dimensional (3D) fragmentation amplitudes are strongly correlated and the mappings from opposite- to same-charge correlation functions are quantitatively similar. Thus, the same 8.5\% attenuation factor for $\mu$ derived from \Hijing for the invariant mapping is used for the three-dimensional fits as well.

The numerical values used for mapping the amplitude \lbkgdosl, fragmentation width colinear with the jet axis \Rbkgdout, and fragmentation width transverse to the jet axis \Rbkgdsl are given by the following parameterizations (\kt in \GeV\ and \Rbkgd in $\GeV^{-1}$):
\begin{align}
\ln \mu \left(\kt\right) &= -3.9 + 9.5 \kt - 6.4 \kt^2 \label{eqn:mu_osl}\\
\nu \left(\kt\right) &= 0.03 + 2.6 \kt - 1.6 \kt^2 \label{eqn:nu_osl}\\
\Delta \Rbkgd^\mathrm{out} \left(\kt\right) &= 0.43  - 0.49 \kt \label{eqn_rbkgdout}\\
\Delta \Rbkgd^\mathrm{sl} \left(\kt\right) &= \frac{0.51}{1 + \left(1.30 \kt\right)^2} \, . \label{eqn_rbkgdsl}
\end{align}

The jet fragmentation parameters of the \pPb data depend on centrality and \kt.
The same-charge amplitude of the background ranges from being negligible at low \kt up to a maximum of roughly 0.25 at the largest measured \kt of 0.8 \GeV\ for the most peripheral events.
The widths of the same-charge fragmentation correlation have length scales which are typically in a range of 0.3--0.5 fm at the largest \kt where they are most relevant.
The \pPb femtoscopic measurement is most challenging at high \kt in peripheral events, where the fragmentation background amplitude is a significant fraction of the Bose-Einstein amplitude and the HBT radii are smaller and closer in magnitude to the length scale of the jet correlation.

\subsection{Fitting procedure}
\label{subsec:fit_procedure}

The bin contents of the histogram representations of $A(q)$ and $B(q)$ are assumed to be Poisson distributed. The correlation function $C(q)$ is assumed to be fit best by the ratio of their means, and a flat Bayesian prior is assumed for both means.
A corresponding $\chi^2$ analog \cite{Baker:1983tu}, the negative log-likelihood ratio $\mathcal{L}$ \cite{PhysRevC.66.054906}, is minimized using the Minuit package \cite{James:1975dr}:
\begin{align}
-2 \ln \mathcal{L} = 2\sum_{i} \left\{ A_i \ln \left[\frac{(1+C_i)A_i}{C_i(A_i+B_i+2)} \right] + (B_i+2) \ln \left[ \frac{(1+C_i)(B_i+2)}{A_i+B_i+2} \right] \right\} \,. \label{eq:loglikedef}
\end{align}
Here $A$ and $B$ are the signal and background relative momentum distributions in \Eqn{\ref{eq:same_over_mixed}} when represented as histograms, such that $A_i$ and $B_i$ are the contents in bin $i$, $C_i$ is shorthand for $C(q_i)$ where $q_i$ is the bin center, and $C(q)$ is the fitting function describing the correlation.
The multiplicative factor of $-2$ causes this statistic to approach $\chi^2$ as the sample size increases.
The $1\sigma$ statistical uncertainties in the fit parameters of $C(q)$ are evaluated using the MINOS routine, and are selected from the points in the parameter space where $-2 \ln \mathcal{L} = \textrm{min}\left(-2 \ln \mathcal{L}\right) + 1$.

The full form of the invariant-correlation-function fit to like-charge track pair data including the hard-process background description is
\begin{equation}
C(q) = \mathcal{N} \left[1-\lambda + \lambda K(\qinv)C_{\textrm{BE}}(q)\right] \Omega(q) \,,\label{eq:correlation_function_full}
\end{equation}
where $C_\mathrm{BE}(q)$ is given by \Eqn{\ref{eqn:cbe_qinv}} or \Eqn{\ref{eqn:cbe_qosl}}, $K(\qinv)$ is given by \Eqn{\ref{eqn:kdef}}, and $\Omega(q)$ is given by \Eqn{\ref{eq:bkgd_form}} or \Eqn{\ref{eqn:jet_frag_form_3d}}.

As discussed in \Sect{\ref{subsec:hard_process}}, the opposite-charge correlation functions are fit in the regions where \qinv (or $|\mathbf{q}|$ in 3D) is greater than 100~\MeV. The opposite-charge parameters are highly insensitive to the choice of cutoff, as the $q$ distributions contribute more statistical weight at larger $q$. The same-charge correlation functions are fit in the regions $\qinv > 30~\MeV$ for the invariant fits and $|\mathbf{q}| > 25~\MeV$ in three dimensions.

\FloatBarrier

\section{Systematic uncertainties}
\label{sec:systematics}

\subsection{Sources of systematic uncertainty}
\label{subsec:syst_description}

The systematic uncertainties in the extracted parameter values have contributions from several sources: the jet fragmentation description, PID, the effective Coulomb-correction size \Reff, charge asymmetry, and particle reconstruction effects.

One of the largest sources of uncertainty originates from the description of the background correlations $\Omega(q)$ from jet fragmentation.
For the uncertainty in the hard-process contribution, three effects are considered.
First, the extrapolation from a \pp to a \pPb system is represented with an uncertainty in the background amplitude.
Also, to investigate the uncertainty in the Monte Carlo description of jet fragmentation, the amplitude of $C^{+-}(\qinv) \big/ C^{\pm\pm}(\qinv)$ is studied in both \Pythia and \Herwig.
\Herwig does not predict enough difference between $+-$ and $\pm\pm$ correlations to describe the data.
Thus, instead of using the ratio of the predicted scalings of the two generators, the standard deviation of the ratio amplitude (across a selection of \kt and multiplicity intervals) is used as a variation reflecting this systematic uncertainty.
The hard-process amplitude \lbkgd is scaled up and down by 12.3\%, the quadrature sum of the relative variation from the difference between the \pp and \pPb systems (4.1\%) and from the generator difference (11.6\%).
The widths of the background description are highly correlated with the amplitude in the \Pythia fit results, so varying the widths in addition to the amplitude would overstate the uncertainty.
The choice of varying the amplitude instead of the width is found to provide a larger and more consistent variation in the radii, so only the amplitude of the background is varied.
The variation from the combination of the generator and the collision system are indicated by a label of ``Gen $\oplus$ Sys'' in the figures of \Sect{\ref{subsec:syst_plots}}.
Additionally, the procedure described in \Sect{\ref{subsec:hard_process}} to control the jet fragmentation correlations is repeated in both the central ($|\kys| < 1$) and forward ($-2 < \kys < -1$) rapidity intervals. While the relationship between the fragmentation widths (\Rbkgdinv, \Rbkgdsl, and \Rbkgdout) is fairly robust, the mappings of the amplitudes (\lbkgdinv and \lbkgdosl) from opposite- to same-charge correlations vary between the two rapidity intervals.
This variation represents the breakdown of the assumptions used to describe the jet fragmentation.
The mapping procedure is repeated with the results from each rapidity interval, and the variation is used as an additional systematic uncertainty in the amplitude.
The HBT radii and amplitudes are both highly correlated with the amplitude of the jet background, so varying \lbkgd is a robust method of evaluating the uncertainties from the background description procedure.
This systematic variation is represented by the ``Jet \kys'' label in the figures of \Sect{\ref{subsec:syst_plots}}.

The analysis is repeated at both a looser and a tighter PID selection than the nominal definition, and the variations are included as a systematic uncertainty.
The effect on the radii is at the 1--2\% level for the lower \kt intervals, but becomes more significant at higher momentum, where there are relatively more kaons and protons and the \dEdx separation is not as large. In the highest \kt intervals studied, variations are typically in the range of 5--30\%.
The PID systematic variation is labeled by ``PID'' in the figures of \Sect{\ref{subsec:syst_plots}}.

The nonzero effective size of the Coulomb correction \Reff should only cause a bin-by-bin change of a few percent in the correlation function, even with a value up to several femtometers, since the Bohr radius of pion pairs is nearly 400 fm.
However, since this parameter changes the width in \qinv over which the Coulomb correction is applied, varying this parameter can affect the source radii measurably.
The effective size is assumed to scale with the size of the source itself, so a scaling constant $\xi$ is chosen such that $\Reff = \xi \Rinv$.
 The nominal value of $\xi$ is taken to be equal to 1 and the associated systematic uncertainty is evaluated by varying this between 1/2 and 2.
The Coulomb size systematic variation is indicated by a label of ``\Reff'' in the figures of \Sect{\ref{subsec:syst_plots}}.

\begin{figure}[tb]
\centering
\subfloat[]{
\includegraphics[width=0.49\linewidth]{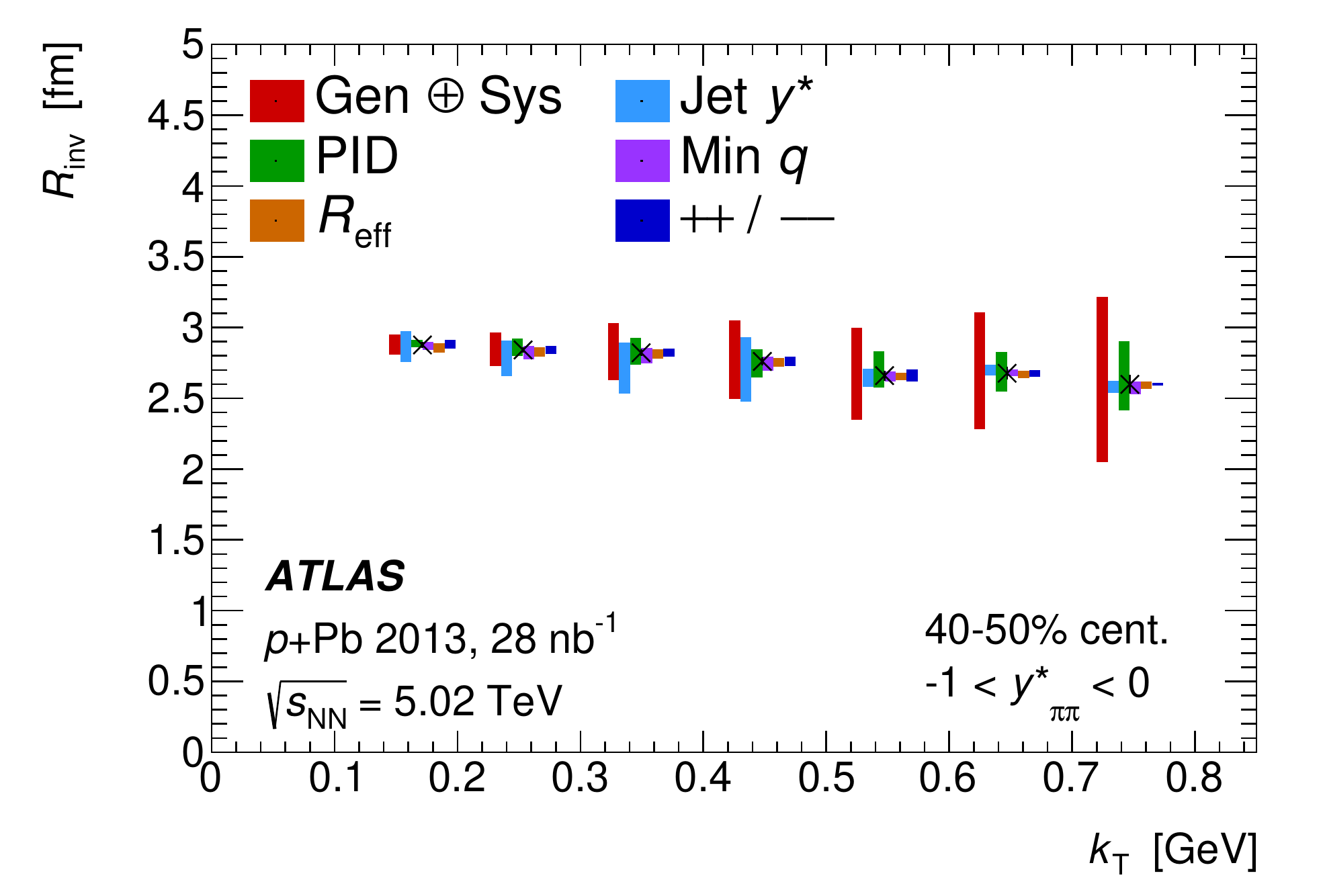}
}
\subfloat[]{
\includegraphics[width=0.49\linewidth]{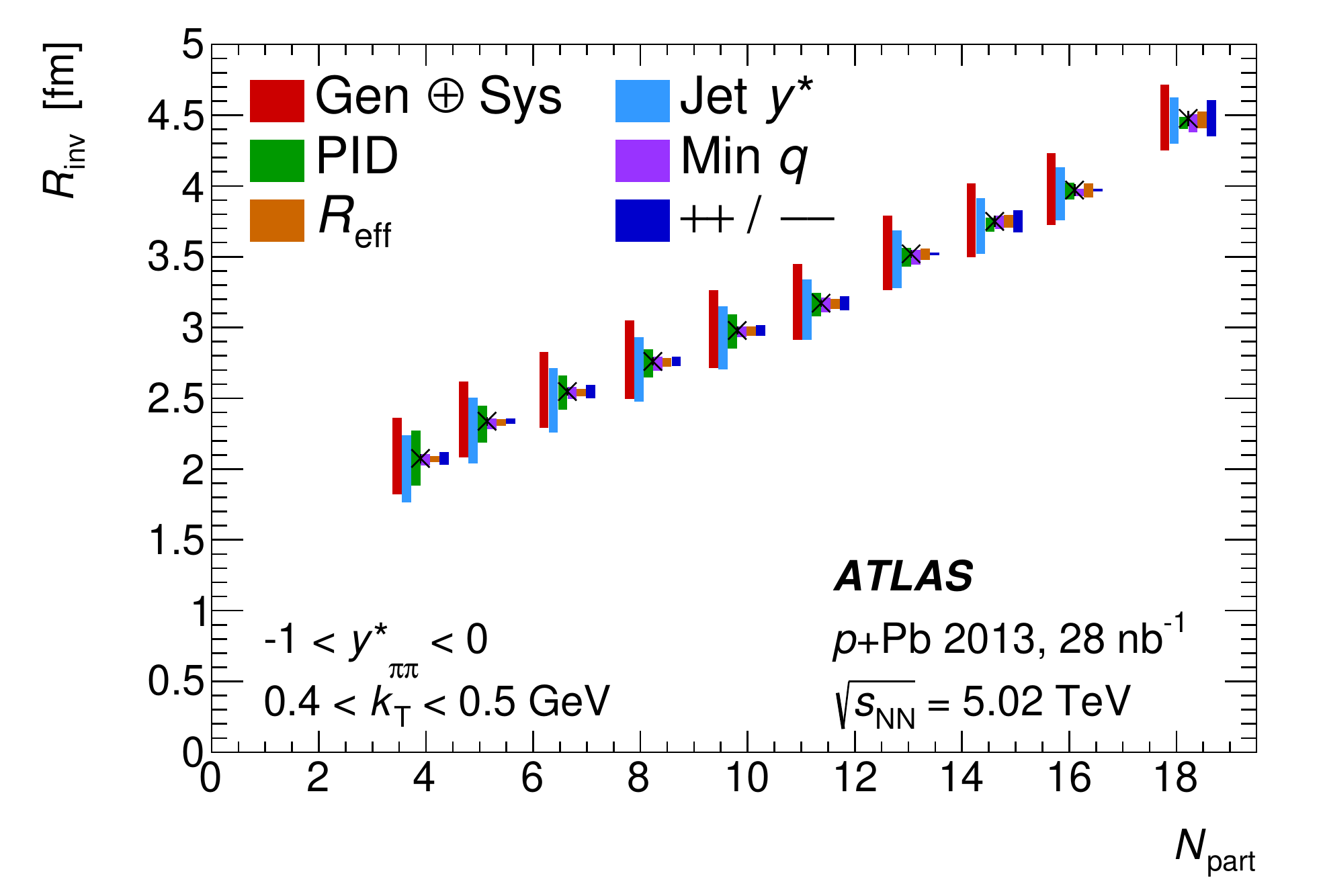}
}\\
\caption{
The contributions of the various sources of systematic uncertainty to the invariant radius \Rinv.
The typical trends with the pair's average transverse momentum \kt are shown in (a) and the trends with the number of nucleon participants \Npart are shown in (b).
The black crosses indicate the nominal results.
}
\label{fig:syst_qinv_R}
\end{figure}
\begin{figure}[t]
\centering
\subfloat[]{
\includegraphics[width=0.49\linewidth]{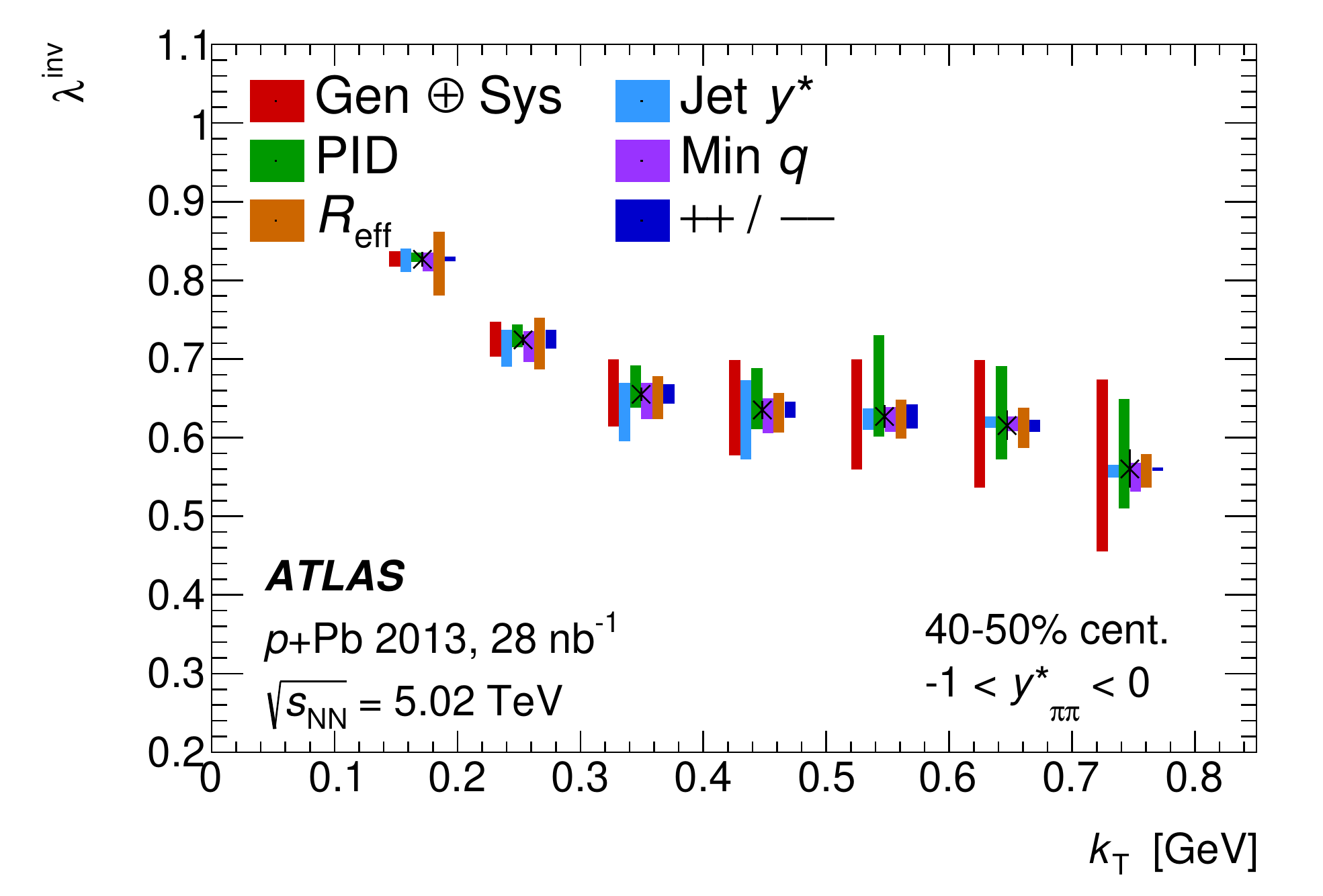}
}
\subfloat[]{
\includegraphics[width=0.49\linewidth]{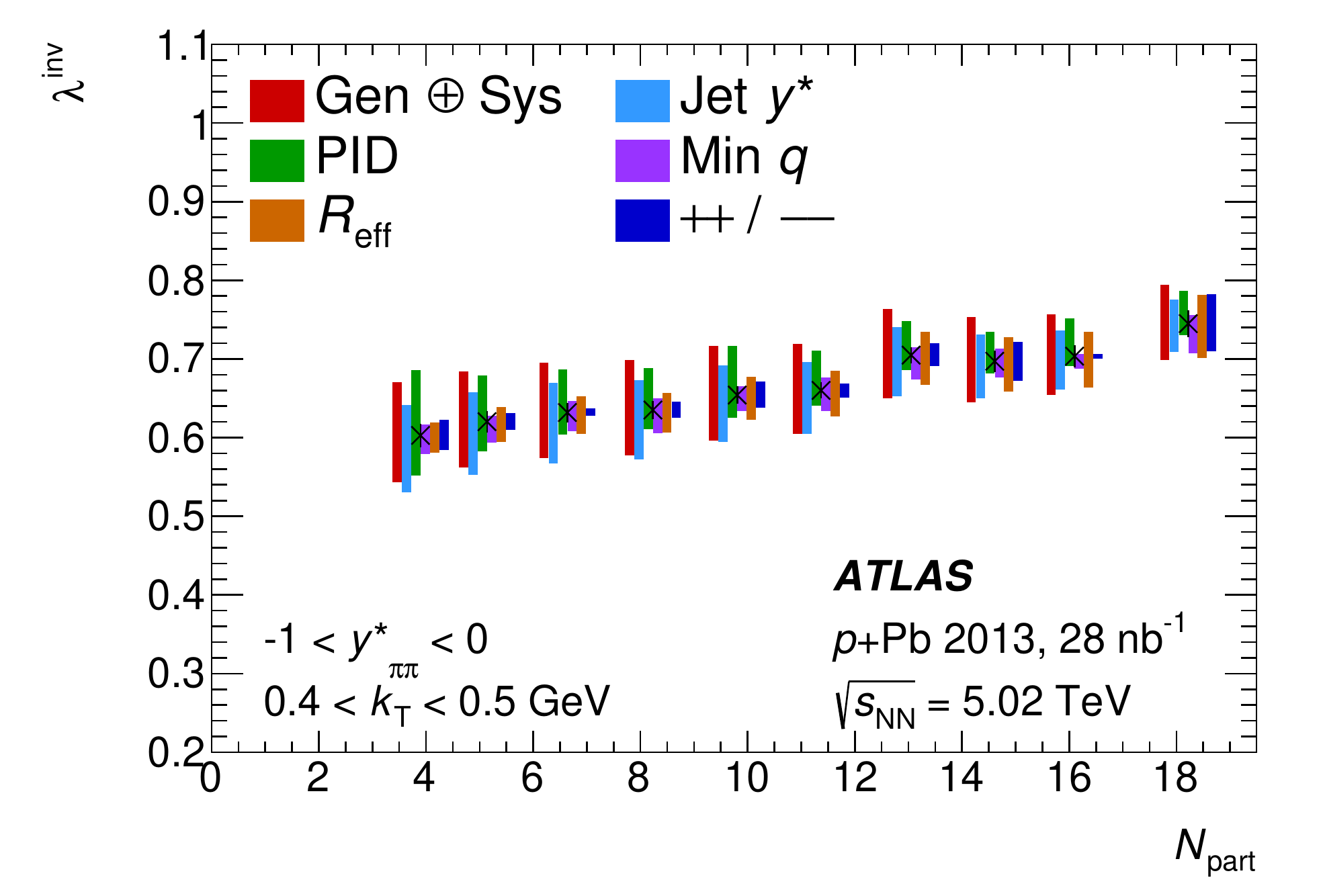}
}\\
\caption{
The contributions of the various sources of systematic uncertainty to the invariant Bose-Einstein amplitude \linv.
The typical trends with the pair's average transverse momentum \kt are shown in (a) and the trends with the number of nucleon participants \Npart are shown in (b).
The black crosses indicate the nominal results.
}
\label{fig:syst_qinv_x}
\end{figure}

A small difference between positive and negative charge pairs is observed, attributable to detector effects such as the orientation of the azimuthal overlap of the inner detector's component staves.
The nominal results use all of the same-charge pairs, and a systematic variation accounting for this charge asymmetry is assigned which covers the results for both of the separate charge states.
The variation from this effect is labeled by ``$++ / --$'' in the figures of \Sect{\ref{subsec:syst_plots}}.

Single-particle correction factors for track reconstruction efficiency cancel in the ratio $A(q) / B(q)$.
However, two-particle effects in the track reconstruction can affect the correlation function at small relative momentum.
Single- and multi-track reconstruction effects are both studied with the fully simulated \Hijing sample.
The generator-level and reconstructed correlation functions are compared, and a deficit in the latter, due to the impact of the two-particle reconstruction efficiency, is observed at \qinv below approximately $50~\MeV$.
At larger $\qinv$ the two-particle reconstruction efficiency is found to not depend on \qinv within statistical uncertainties.
A minimum $q$ cutoff is applied in the fits to minimize the impact of these detector effects.
The sensitivity of the results to this cutoff is checked by taking $\qinv^{\textrm{min}} = 30 \pm 10~\MeV$ in the one-dimensional fits, and symmetrizing the effect of the variation from $|\mathbf{q}|^{\textrm{min}} = 25~\textrm{to}~50~\MeV$ in the 3D fits.
Because this variation has only a small effect on the radii, this procedure is taken to be sufficient to account for two-particle reconstruction effects.
The effects of this variation have the label ``Min $q$'' in the figures of \Sect{\ref{subsec:syst_plots}}.

\begin{figure}[tb]
\centering
\subfloat[]{
\includegraphics[width=0.49\linewidth]{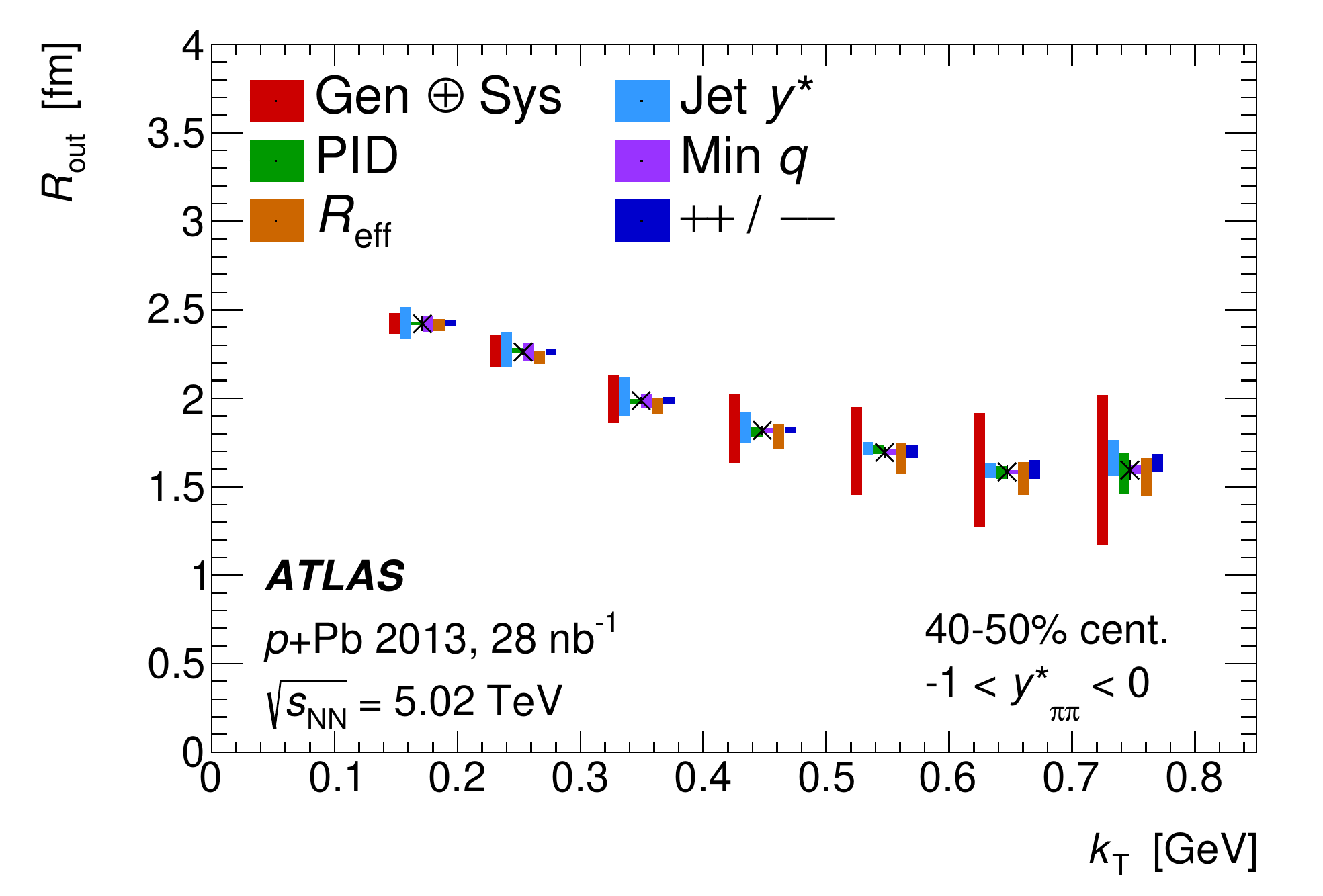}
}
\subfloat[]{
\includegraphics[width=0.49\linewidth]{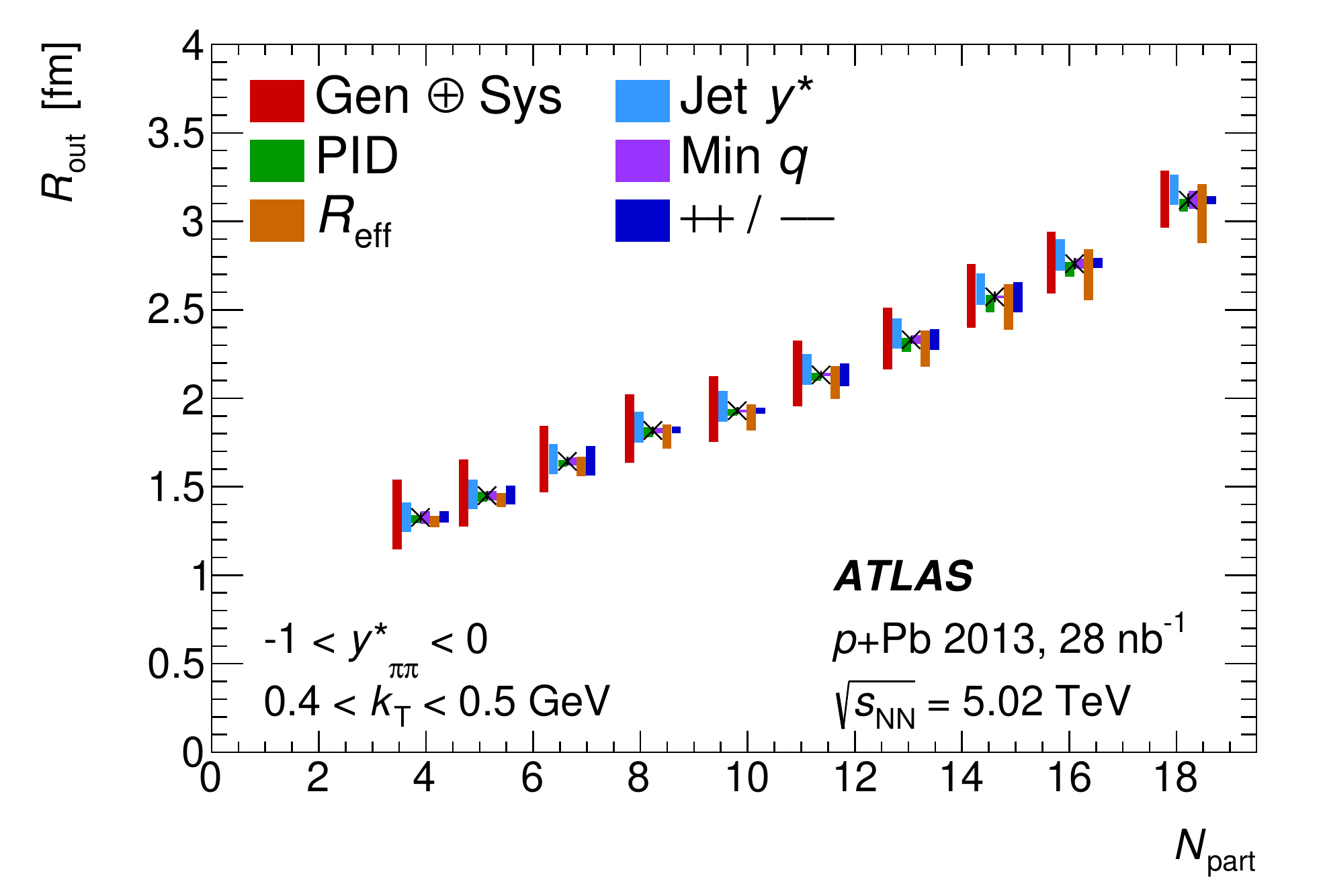}
}\\
\caption{
The contributions of the various sources of systematic uncertainty to the three-dimensional radius \Rout.
The typical trends with the pair's average transverse momentum \kt are shown in (a) and the trends with the number of nucleon participants \Npart are shown in (b).
The black crosses indicate the nominal results.
}
\label{fig:syst_qosl_Rout}
\end{figure}

\begin{figure}[tb]
\centering
\subfloat[]{
\includegraphics[width=0.49\linewidth]{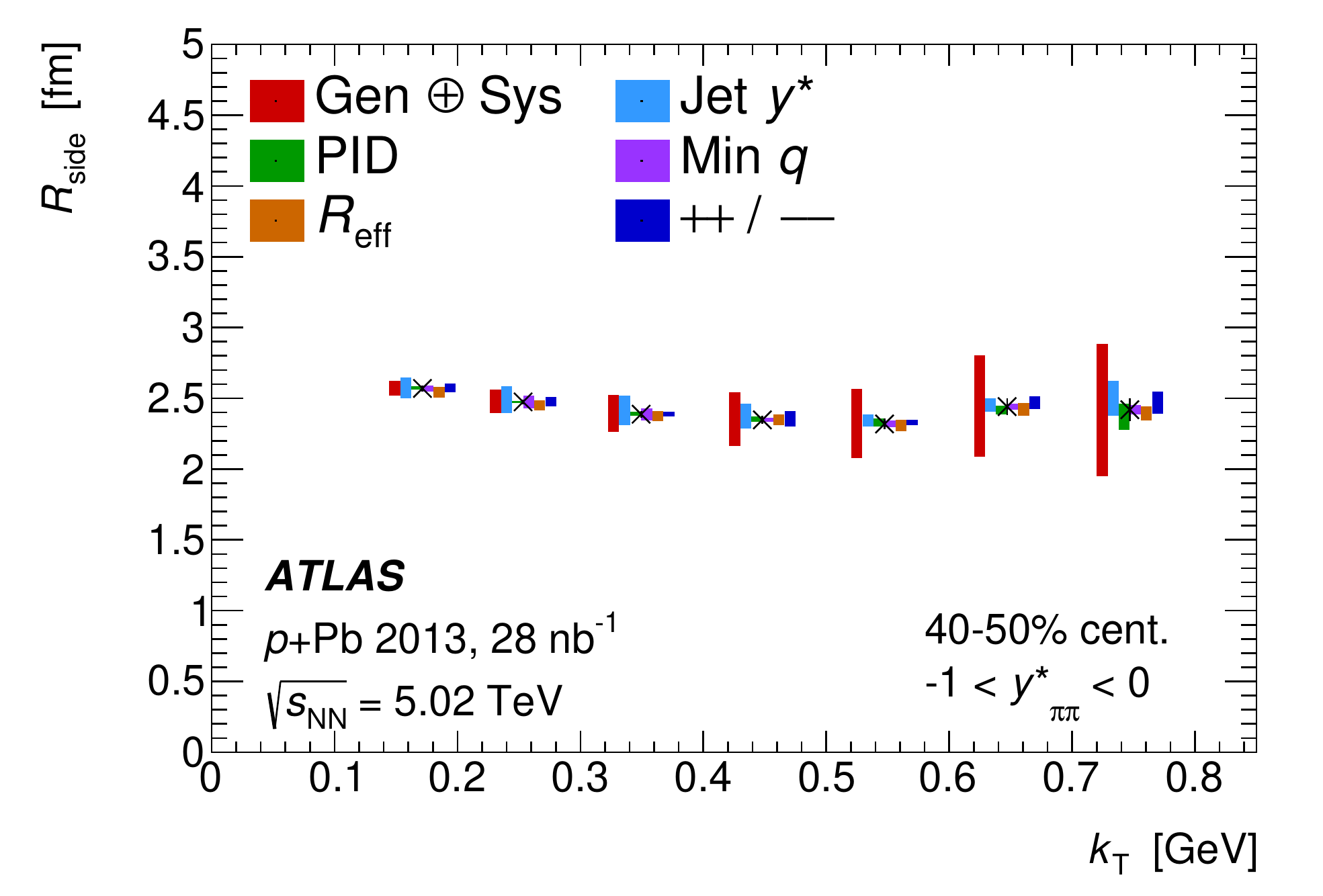}
}
\subfloat[]{
\includegraphics[width=0.49\linewidth]{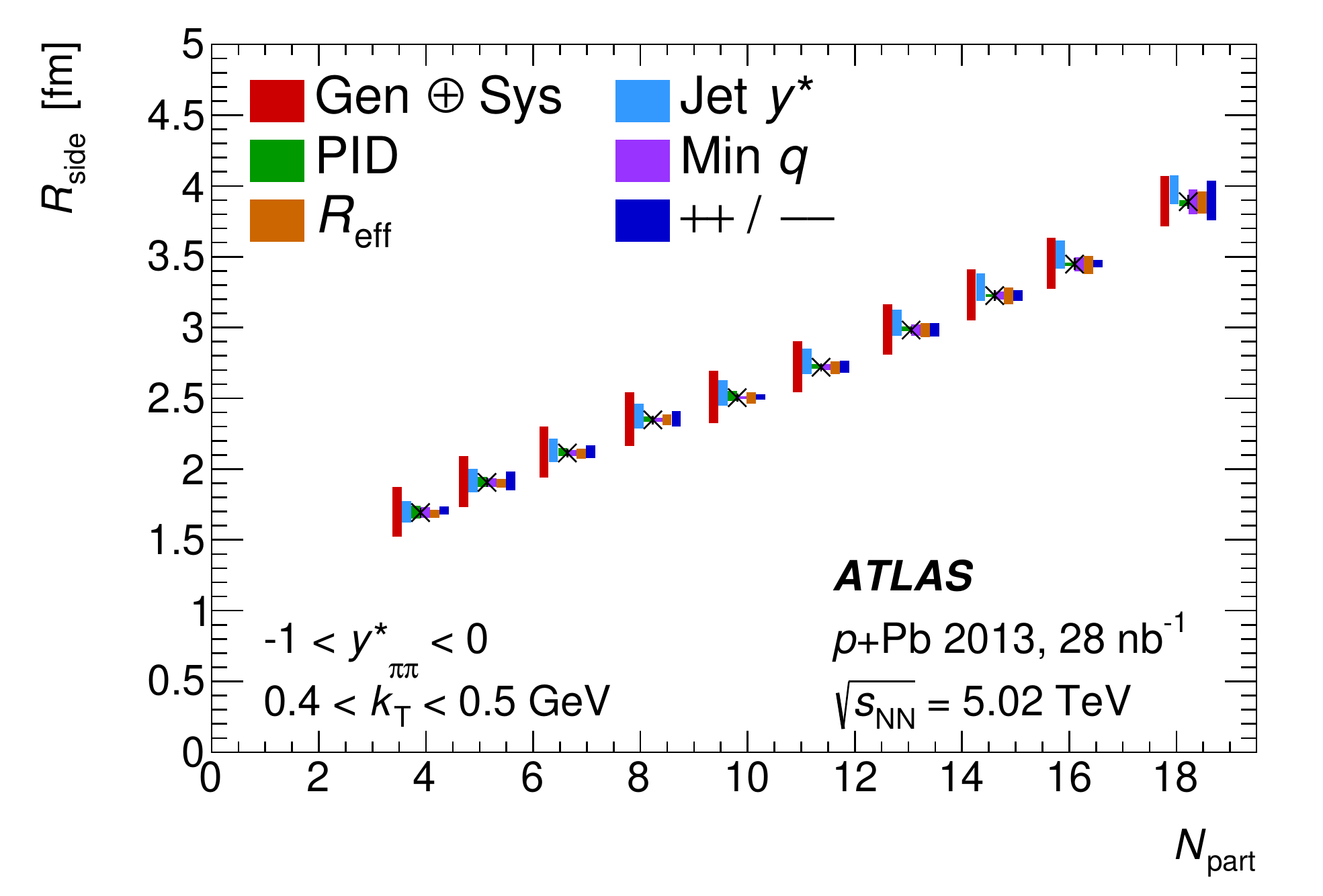}
}\\
\caption{
The contributions of the various sources of systematic uncertainty to the three-dimensional radius \Rside.
The typical trends with the pair's average transverse momentum \kt are shown in (a) and the trends with the number of nucleon participants \Npart are shown in (b).
The black crosses indicate the nominal results.
}
\label{fig:syst_qosl_Rside}
\end{figure}

\FloatBarrier

\subsection{Magnitude of systematic effects}
\label{subsec:syst_plots}

In this section the contributions of each source of systematic uncertainty are illustrated.
Examples of the systematic uncertainties in the invariant parameters \Rinv and \linv are shown as a function of \kt and centrality in \Fig{\ref{fig:syst_qinv_R} and \ref{fig:syst_qinv_x}}.

\begin{figure}[t]
\centering
\subfloat[]{
\includegraphics[width=0.49\linewidth]{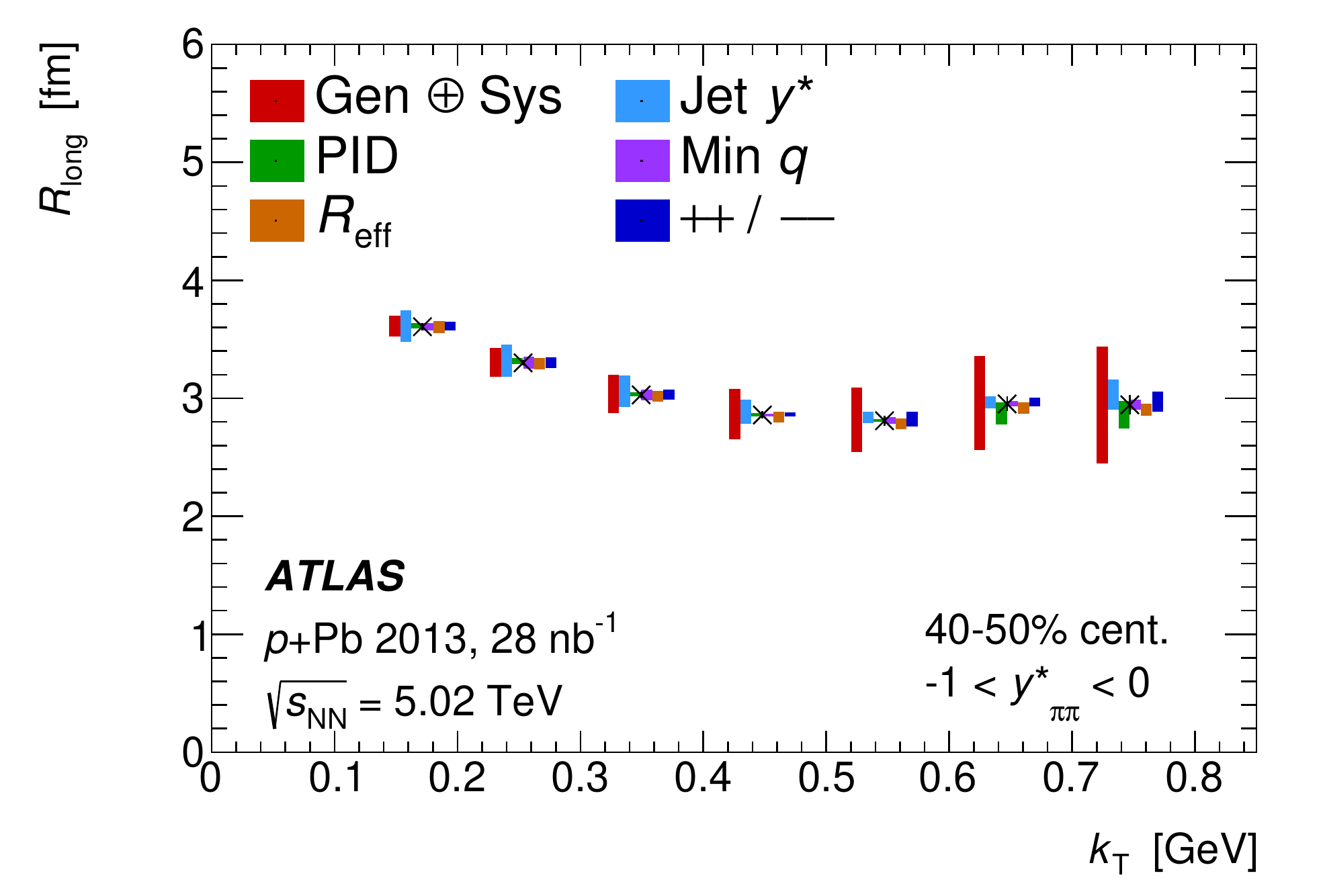}
}
\subfloat[]{
\includegraphics[width=0.49\linewidth]{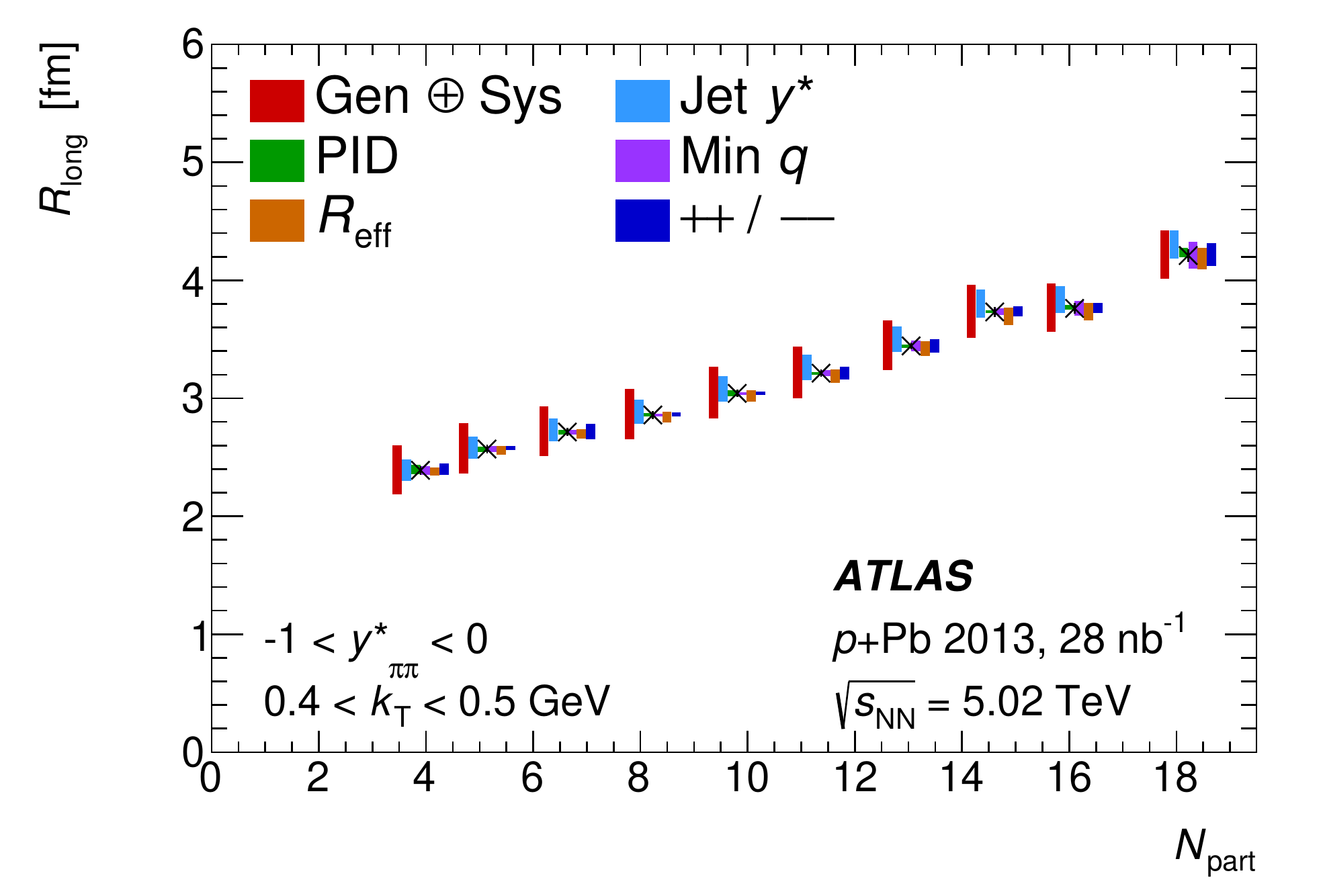}
}
\caption{
The contributions of the various sources of systematic uncertainty to the three-dimensional radius \Rlong.
The typical trends with the pair's average transverse momentum \kt are shown in (a) and the trends with the number of nucleon participants \Npart are shown in (b).
The black crosses indicate the nominal results.
}
\label{fig:syst_qosl_Rlong}
\end{figure}

\begin{figure}[t]
\centering
\subfloat[]{
\includegraphics[width=0.49\linewidth]{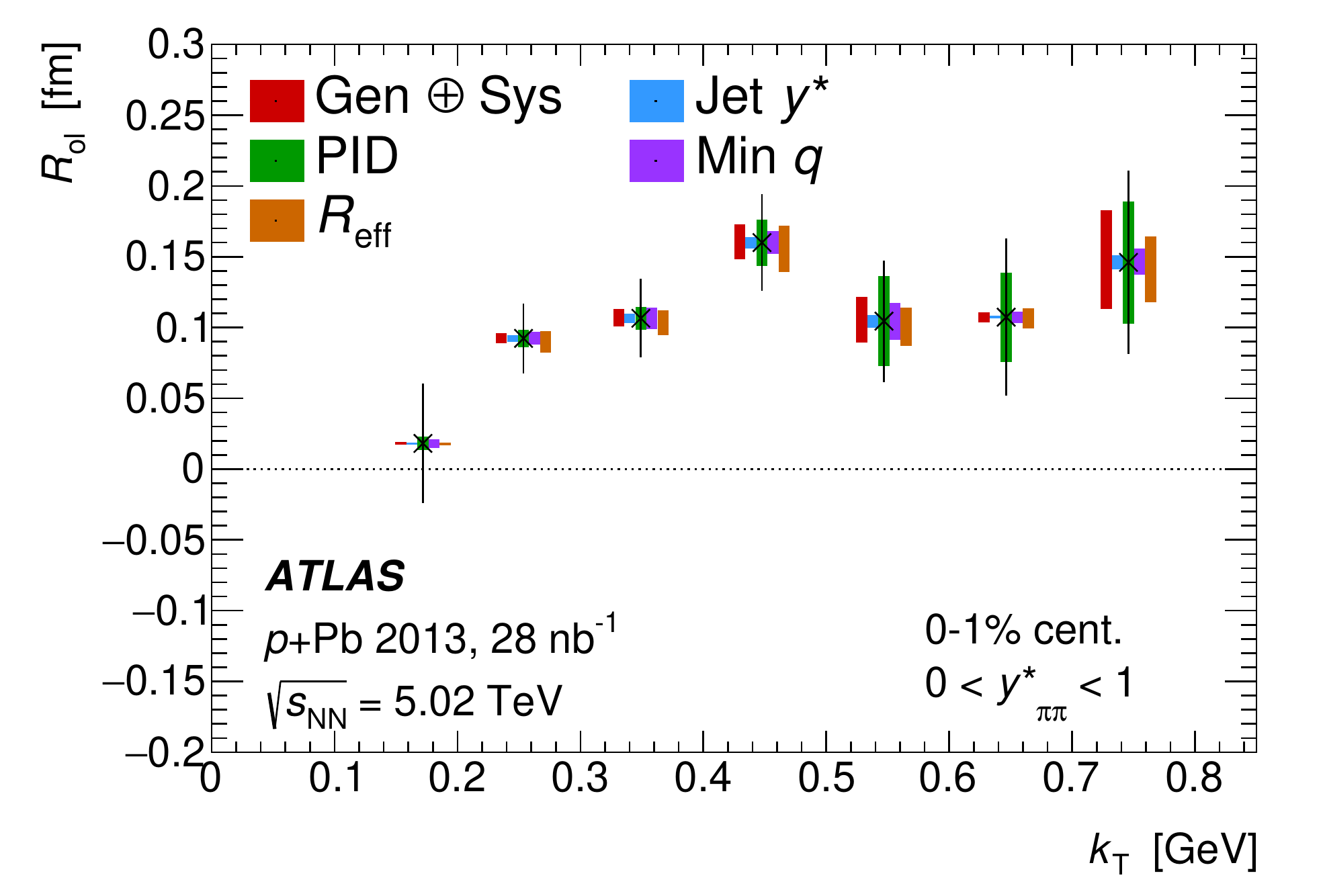}
}
\subfloat[]{
\includegraphics[width=0.49\linewidth]{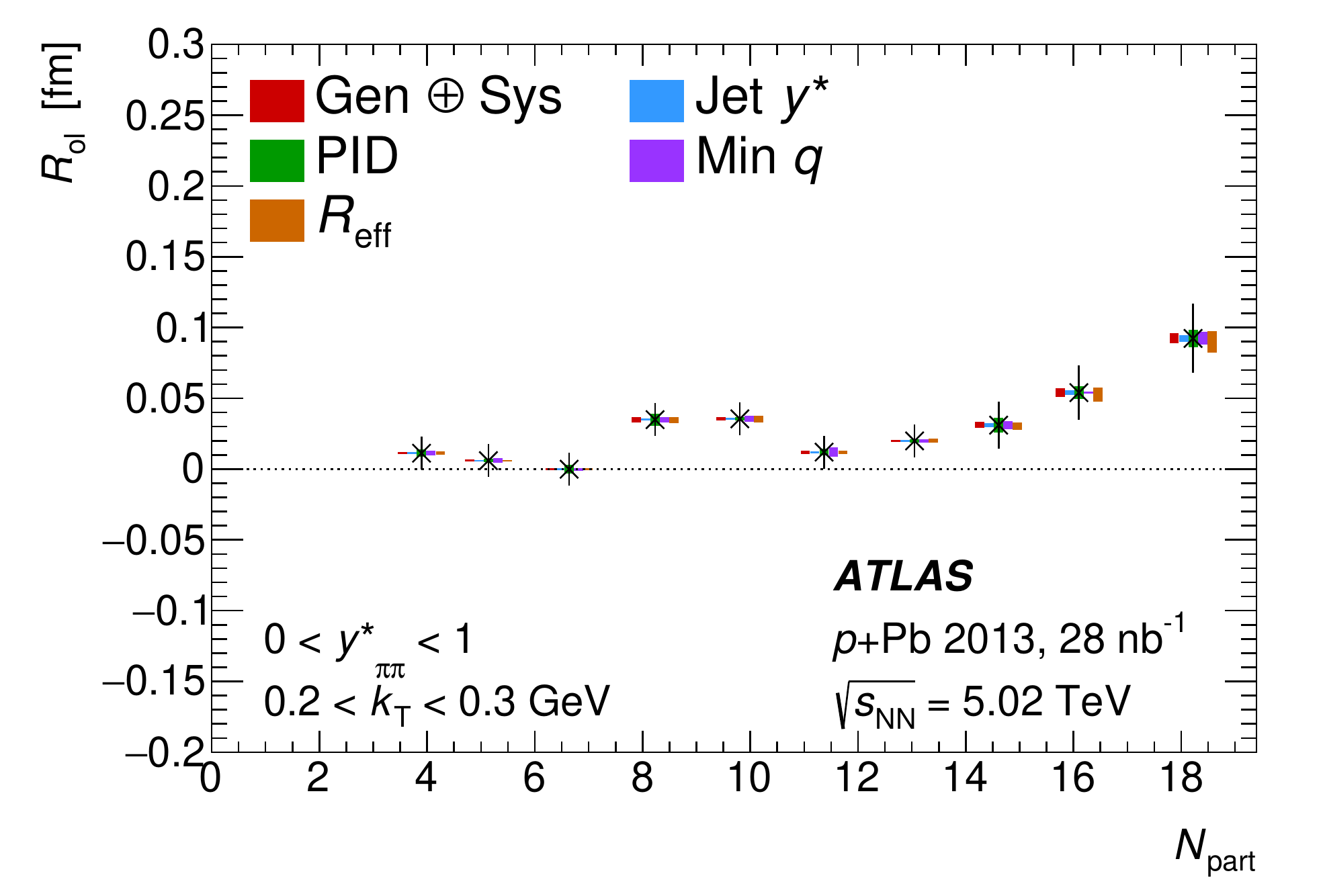}
}
\caption{
The contributions of the various sources of systematic uncertainty to the three-dimensional radius \Rol.
The typical trends with the pair's average transverse momentum \kt are shown in (a) and the trends with the number of nucleon participants \Npart are shown in (b).
The large uncertainties from PID at high \kt are mostly the result of statistical fluctuations, and including them in the reported uncertainty is a conservative choice.
The uncertainties for Jet \ystar and PID are explicitly symmetrized.
The black crosses indicate the nominal results, and the dotted line at $\Rol = 0$ is drawn for visibility.
}
\label{fig:syst_qosl_Rol}
\end{figure}

\begin{figure}[ht]
\centering
\subfloat[]{
\includegraphics[width=0.49\linewidth]{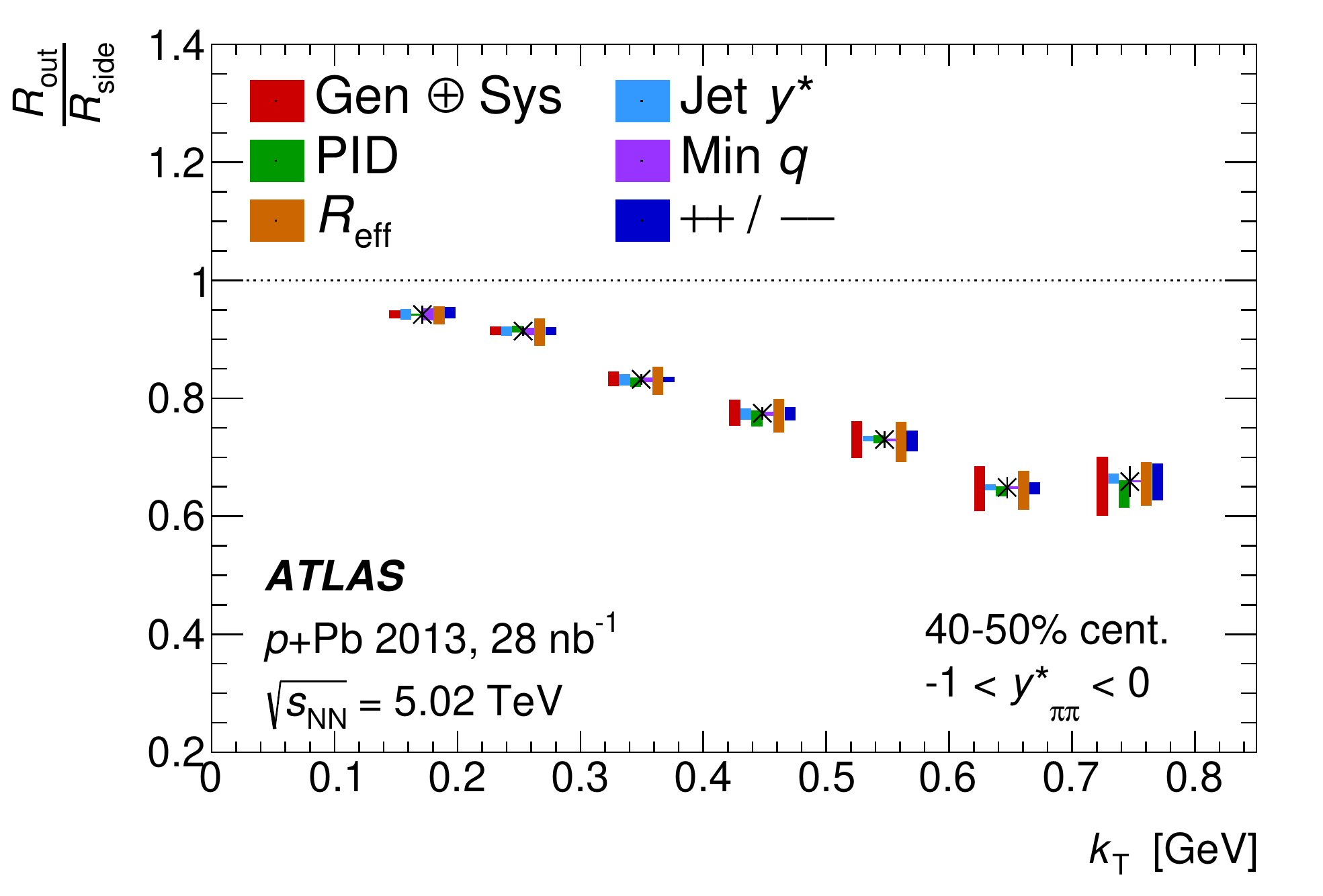}
}
\subfloat[]{
\includegraphics[width=0.49\linewidth]{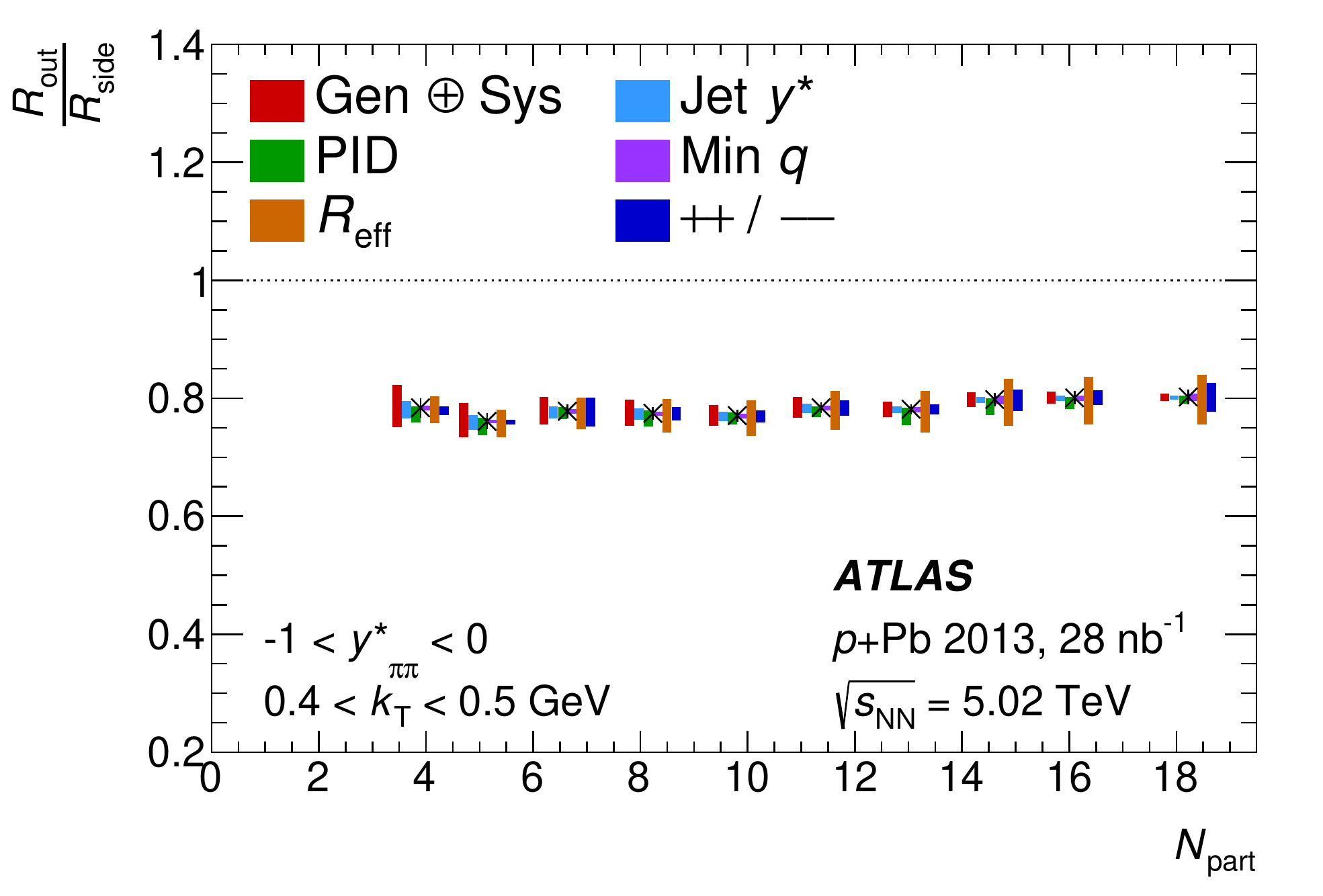}
}
\caption{
  The contributions of the various sources of systematic uncertainty to the ratio $\Rout / \Rside$.
  The typical trends with the pair's average transverse momentum \kt are shown in (a) and the trends with the number of nucleon participants \Npart are shown in (b).
  The black crosses indicate the nominal results, and the dotted line at $\Rout/\Rside = 1$ is drawn for visibility.
}
\label{fig:syst_qosl_RoutOverRside}
\end{figure}

\begin{figure}[ht]
\centering
\subfloat[]{
\includegraphics[width=0.49\linewidth]{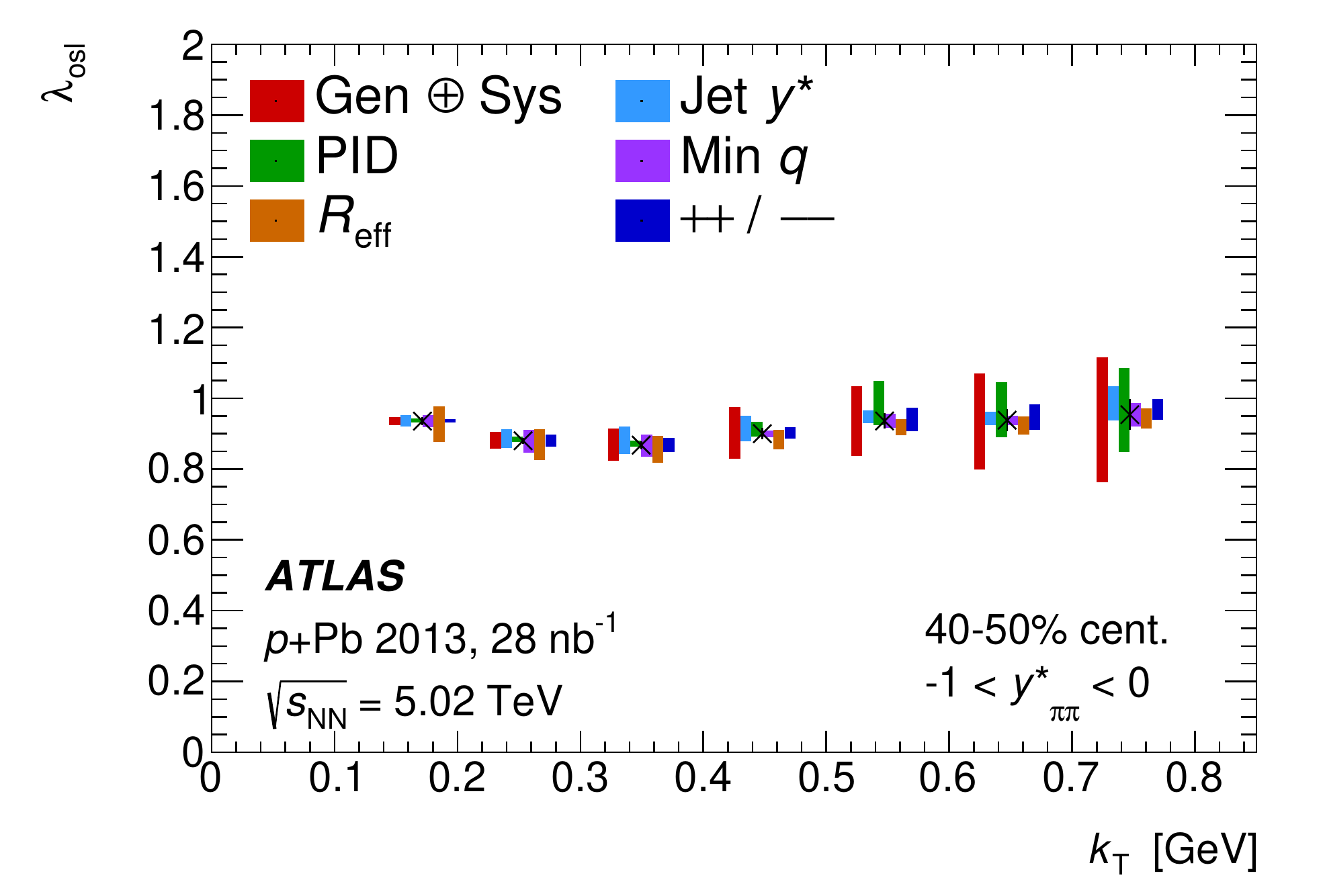}
}
\subfloat[]{
\includegraphics[width=0.49\linewidth]{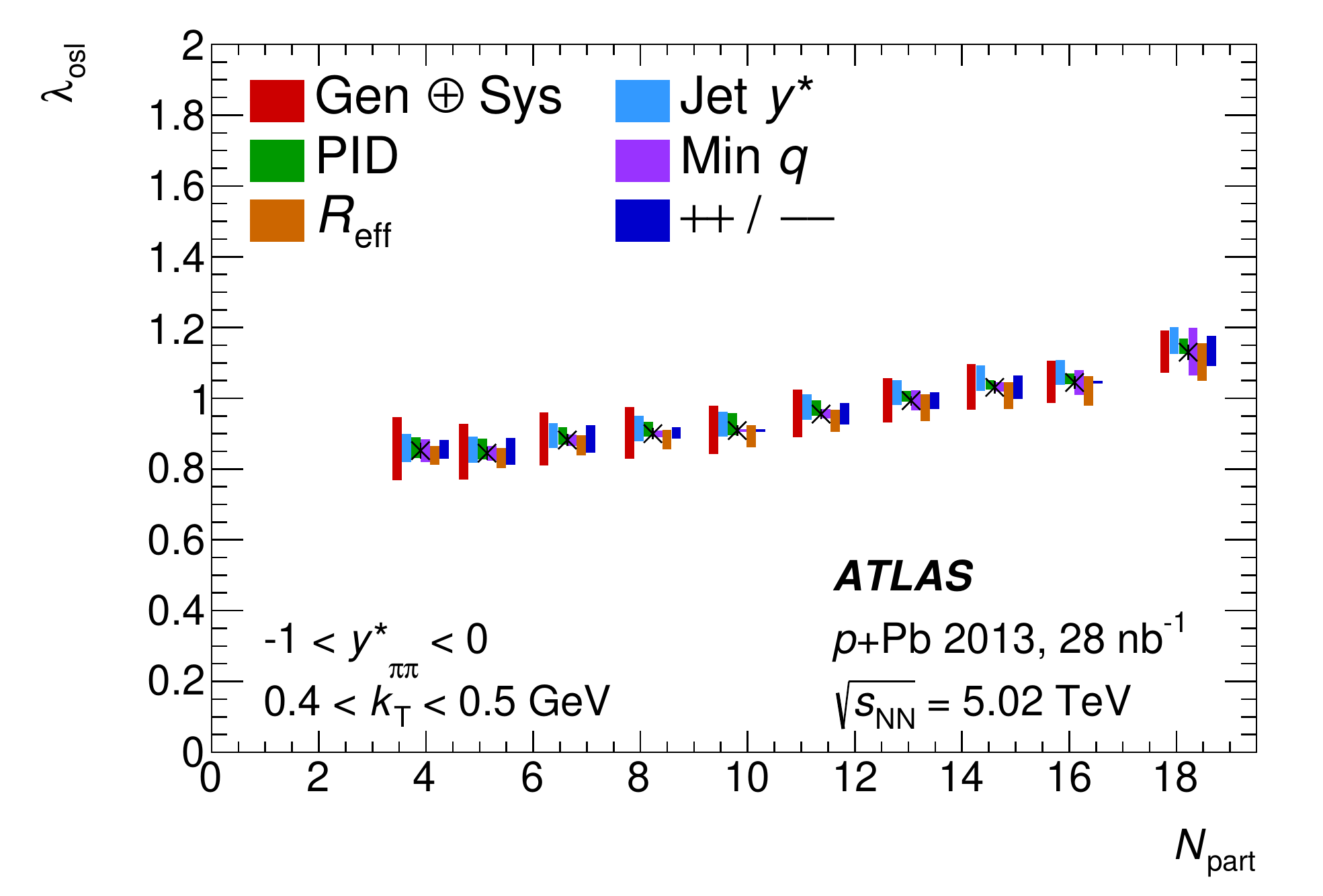}
}
\caption{
  The contributions of the various sources of systematic uncertainty to the three-dimensional Bose-Einstein amplitude \losl.
  The typical trends with the pair's average transverse momentum \kt are shown in (a) and the trends with the number of nucleon participants \Npart are shown in (b).
  The black crosses indicate the nominal results.
}
\label{fig:syst_qosl_x}
\end{figure}

Systematic uncertainties are also shown for the 3D radii \Rout (\Fig{\ref{fig:syst_qosl_Rout}}), \Rside (\Fig{\ref{fig:syst_qosl_Rside}}), \Rlong (\Fig{\ref{fig:syst_qosl_Rlong}}), and \Rol (\Fig{\ref{fig:syst_qosl_Rol}}), as well as the ratio $\Rout / \Rside$ (\Fig{\ref{fig:syst_qosl_RoutOverRside}}) and the amplitude (\Fig{\ref{fig:syst_qosl_x}}).
These are all shown for typical choices of centrality, \kt, and \kys so that they represent standard, rather than exceptional, values of the uncertainties.

The uncertainties in the HBT radii (\Figs{\ref{fig:syst_qinv_R}, \ref{fig:syst_qosl_Rout}, \ref{fig:syst_qosl_Rside}}{\ref{fig:syst_qosl_Rlong}}) are dominated by the jet background description.
At larger \kt the generator (\Pythia vs.\ \Herwig) and system (\pp vs.\ \pPb) contributions constitute the larger portion of this, and at lower \kt the variation of the mapping over \kys is more significant.

The Bose-Einstein amplitudes \linv (\Fig{\ref{fig:syst_qinv_x}}) and \losl (\Fig{\ref{fig:syst_qosl_x}}) are also affected strongly by the jet fragmentation description, but at sufficiently large \kt ($\gtrsim 0.4~\GeV$) pion identification contributes a comparable systematic uncertainty.
This is expected because other particles misidentified as pions do not exhibit Bose-Einstein interference with real pions, and the most significant effect of their inclusion in the correlation function is to decrease the amplitude of the Bose-Einstein enhancement.

The systematic uncertainties in the ratio $\Rout / \Rside$ (\Fig{\ref{fig:syst_qosl_RoutOverRside}}) are estimated by evaluating the ratio after each variation reflecting a systematic uncertainty and taking the difference from the nominal value.
Thus, the uncertainties that are correlated between \Rout and \Rside cancel properly in the ratio.
The uncertainties in the ratio are not universally dominated by any single effect.
At sufficiently large \kt in central events, the effective Coulomb size becomes the largest contributor.
This is understandable because the Coulomb correction is applied as a function of \qinv, and as a result it is applied over a wider range in \qout than in \qside at larger \kt.

Similarly, systematic effects in the cross-term \Rol (\Fig{\ref{fig:syst_qosl_Rol}}) are not dominated by any one source, and in fact the uncertainties in this quantity are predominantly statistical.
At large \kt the systematic uncertainties from PID appear large.
However, these are mostly a result of statistical fluctuations that arise because the fit in the tight PID selection is performed on a sample even smaller than the nominal dataset.
Therefore, the reported systematic uncertainties are overly conservative for this quantity.
For a similar reason, the systematic uncertainty for charge-asymmetric detector effects is not included in the error bars for \Rol.
No systematic dependence of \Rol is observed when measuring $++$ and $--$ correlations independently, so they are excluded from the total systematic uncertainties in order not to include additional statistical fluctuations as systematic effects.


\section{Results}
\label{sec:results}

This section shows examples of one- and three-dimensional fits to correlation functions, then presents results for extracted invariant and 3D source radii. The results are shown as a function of \kt, which can illustrate the time-dependence of the source size. They are also shown as a function of \kys, showing any variations in source size along the collision axis, and against several quantities related to multiplicity and centrality. These results show the freeze-out density and the evolution of the source with the size of the initial geometry.

\subsection{Performance of fit procedure}

\begin{figure}[tb]
\centering
\includegraphics[width=0.98\linewidth]{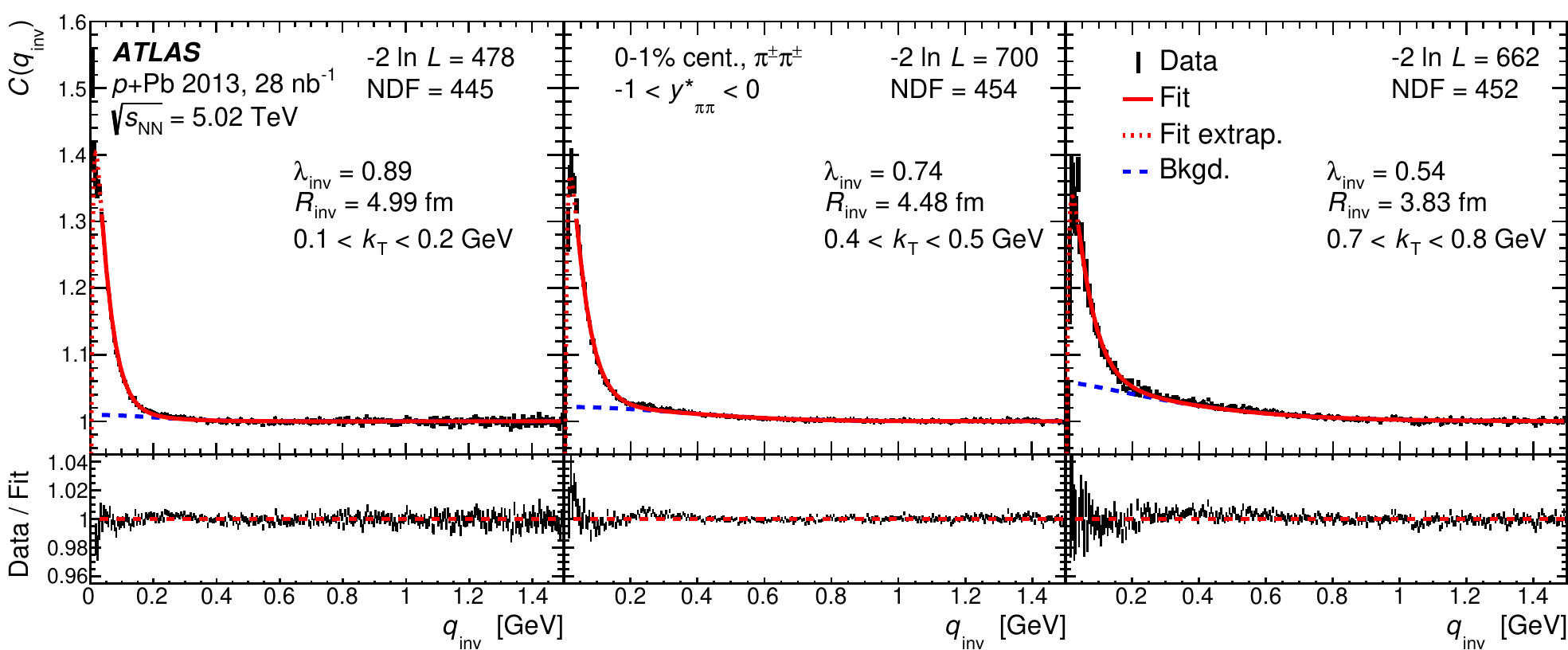}
\caption{Results of the fit to the one-dimensional correlation function in very central (0--1\%) events in three \kt intervals.
The dashed blue line indicates the description of the contribution from jet fragmentation and the red line shows the full correlation function fit.
The dotted red line indicates the extrapolation of the fit function beyond the interval over which the fit is performed.
}
\label{fig:cqinv_cent0}
\end{figure}

\begin{figure}[tb]
\centering
\includegraphics[width=0.98\linewidth]{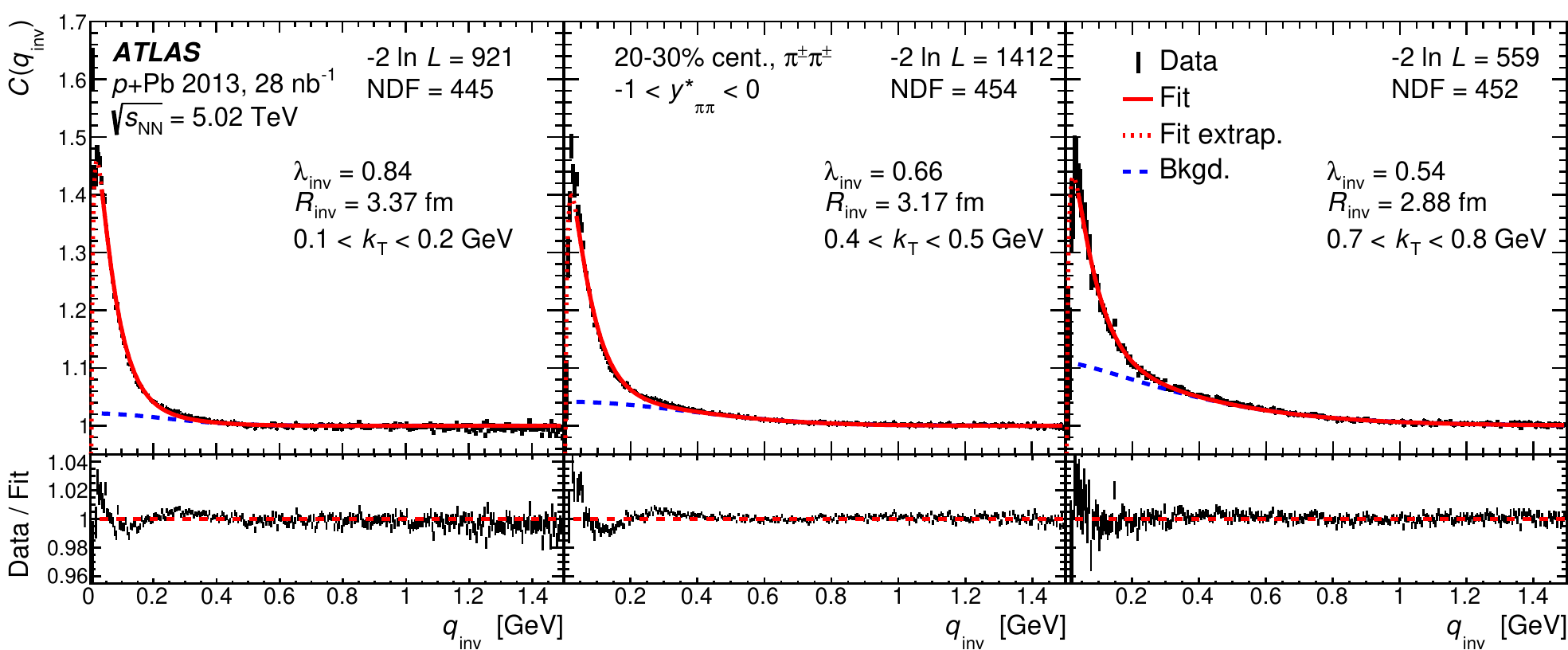}
\caption{Results of the fit to the one-dimensional correlation function in semi-central (20--30\%) events in three \kt intervals.
The dashed blue line indicates the description of the contribution from jet fragmentation and the red line shows the full correlation function fit.
The dotted red line indicates the extrapolation of the fit function beyond the interval over which the fit is performed.
}
\label{fig:cqinv_cent4}
\end{figure}
\begin{figure}[tb]
\centering
\includegraphics[width=0.98\linewidth]{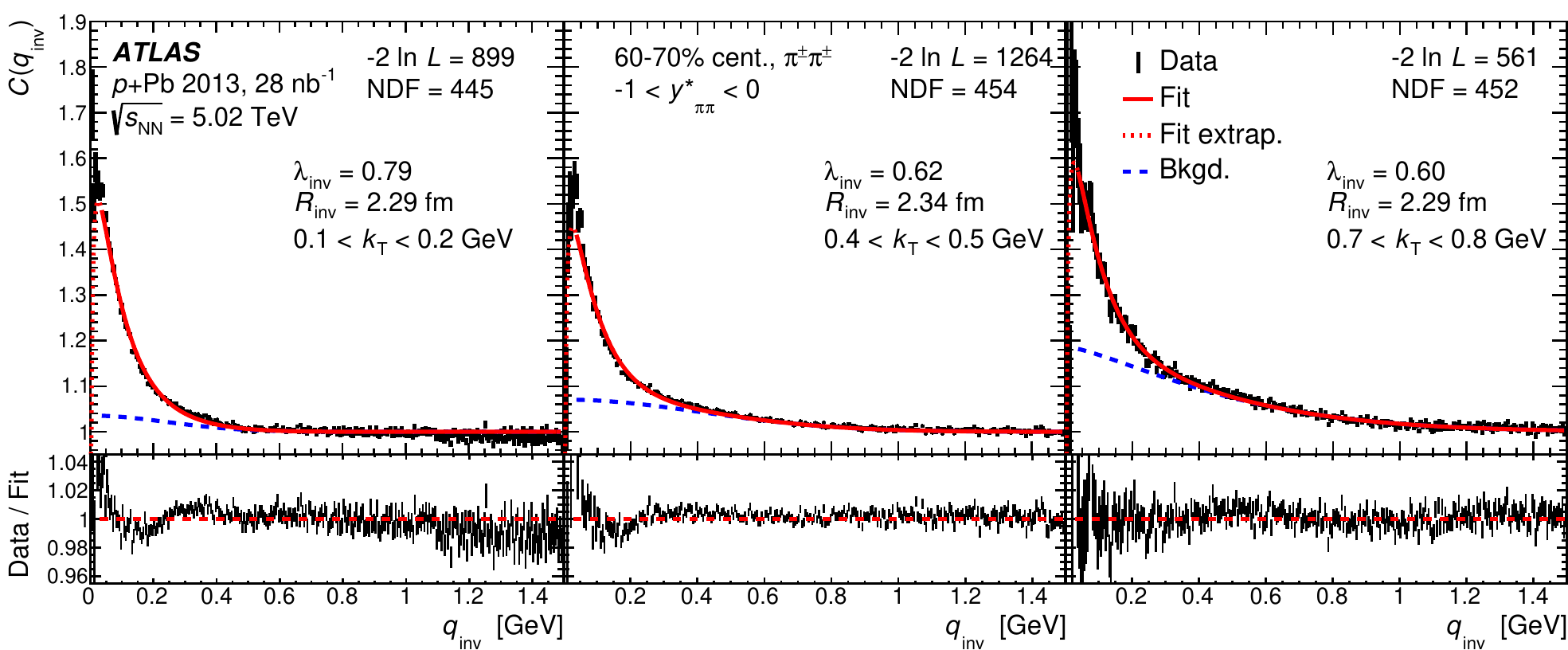}
\caption{Results of the fit to the one-dimensional correlation function in relatively peripheral (60--70\%) events in three \kt intervals.
The dashed blue line indicates the description of the contribution from jet fragmentation and the red line shows the full correlation function fit.
The dotted red line indicates the extrapolation of the fit function beyond the interval over which the fit is performed.
}
\label{fig:cqinv_cent8}
\end{figure}

An example of a one-dimensional fit to $C(\qinv)$ using the functional form of \Eqn{\ref{eq:correlation_function_full}}  is included in \Fig{\ref{fig:hp_example}}.
Additional examples of one-dimensional fits for different \kt\
intervals are shown in \Figrange{\ref{fig:cqinv_cent0}}{\ref{fig:cqinv_cent8}} for very central
(0--1\%), semi-central (20--30\%), and peripheral (60--70\%) centrality intervals, respectively.
The test statistic $-2 \ln \mathcal{L}$, defined in \Eqn{\ref{eq:loglikedef}}, is displayed on these figures.
The values of this $\chi^2$ analog indicate that the fits to the same-charge correlation functions generally describe the data well when compared to the number of degrees of freedom (NDF), with only small departures from an exponential description.

\begin{figure}[!h]
\centering
\includegraphics[width=0.98\linewidth]{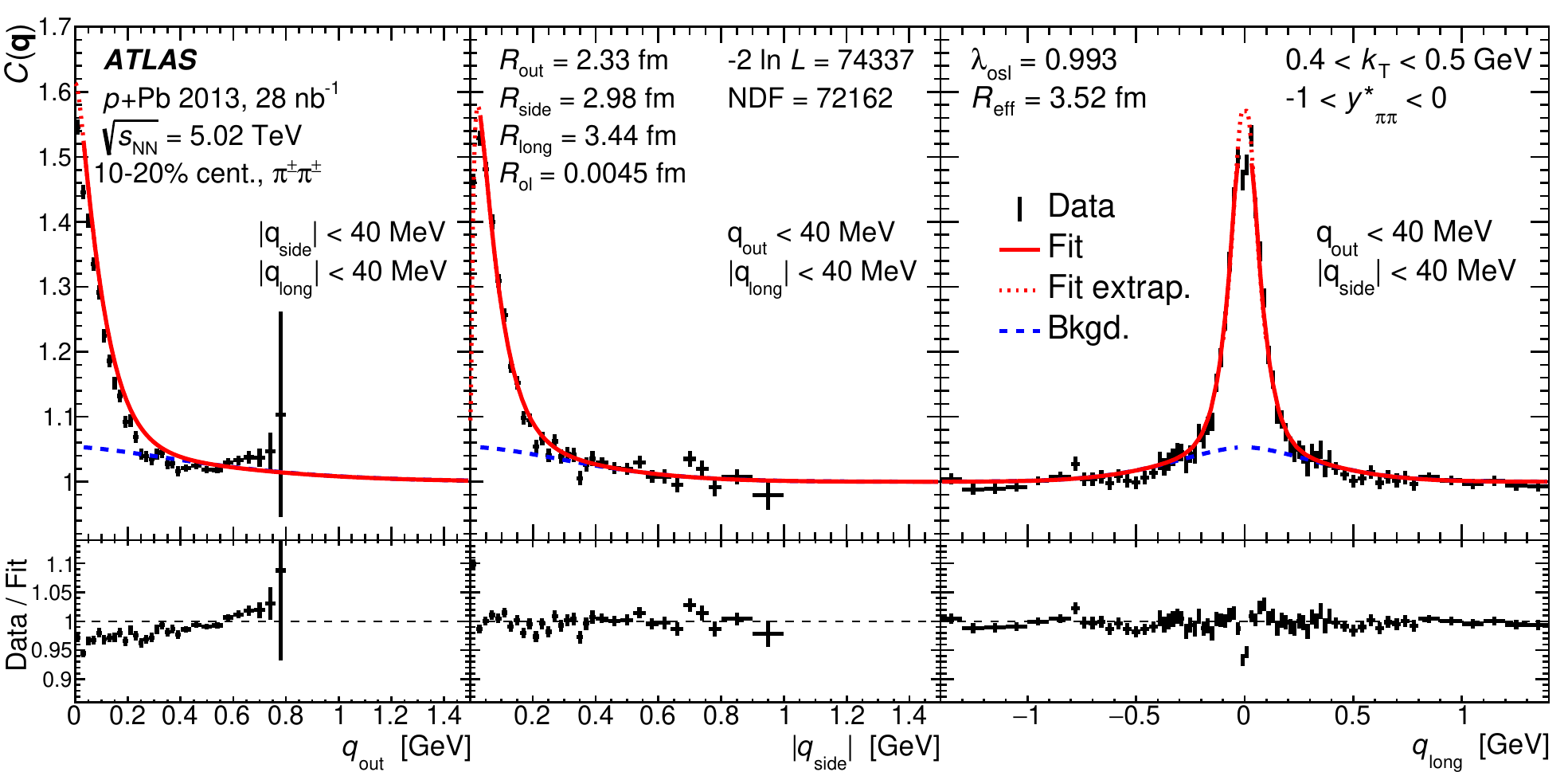}
\caption{Results of the 3D fit to the correlation function in the $0.4 < \kt < 0.5$~\GeV, $-1 < \kys < 0$ kinematic intervals and for the 10--20\% centrality interval.
The left, middle, and right panels show the distributions versus \qout, \qside, and \qlong, respectively, with limits on the other two components of $\mathbf{q}$ such that $|q_i| < 40~\MeV$.
The dashed blue line indicates the description of the contribution from hard processes and the red line shows the full correlation function fit.
The dotted red line indicates the extrapolation of the fit function beyond the interval over which the fit is performed.
}
\label{fig:cqosl_slices}
\end{figure}

Slices of a three-dimensional fit of $C(\mathbf{q})$ to the three-dimensional variant of \Eqn{\ref{eq:correlation_function_full}} are shown in \Fig{\ref{fig:cqosl_slices}}.
The apparently imperfect fit along the \qout axis is characteristic of $\qside \approx \qlong \approx 0$, and away from this slice the  fit agrees better with the data (the test statistic per degree of freedom is 1.03 for the fit shown).

\subsection{One-dimensional results}
\label{subsec:invariant_results}

\begin{figure}[t]
\centering
\subfloat[]{
\includegraphics[width=0.49\linewidth]{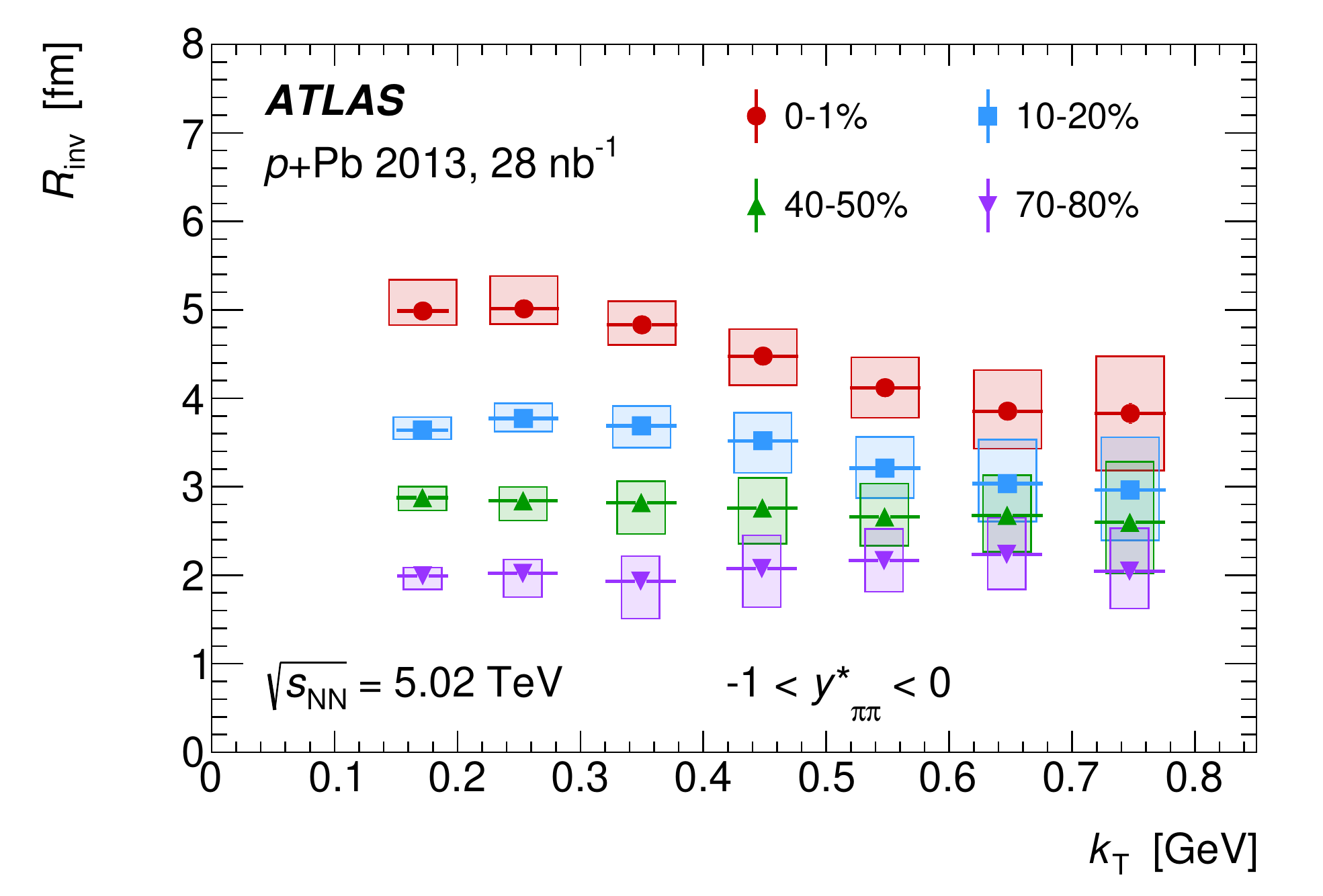}
}
\subfloat[]{
\includegraphics[width=0.49\linewidth]{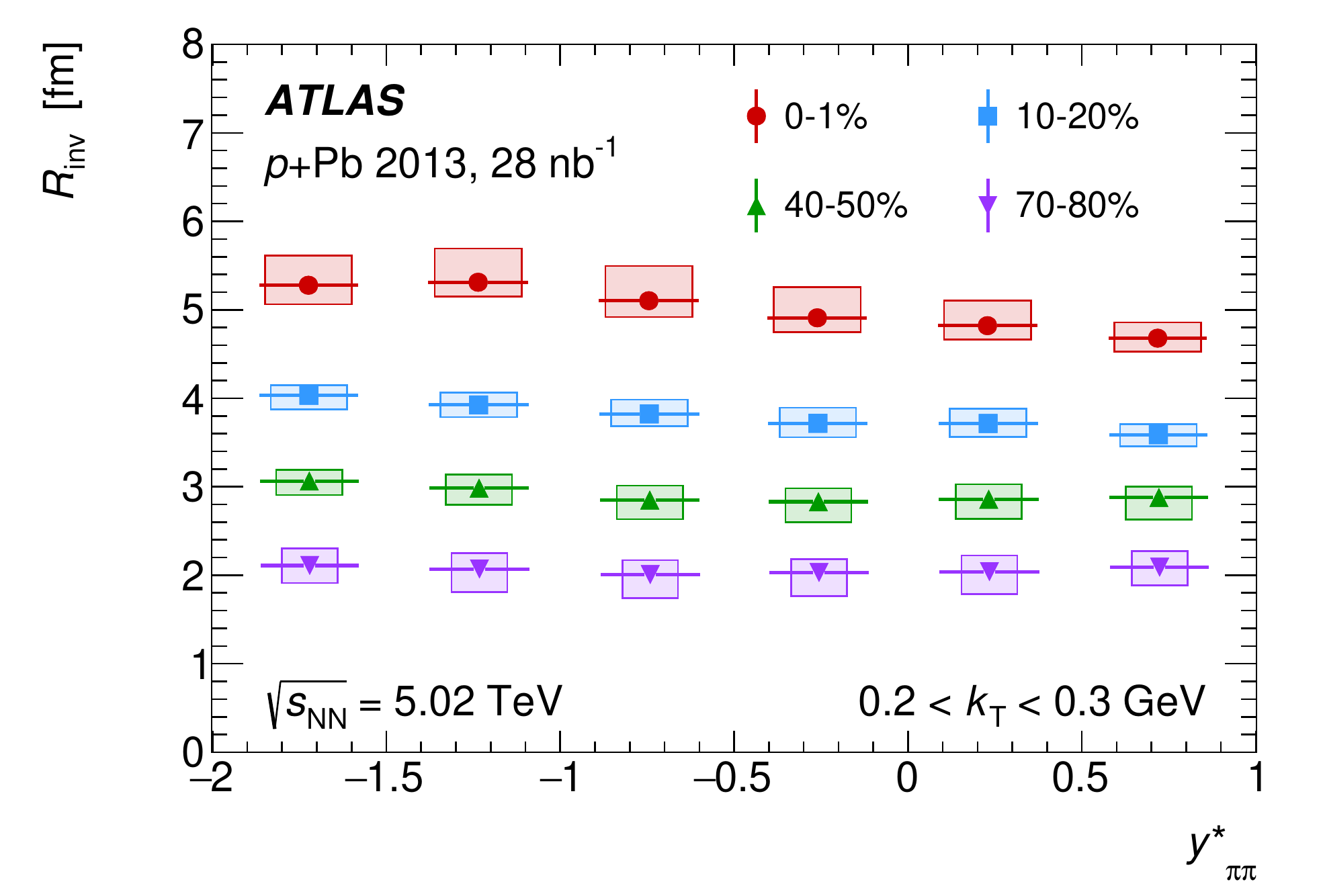}
}
\caption{The exponential invariant radii, \Rinv, obtained from
one-dimensional fits to
the \qinv\ correlation functions shown as a function of pair transverse momentum, \kt, (a) and rapidity, \kys (b). Four non-adjacent centrality intervals are shown. The vertical size of each box represents the quadrature sum of the systematic uncertainties described in \Sect{\ref{sec:systematics}}, and statistical uncertainties are shown with vertical lines. The horizontal positions of the points are the average \kt or \kys in each interval, and the horizontal lines indicate the standard deviation of \kt or \kys. The widths of the boxes differ among centrality intervals only for visual clarity.}
\label{fig:results_Rinv_kt}
\end{figure}
\begin{figure}[h!]
\centering
\subfloat[]{
\includegraphics[width=0.49\linewidth]{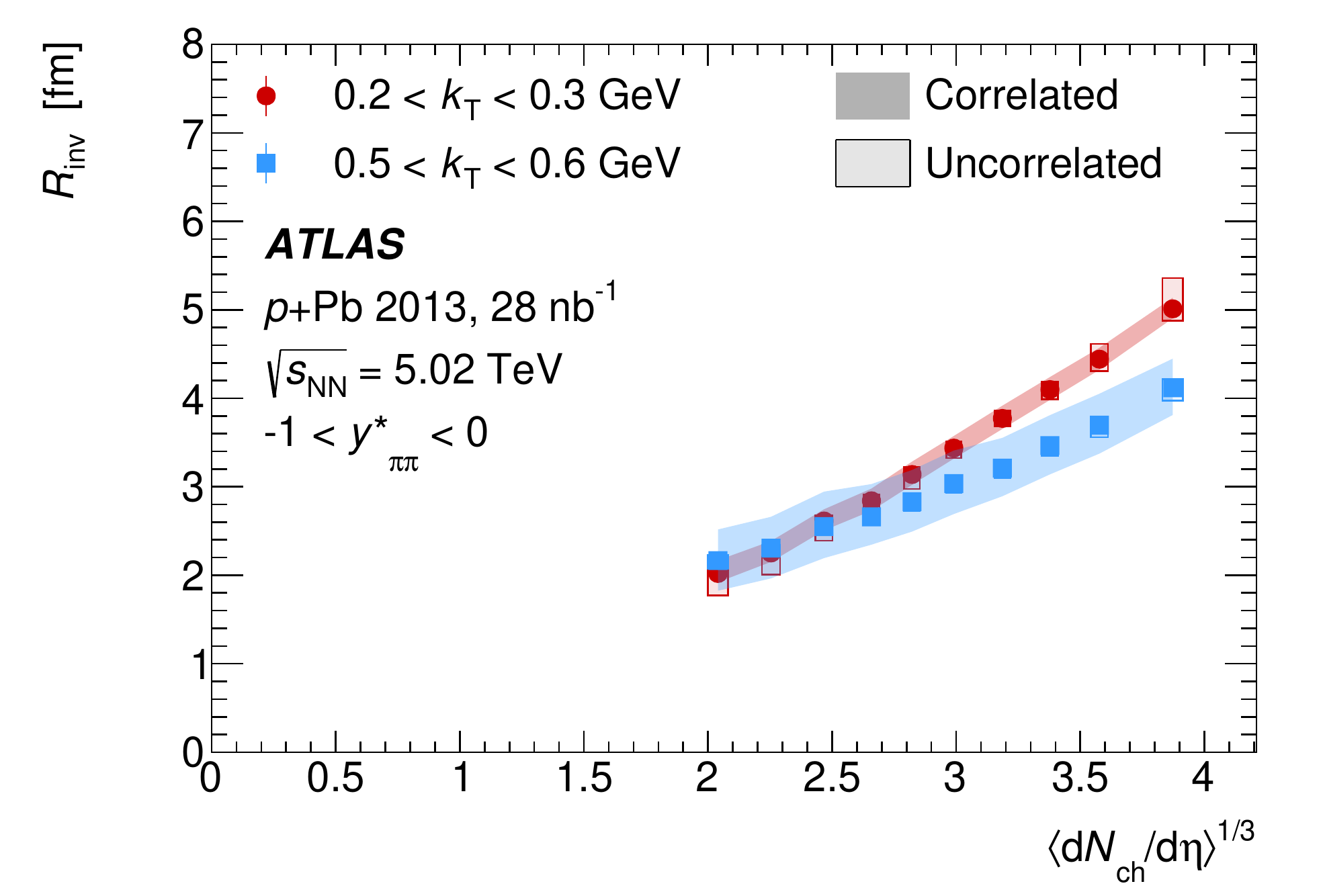}
}
\subfloat[]{
\includegraphics[width=0.49\linewidth]{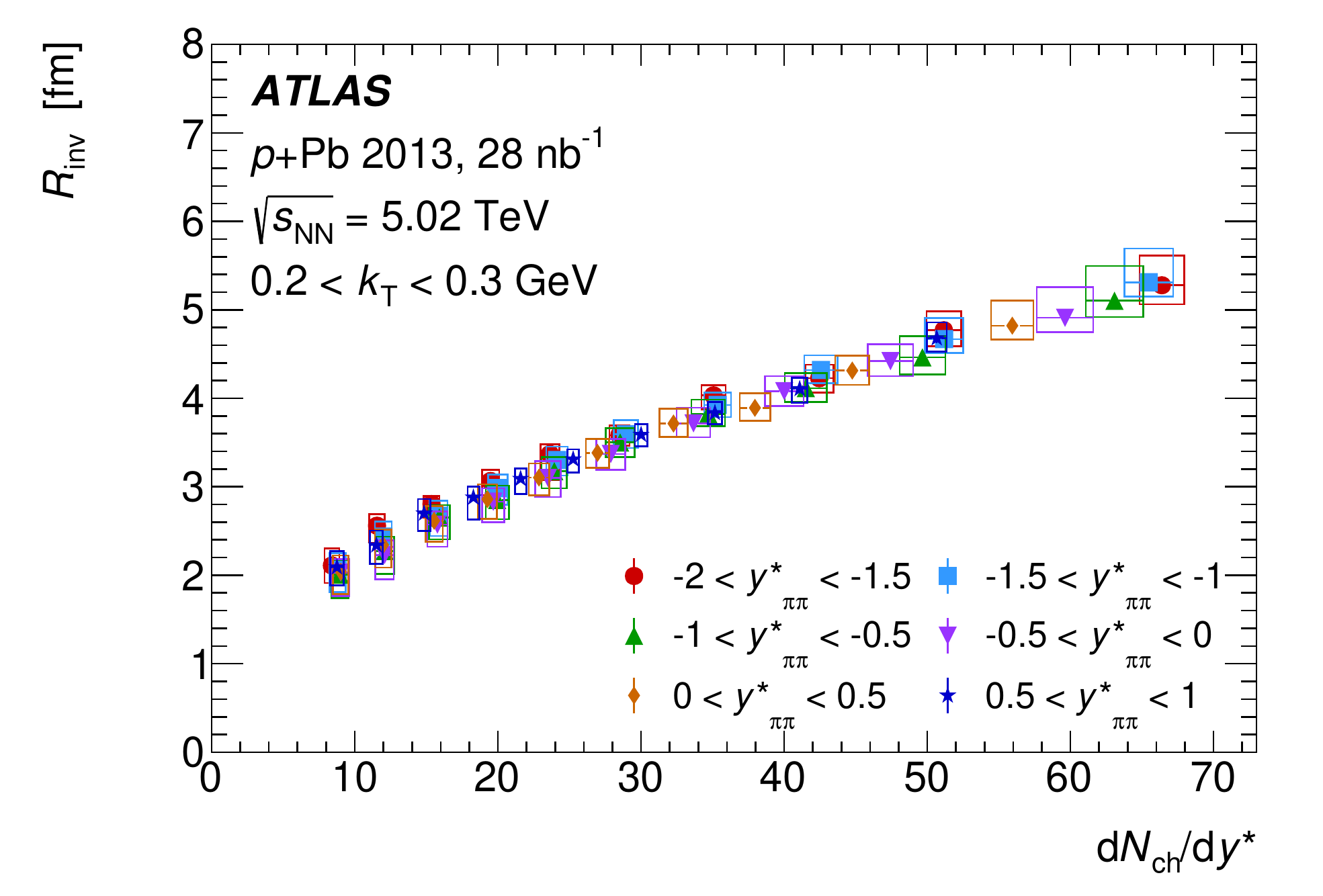}
}
\caption{Exponential fit results for $\Rinv$ as a function of the cube root of average charged-particle multiplicity $\avgdNdeta^{1/3}$ (a), where the average is taken over $|\eta| < 1.5$, and as a function of the local density, \dNdy, over several centrality and rapidity intervals (b). In the left plot the systematic uncertainties from pion identification and from the generator and collision system components of the background amplitude are treated as correlated and shown as error bands, and the systematic uncertainties from charge asymmetry, \Reff, the rapidity variation of the jet fragmentation description, and two-particle reconstruction are treated as uncorrelated and indicated by the height of the boxes. The horizontal error bars indicate the systematic uncertainty from \avgdNdeta or \dNdy.}
\label{fig:results_Rinv_dndeta}
\end{figure}

\begin{figure}[tb]
\centering
\subfloat[]{
\includegraphics[width=0.49\linewidth]{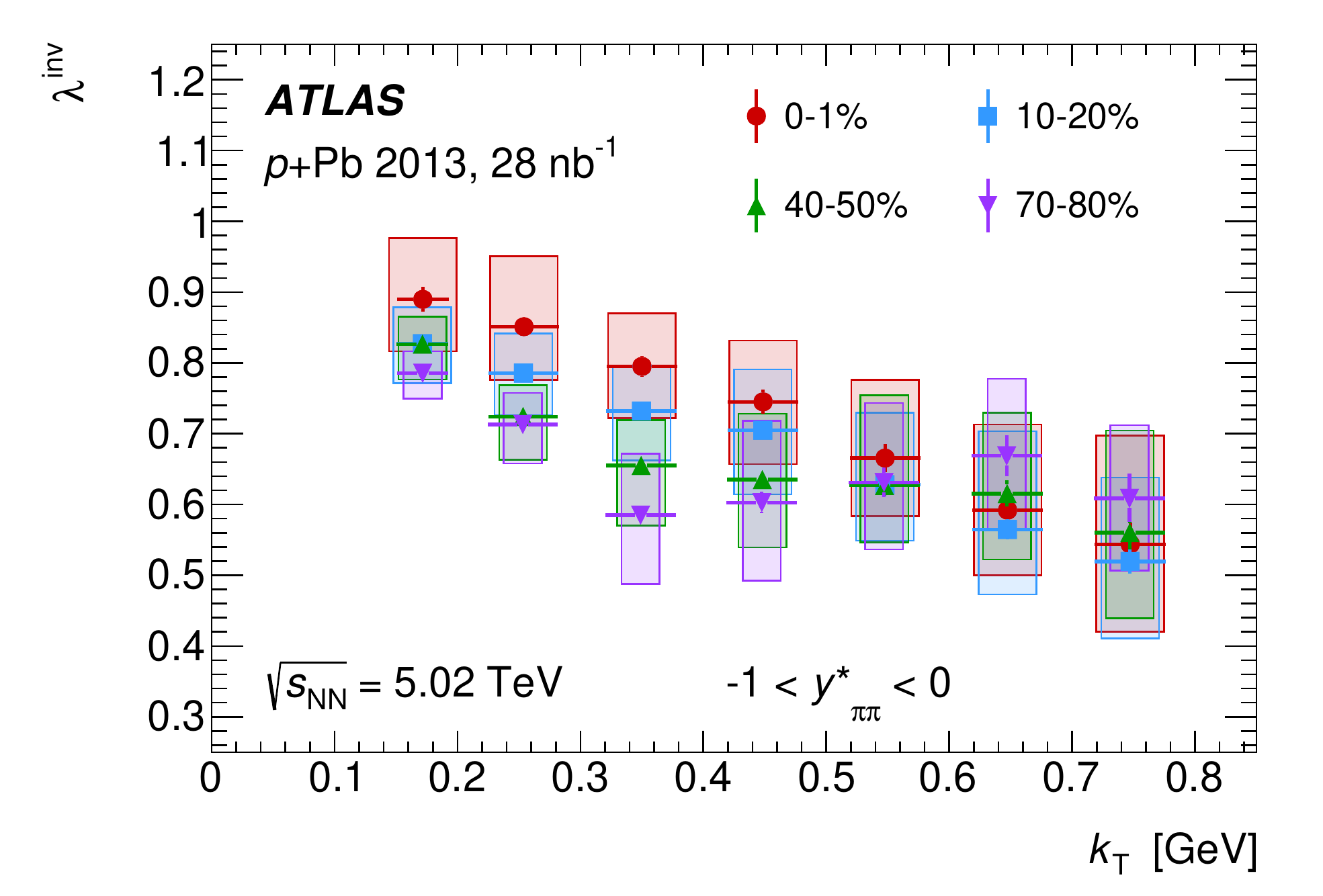}
}
\subfloat[]{
\includegraphics[width=0.49\linewidth]{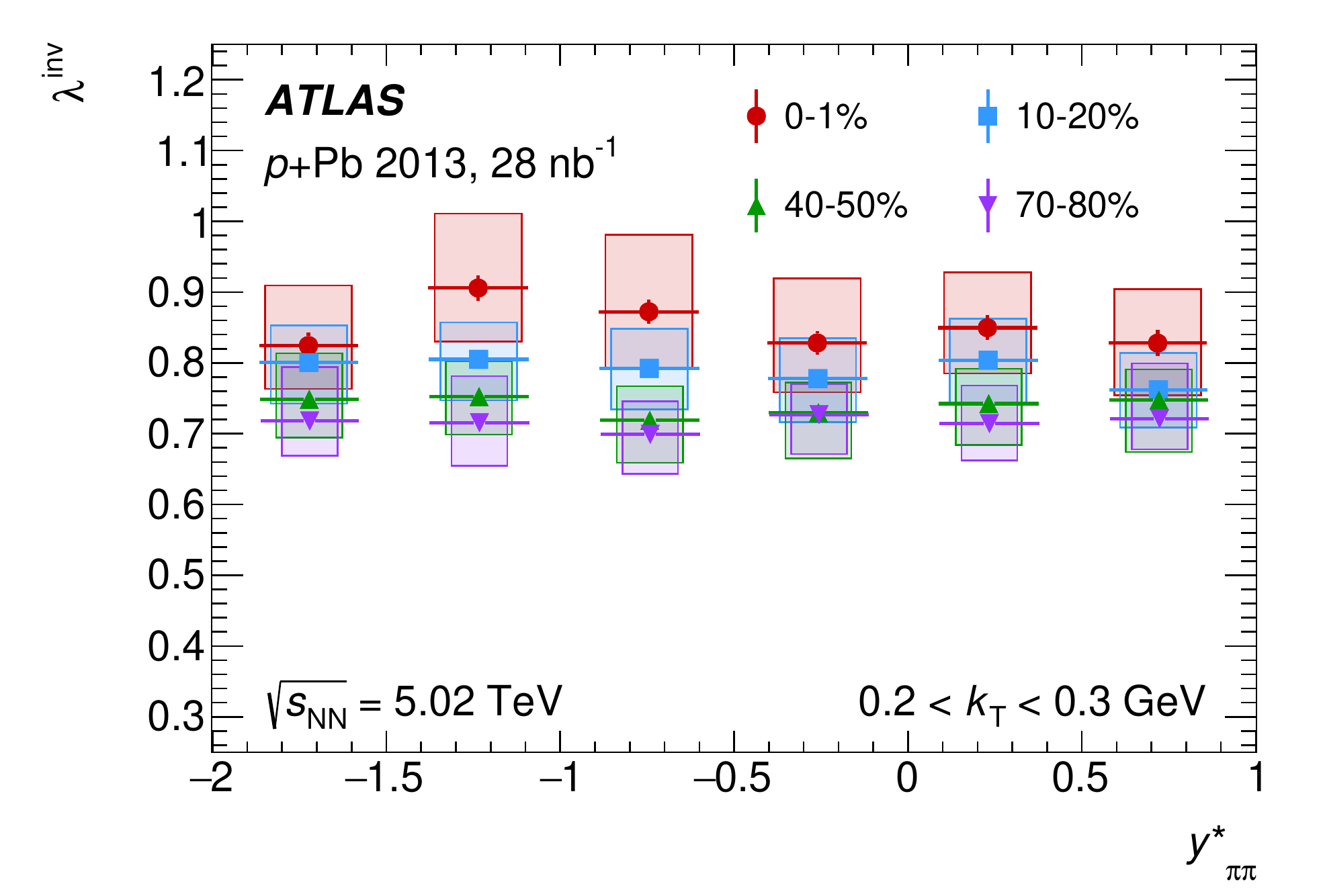}
}
\caption{The Bose-Einstein amplitude, \linv, obtained from one-dimensional fits to
the \qinv\ correlation functions shown as a function of pair transverse momentum, \kt, (a) and rapidity, \kys (b). Four non-adjacent centrality intervals are shown. The vertical size of each box represents the quadrature sum of the systematic uncertainties described in \Sect{\ref{sec:systematics}}, and statistical uncertainties are shown with vertical lines. The horizontal positions of the points are the average \kt or \kys in each interval, and the horizontal lines indicate the standard deviation of \kt or \kys. The widths of the boxes differ among centrality intervals only for visual clarity.}
\label{fig:results_qinv_x}
\end{figure}

The results from fits of $C(\qinv)$ to \Eqn{\ref{eq:correlation_function_full}} for the invariant radius, \Rinv, are shown in \Fig{\ref{fig:results_Rinv_kt}} in four selected centrality intervals.
Only an intermediate rapidity interval $-1 < \kys < 0$ is shown for these and similar results as a function of \kt, as the qualitative behavior is consistent in forward and backward rapidities.
The clear decrease in size with increasing \kt that is observed in central events is not observed in peripheral events.
This is consistent with the interpretation that central events undergo transverse expansion, since in hydrodynamic models higher-\pt particles are more likely to freeze out earlier in the event.
Another way of understanding this trend as evidence for transverse expansion is that there is a smaller homogeneity region for particles with higher \pt \cite{Kolb:2003dz}.
At low \kt, ultra-central (0--1\%) events have an invariant radius significantly greater than peripheral (70--80\%) events by a factor of about 2.6.
This difference becomes less prominent at high \kt.
In central events \Rinv is larger on the lead-going side than on the proton-going side, while in peripheral events the rapidity dependence of the radius becomes constant.

Invariant radii are shown for several centralities in \Fig{\ref{fig:results_Rinv_dndeta}} (left) as a function of the cube root of average \dNdeta. 
For both \kt intervals shown, the scaling of \Rinv with $\avgdNdeta^{1/3}$ is close to linear but with a slightly increasing slope at higher multiplicities.
The invariant radius, \Rinv, has a steeper trend versus multiplicity at lower \kt.
\Fig{\ref{fig:results_Rinv_dndeta}} (right) shows \Rinv in several centrality and rapidity intervals as a function of the local particle density, \dNdy, which is evaluated by taking the average over the same interval used for the pair's rapidity.
The extracted radius and the local particle density are seen to be tightly correlated, such that the radius can be predicted, within uncertainties, by the local density alone.

The Bose-Einstein amplitude of the invariant fits, \linv, is shown
in \Fig{\ref{fig:results_qinv_x}} as a function of \kt and \kys. At low \kt, \linv has values near
unity, and it decreases with rising \kt. In the lower \kt intervals a systematic difference is observed between centrality intervals, with \linv having larger values in central events. In contrast, at larger \kt the amplitudes are indistinguishable between different centralities. The amplitude exhibits no significant variation over rapidity.

\FloatBarrier

\subsection{Three-dimensional results}
\label{subsec:3d_results}

\begin{figure}[tb]
\centering
\subfloat[]{
\includegraphics[width=.49\linewidth]{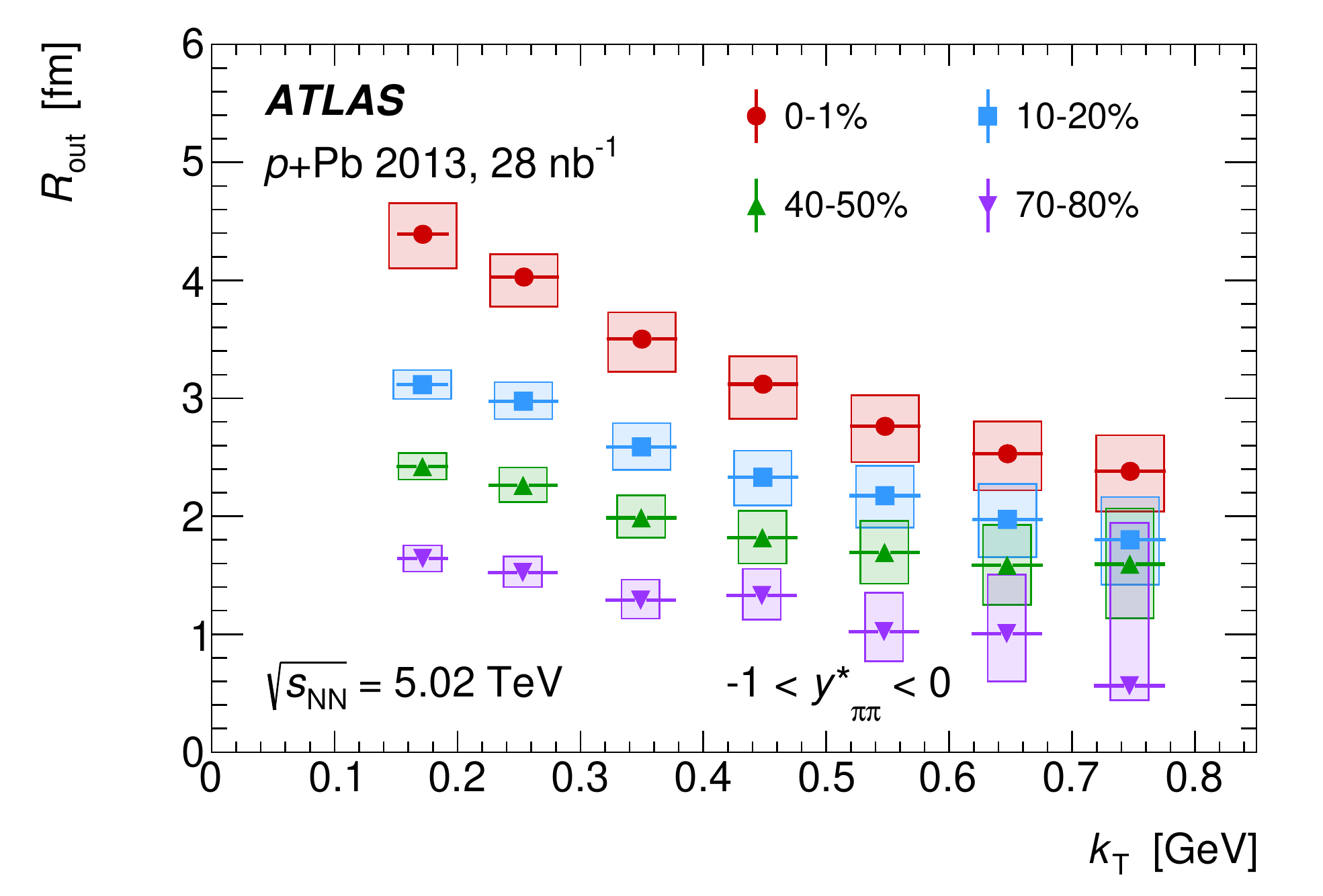}
}
\subfloat[]{
\includegraphics[width=.49\linewidth]{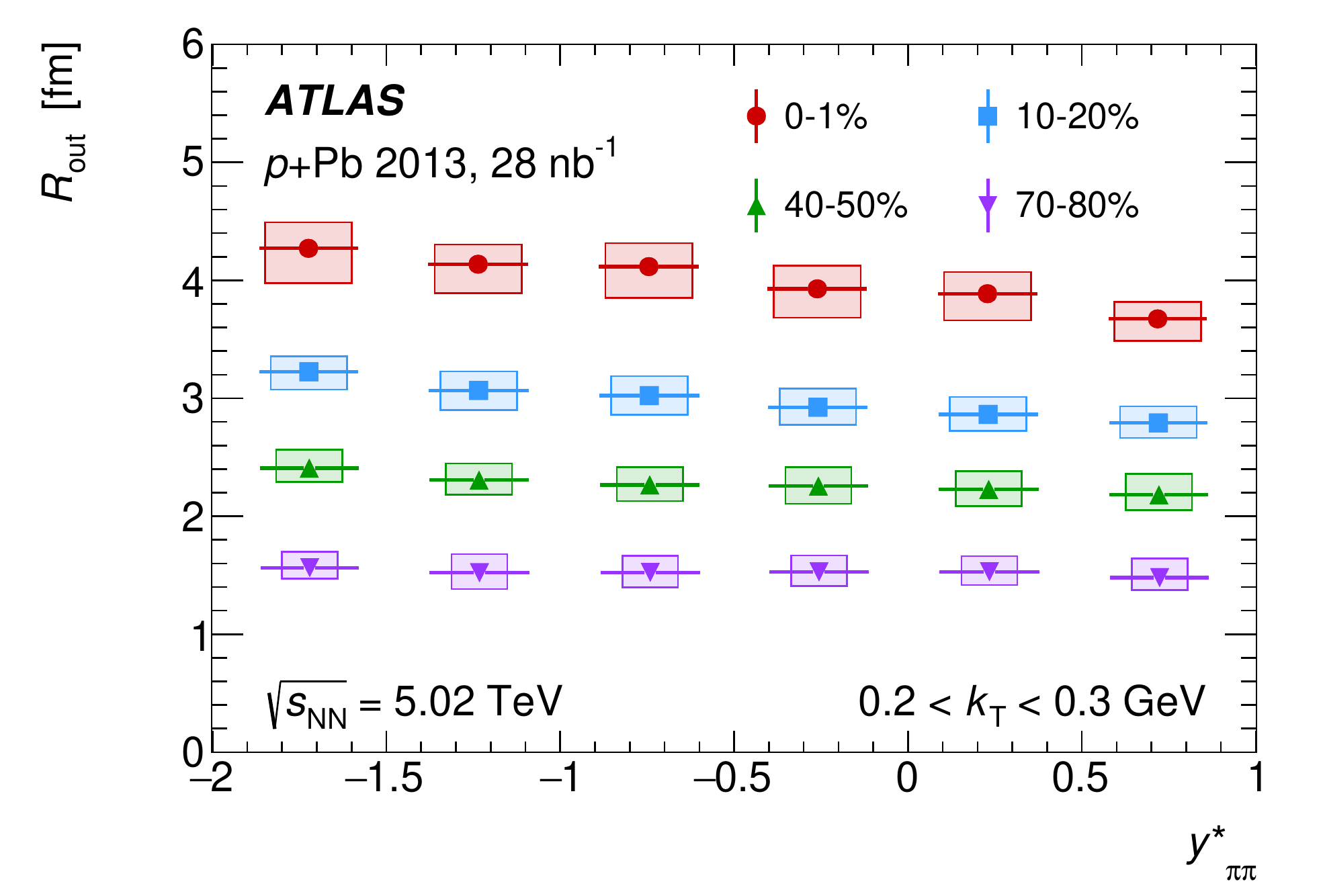}
}
\caption{Exponential fit results for the 3D source radius, \Rout, as a function of pair transverse momentum, \kt, (a) and rapidity, \kys (b). Four non-adjacent centrality intervals are shown. The vertical size of each box represents the quadrature sum of the systematic uncertainties described in \Sect{\ref{sec:systematics}}, and statistical uncertainties are shown with vertical lines. The horizontal positions of the points are the average \kt or \kys in each interval, and the horizontal lines indicate the standard deviation of \kt or \kys. The widths of the boxes differ among centrality intervals only for visual clarity.}
\label{fig:results_Rout}
\end{figure}

\begin{figure}[h!]
\centering
\subfloat[]{
\includegraphics[width=.49\linewidth]{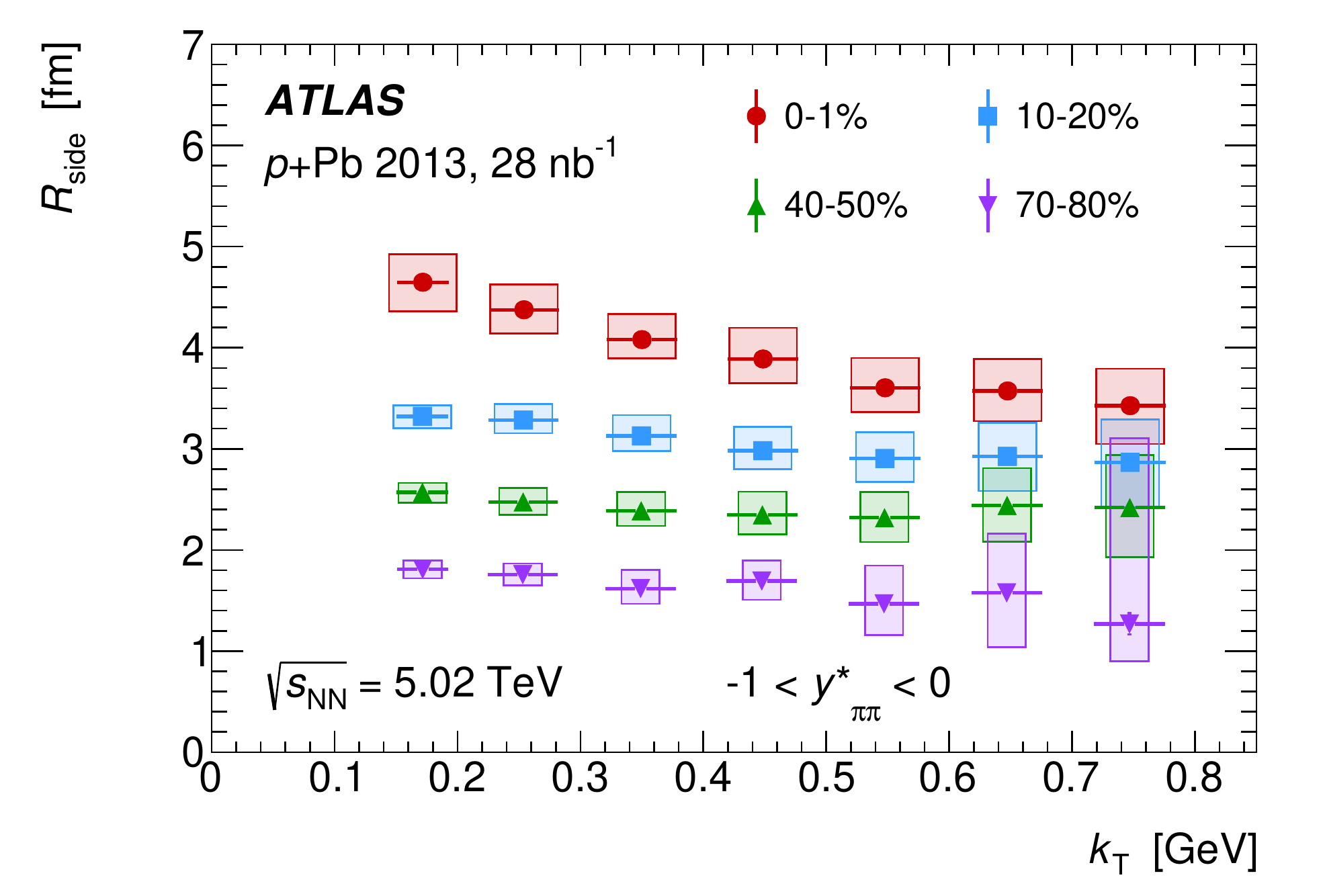}
}
\subfloat[]{
\includegraphics[width=.49\linewidth]{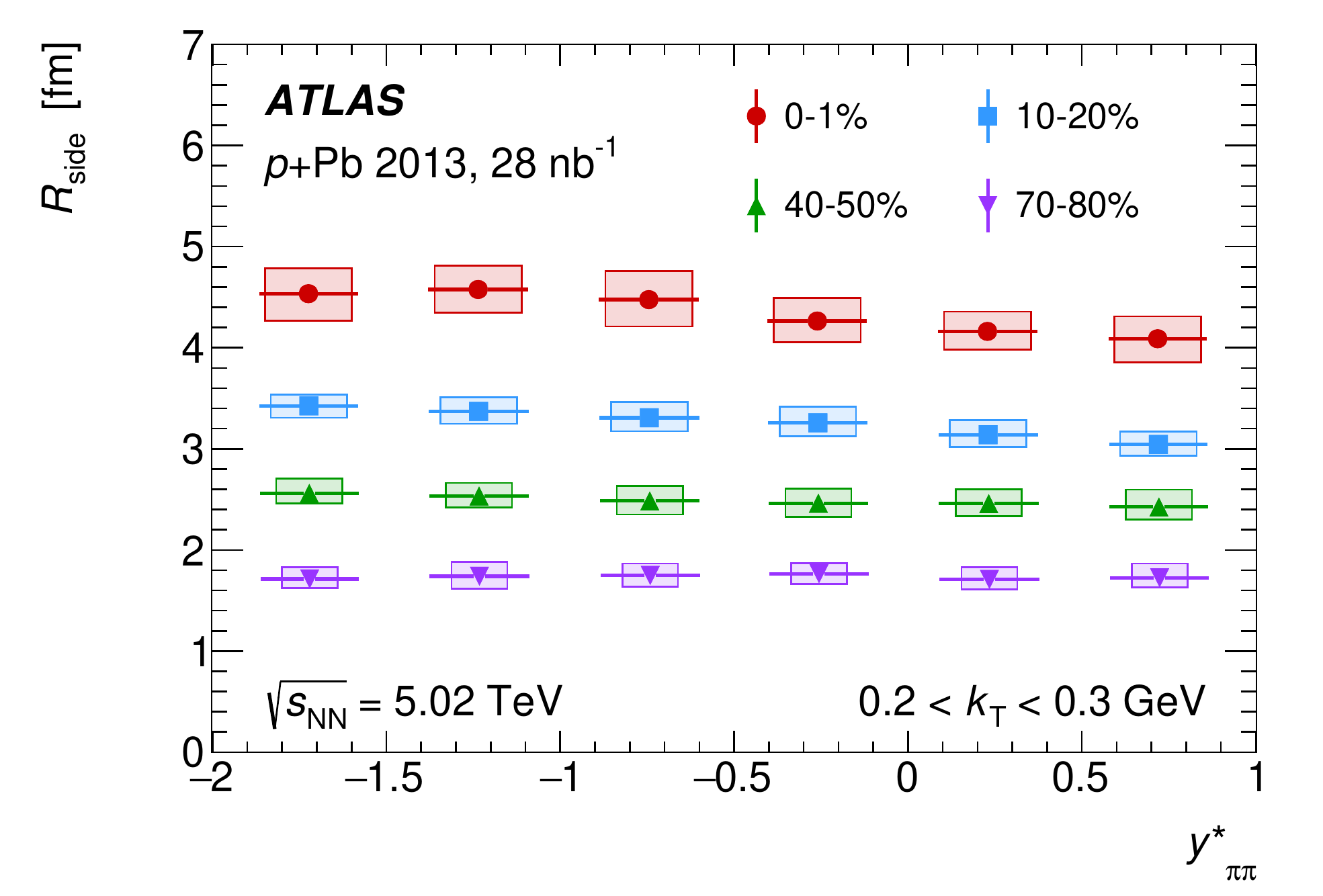}
}
\caption{Exponential fit results for the 3D source radius, \Rside, as a function of pair transverse momentum, \kt, (a) and rapidity, \kys (b). Four non-adjacent centrality intervals are shown. The vertical size of each box represents the quadrature sum of the systematic uncertainties described in \Sect{\ref{sec:systematics}}, and statistical uncertainties are shown with vertical lines. The horizontal positions of the points are the average \kt or \kys in each interval, and the horizontal lines indicate the standard deviation of \kt or \kys. The widths of the boxes differ among centrality intervals only for visual clarity.}
\label{fig:results_Rside}
\end{figure}

\begin{figure}[t]
\centering
\subfloat[]{
\includegraphics[width=.49\linewidth]{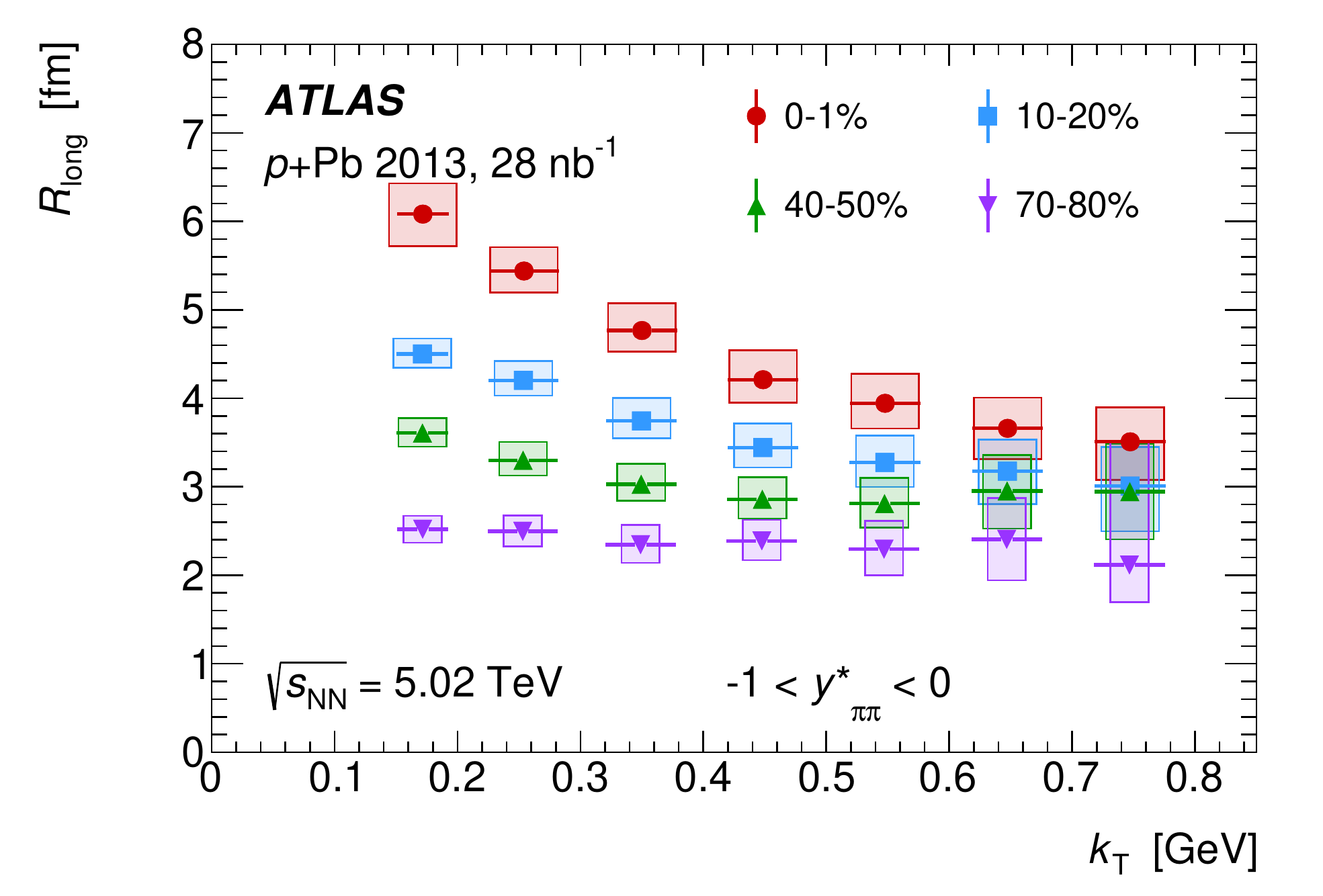}
}
\subfloat[]{
\includegraphics[width=.49\linewidth]{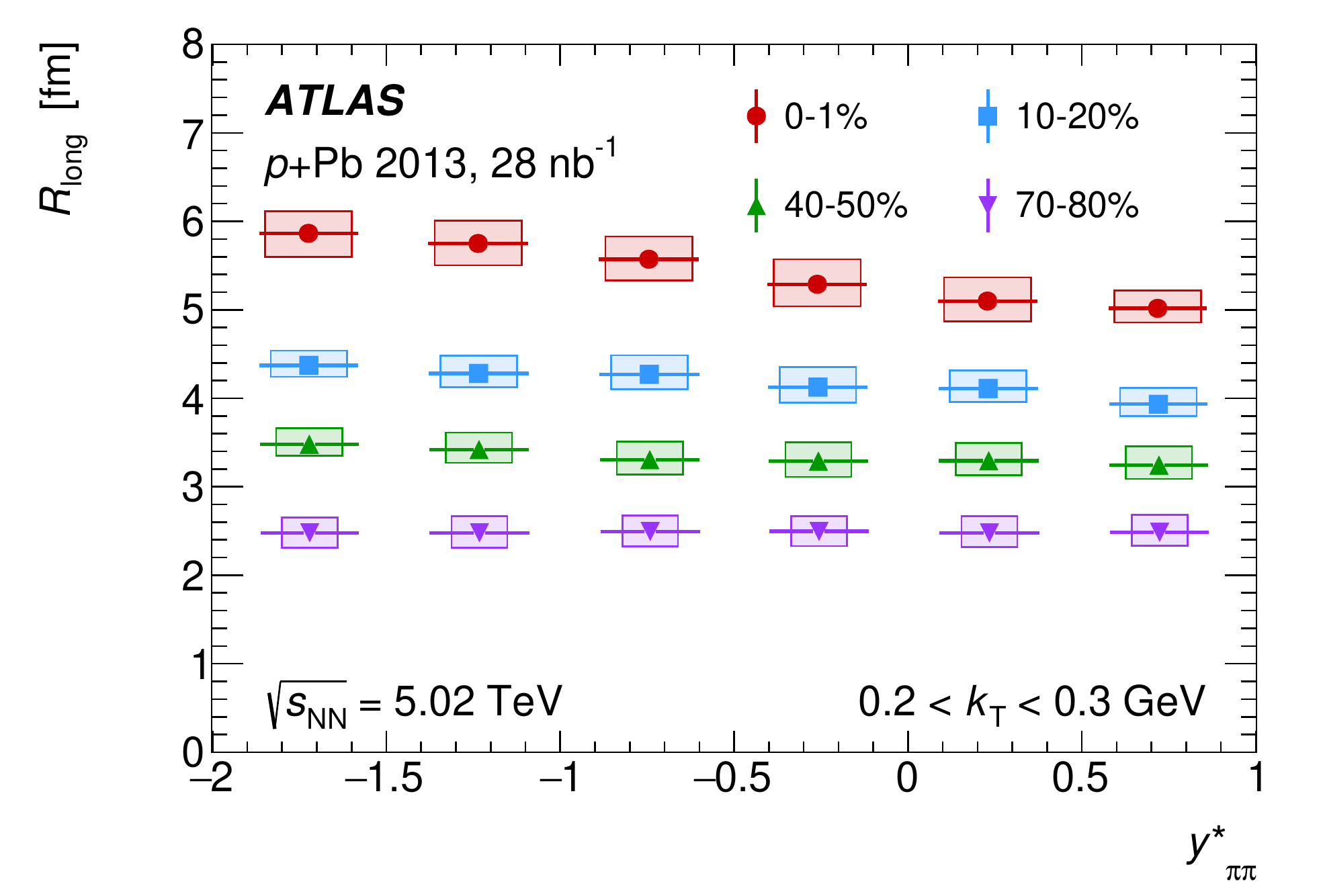}
}
\caption{Exponential fit results for 3D source radius, \Rlong, as a function of pair transverse momentum, \kt, (a) and rapidity, \kys (b). Four non-adjacent centrality intervals are shown. The vertical size of each box represents the quadrature sum of the systematic uncertainties described in \Sect{\ref{sec:systematics}}, and statistical uncertainties are shown with vertical lines. The horizontal positions of the points are the average \kt or \kys in each interval, and the horizontal lines indicate the standard deviation of \kt or \kys. The widths of the boxes differ among centrality intervals only for visual clarity.}
\label{fig:results_Rlong}
\end{figure}

\begin{figure}[h!]
\centering
\subfloat[]{
\includegraphics[width=0.49\linewidth]{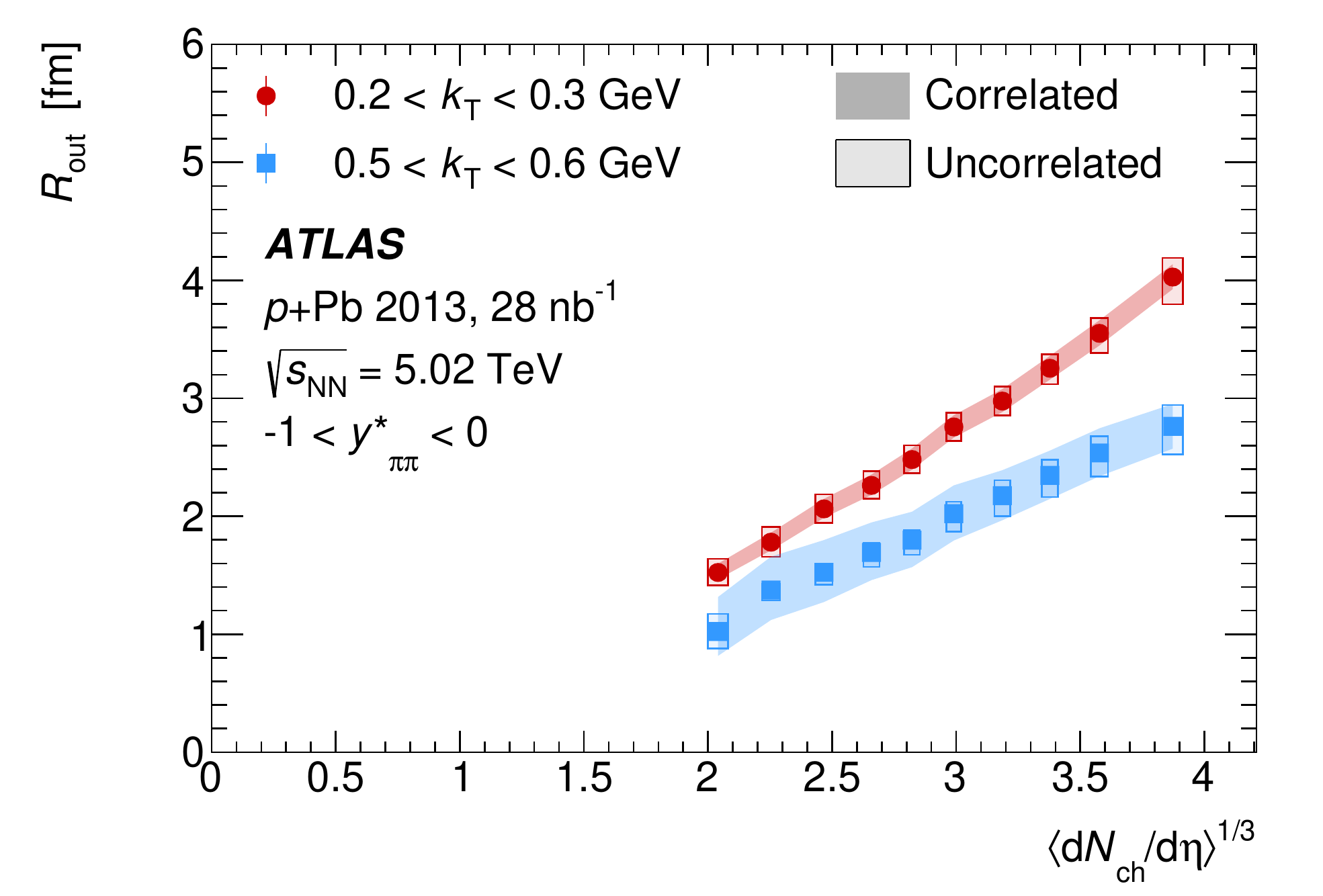}
}
\subfloat[]{
\includegraphics[width=0.49\linewidth]{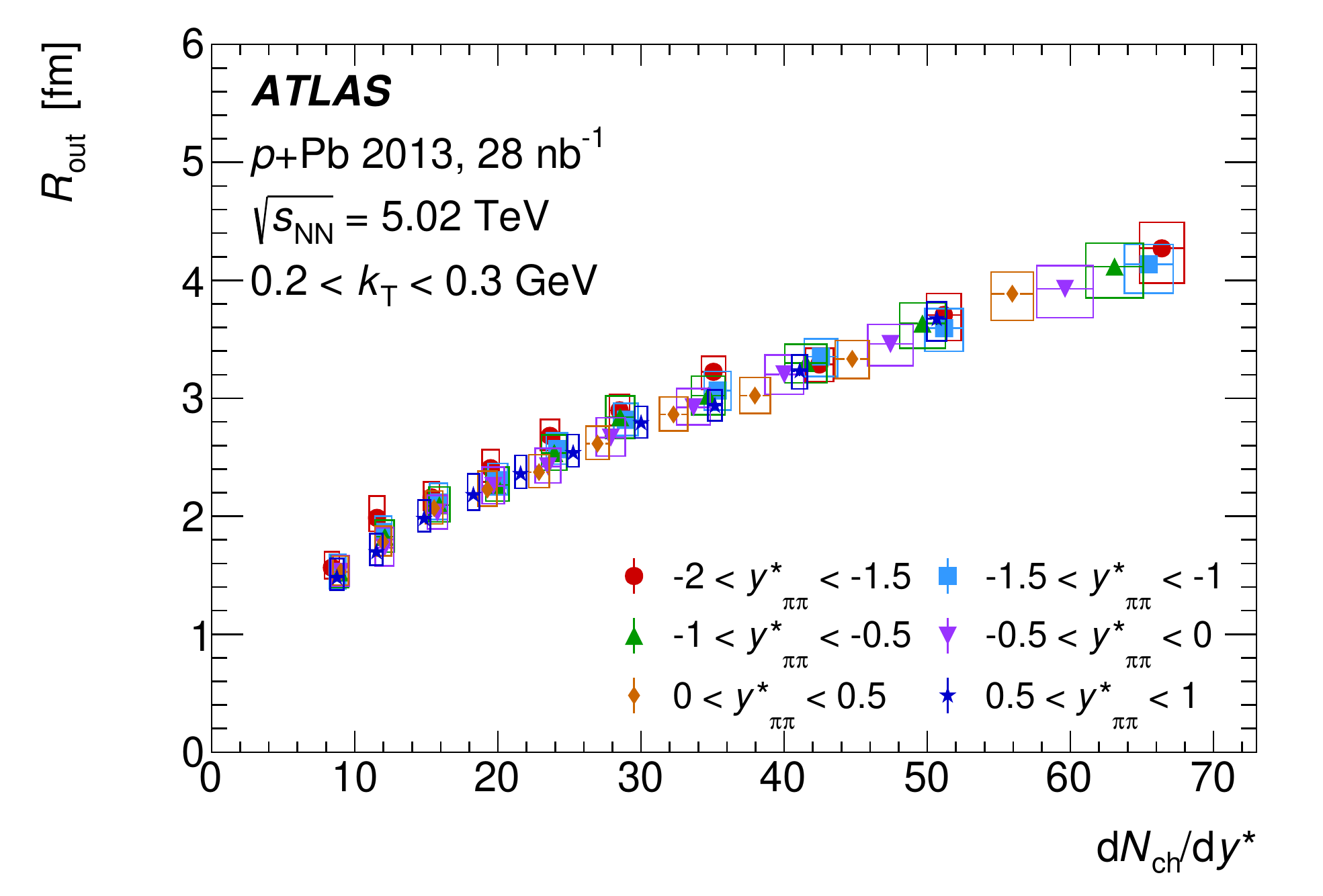}
}
\caption{Exponential fit results for $\Rout$ as a function of (a) the cube root of average charged-particle multiplicity, $\avgdNdeta^{1/3}$, where the average is taken over $|\eta| < 1.5$ and (b) the local density, \dNdy, in intervals of \kys. In the left plot the systematic uncertainties from pion identification and from the generator and collision system components of the background amplitude are treated as correlated and shown as error bands, while the systematic uncertainties from charge asymmetry, \Reff, the rapidity variation of the jet fragmentation description, and two-particle reconstruction are treated as uncorrelated and indicated by the height of the boxes. The horizontal error bars indicate the systematic uncertainty from \avgdNdeta or \dNdy.}
\label{fig:results_Rout_mult}
\end{figure}

The three-dimensional radii \Rout, \Rside, and \Rlong are shown as a
function of \kt and \kys in four selected centrality intervals
in \Figrange{\ref{fig:results_Rout}}{\ref{fig:results_Rlong}}.
In central collisions, the 3D radii exhibit an even steeper decrease with increasing \kt relative to that observed for the invariant radii in \Fig{\ref{fig:results_Rinv_kt}}.
A similar, but weaker trend is present in peripheral events.
Central collisions exhibit larger radii on the backward (Pb-going) side of the event, while peripheral events show no distinguishable variation of the radii with rapidity.

\begin{figure}[ht!]
\centering
\subfloat[]{
\includegraphics[width=0.49\linewidth]{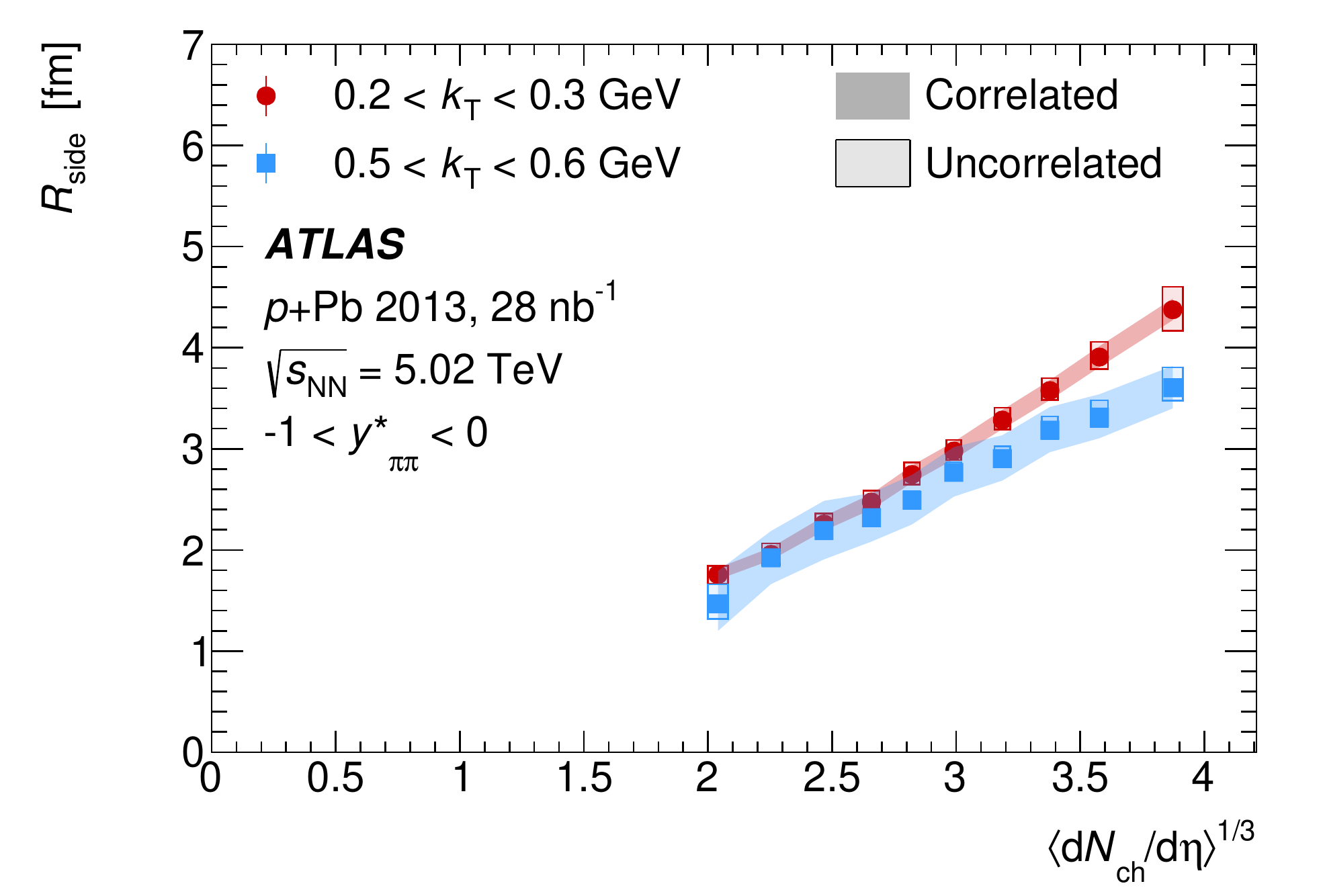}
}
\subfloat[]{
\includegraphics[width=0.49\linewidth]{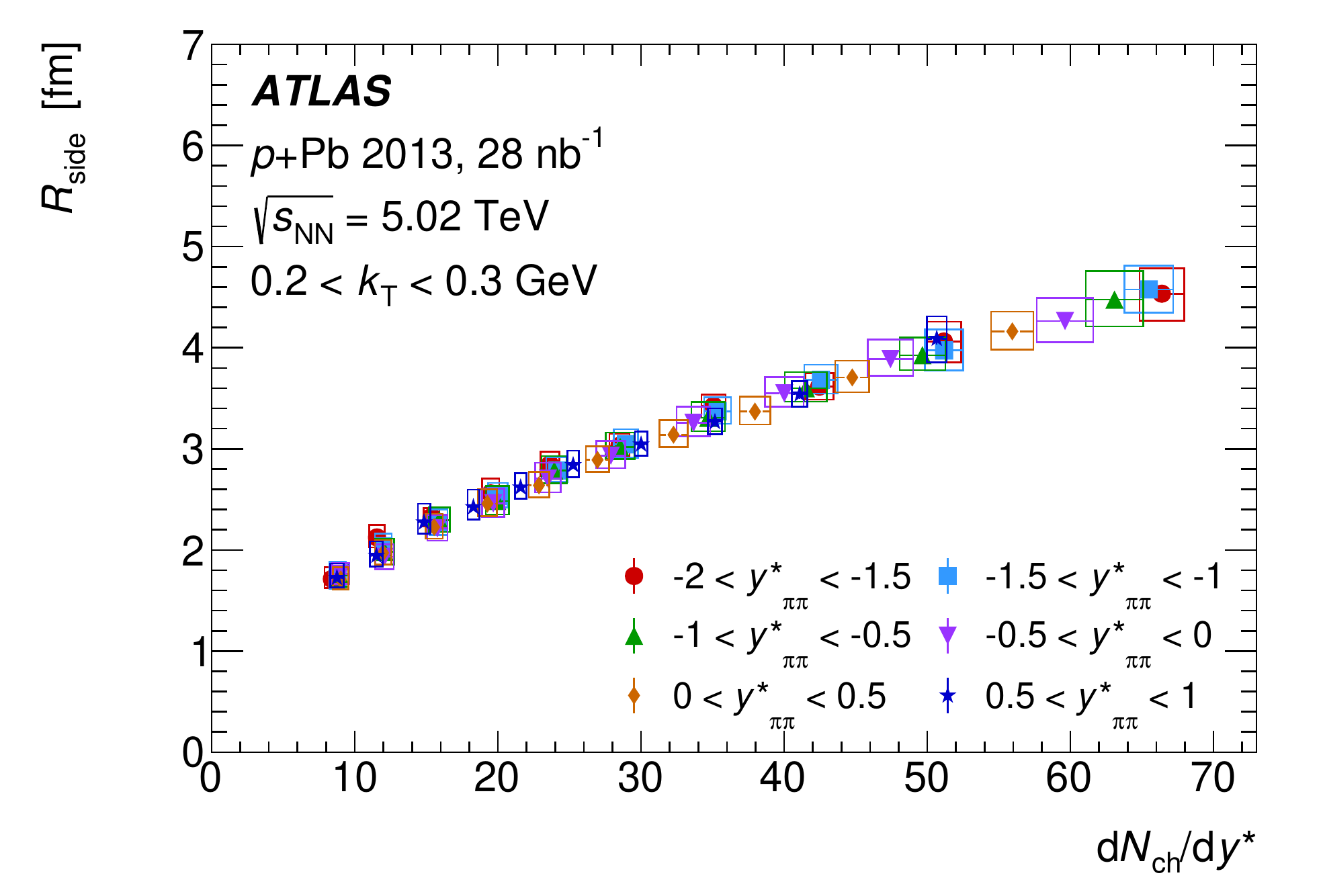}
}
\caption{Exponential fit results for $\Rside$ as a function of (a) the cube root of average charged-particle multiplicity, $\avgdNdeta^{1/3}$, where the average is taken over $|\eta| < 1.5$ and (b) the local density, \dNdy, in intervals of \kys. In the left plot the systematic uncertainties from pion identification and from the generator and collision system components of the background amplitude are treated as correlated and shown as error bands, while the systematic uncertainties from charge asymmetry, \Reff, the rapidity variation of the jet fragmentation description, and two-particle reconstruction are treated as uncorrelated and indicated by the height of the boxes. The horizontal error bars indicate the systematic uncertainty from \avgdNdeta or \dNdy.}
\label{fig:results_Rside_mult}
\end{figure}

\begin{figure}[h!]
\centering
\subfloat[]{
\includegraphics[width=0.49\linewidth]{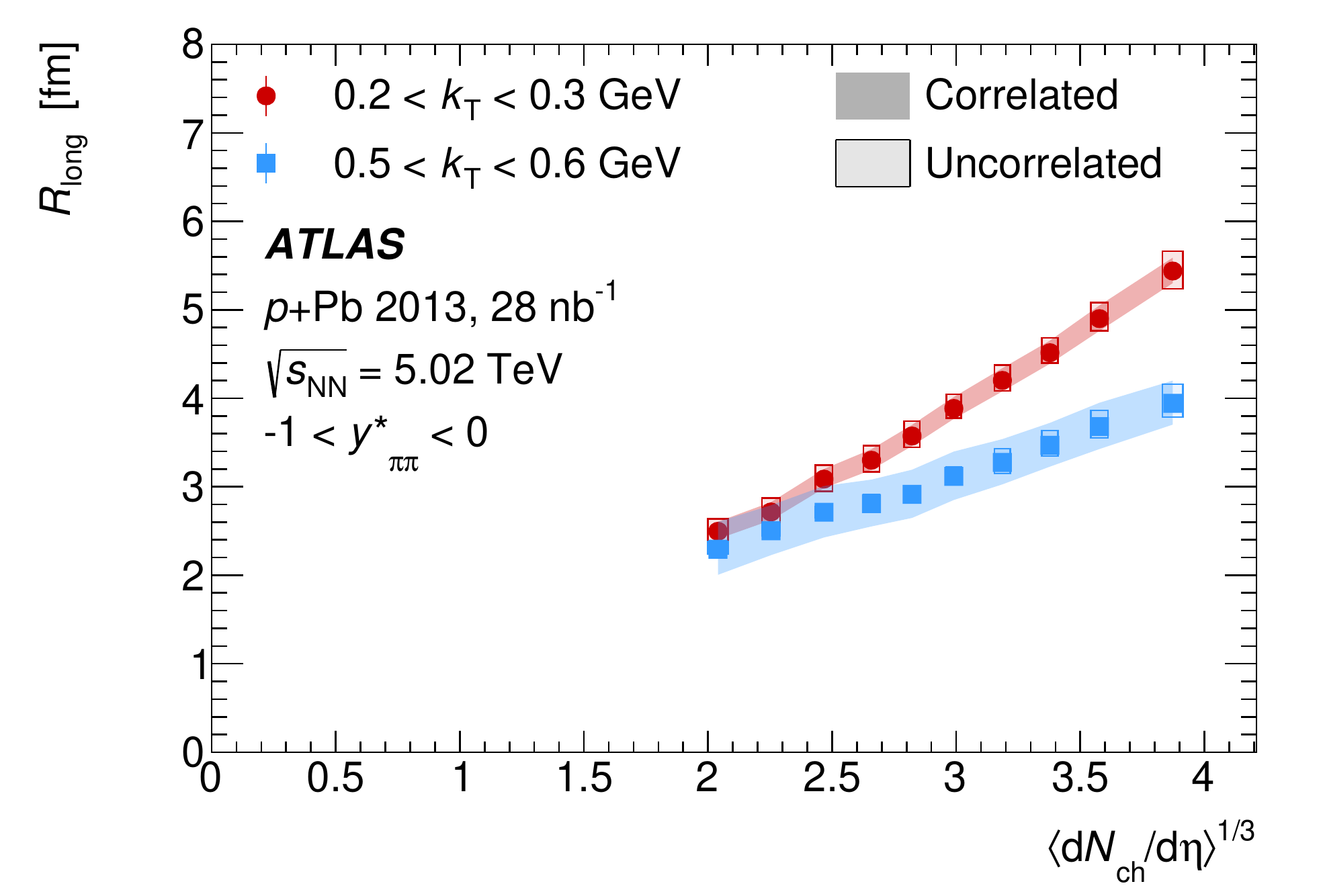}
}
\subfloat[]{
\includegraphics[width=0.49\linewidth]{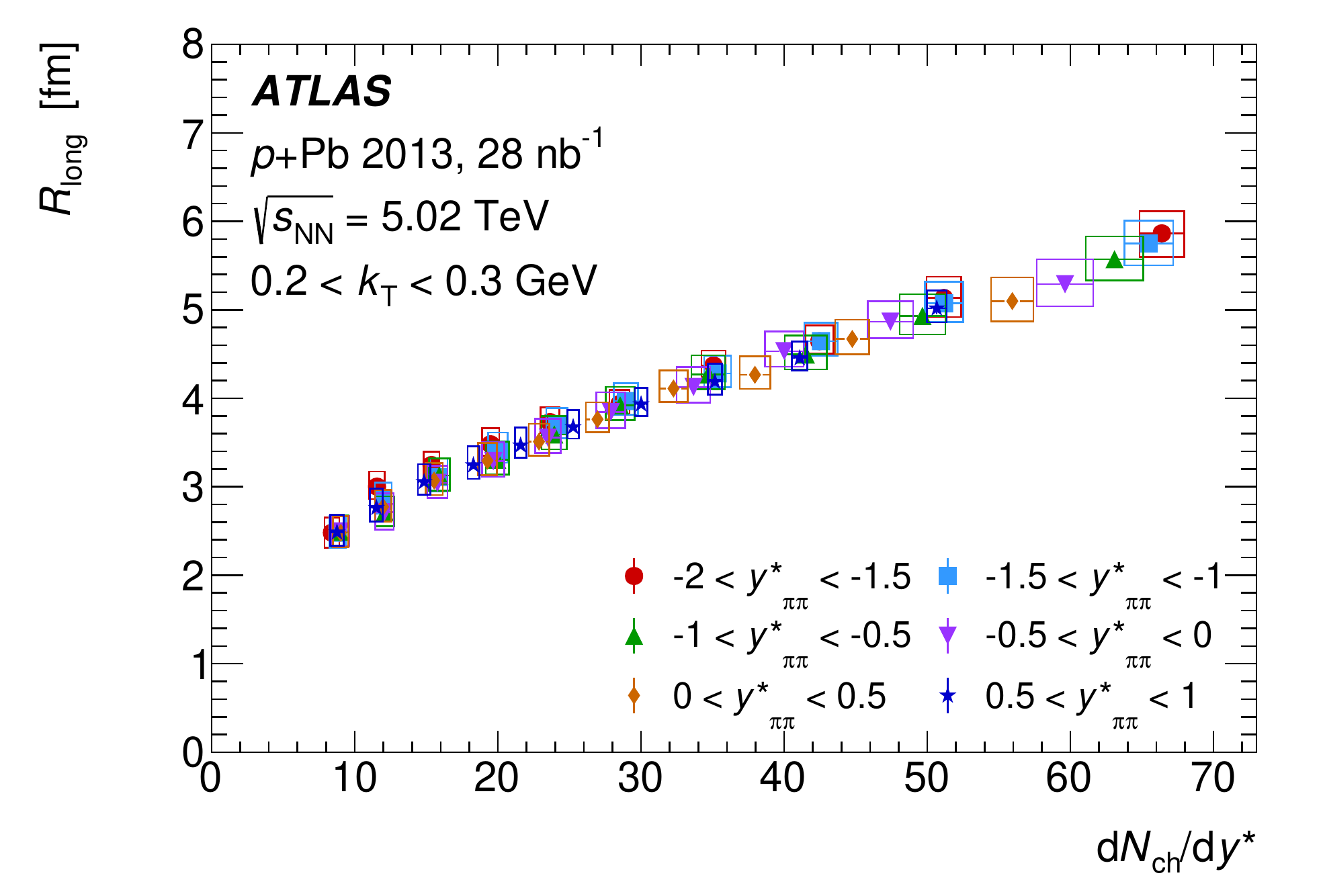}
}
\caption{Exponential fit results for $\Rlong$ as a function of (a) the cube root of average charged-particle multiplicity, $\avgdNdeta^{1/3}$, where the average is taken over $|\eta| < 1.5$  and (b) the local density, \dNdy, in intervals of \kys. In the left plot the systematic uncertainties from pion identification and from the generator and collision system components of the background amplitude are treated as correlated and shown as error bands, while the systematic uncertainties from charge asymmetry, \Reff, the rapidity variation of the jet fragmentation description, and two-particle reconstruction are treated as uncorrelated and indicated by the height of the boxes. The horizontal error bars indicate the systematic uncertainty from \avgdNdeta or \dNdy.}
\label{fig:results_Rlong_mult}
\end{figure}

The 3D radii are also shown as a function of the cube root of both
average event multiplicity and local density
in \Figrange{\ref{fig:results_Rout_mult}}{\ref{fig:results_Rlong_mult}}.
These plots demonstrate the relationship between the size and the density of the source at freeze-out.
All of the radii are seen to be very strongly correlated with the local density.
The scaling of the radii is not far from being linear with the cube root of multiplicity.
This behavior is qualitatively similar to the scaling of \Rinv with \avgdNdeta in \Fig{\ref{fig:results_Rinv_dndeta}}.

\begin{figure}[ht]
\centering
\subfloat[]{
\includegraphics[width=0.49\linewidth]{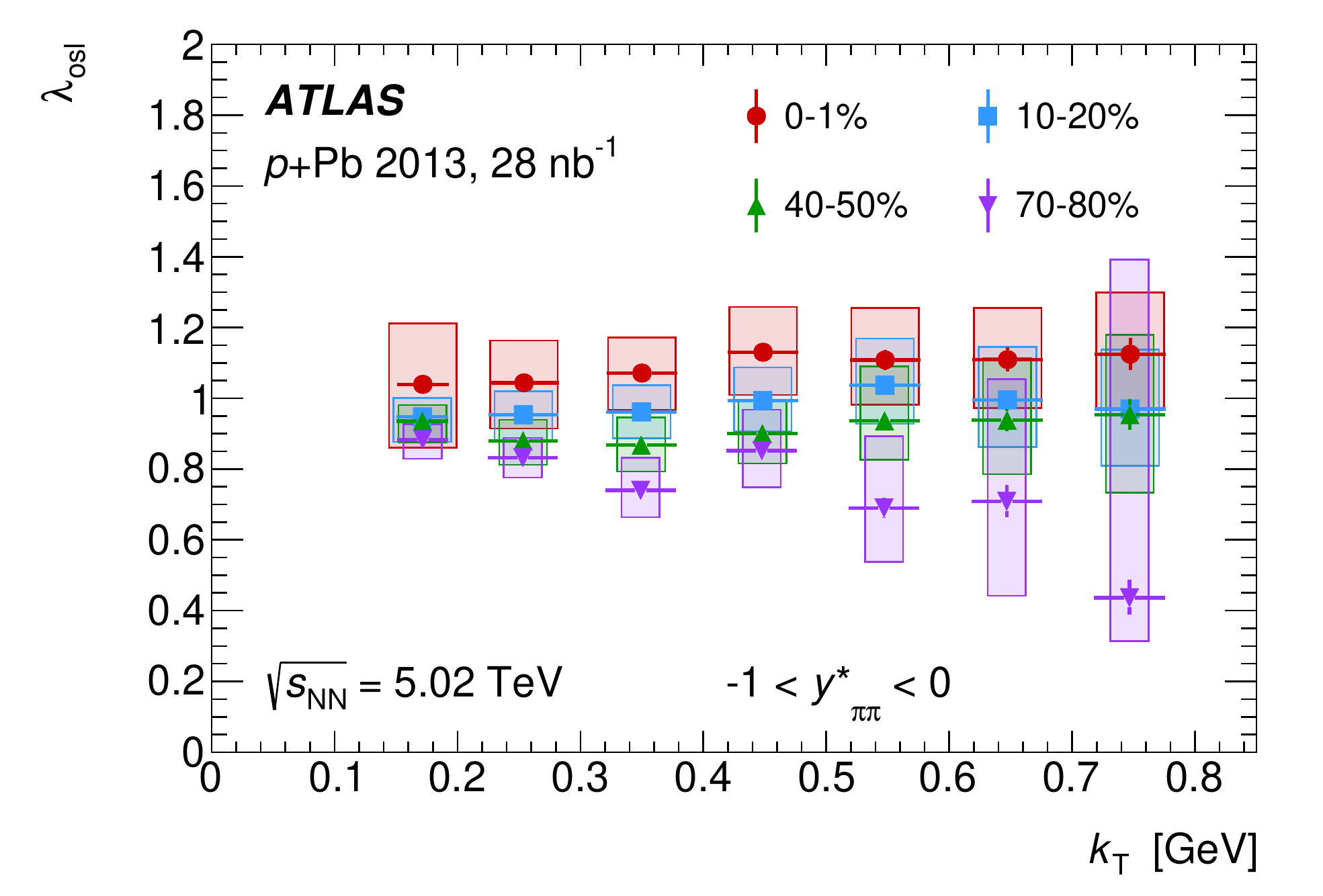}
}
\subfloat[]{
\includegraphics[width=0.49\linewidth]{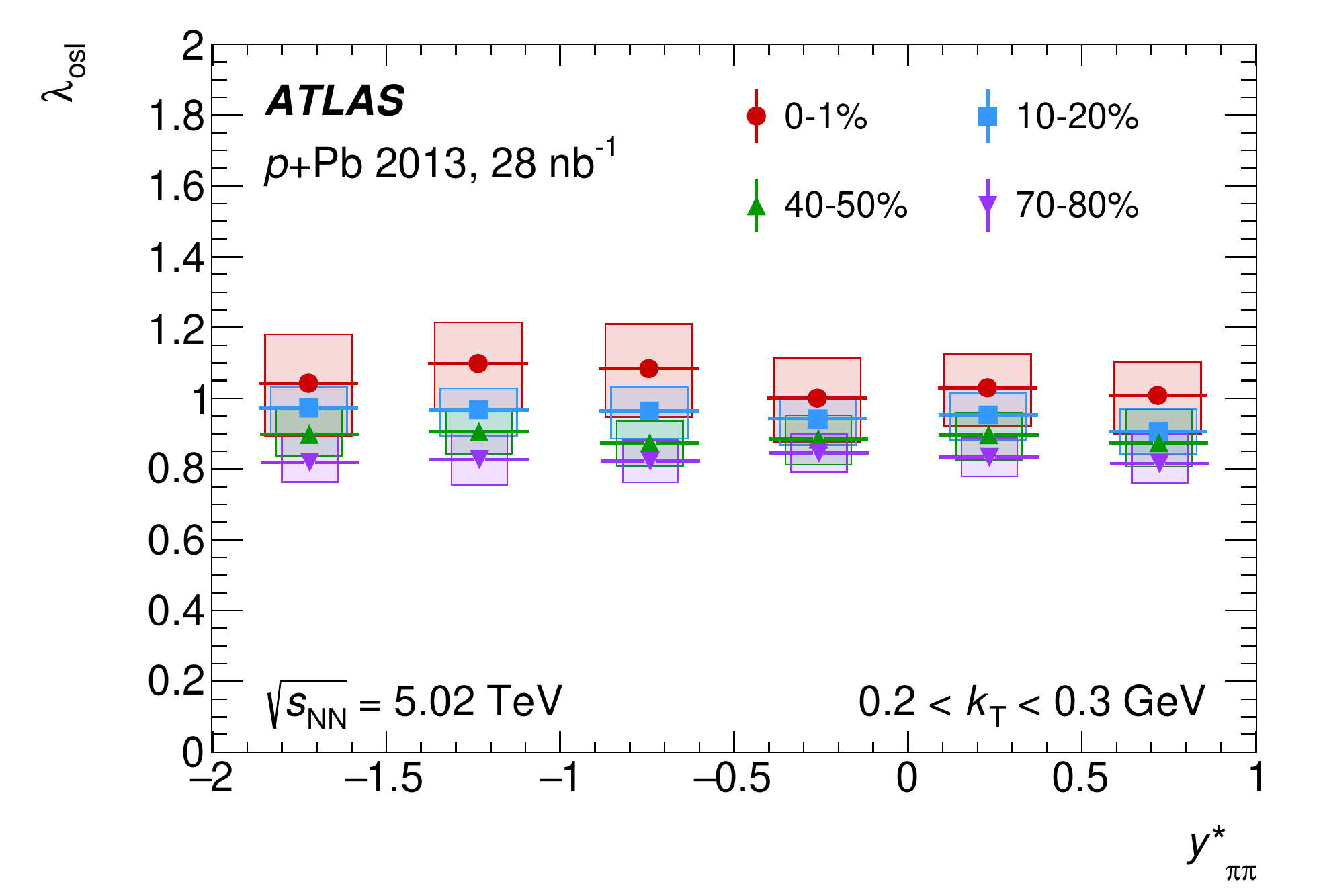}
}
\caption{The Bose-Einstein amplitude, \losl, as a function of pair transverse momentum, \kt, (la) and rapidity, \kys (b). Four non-adjacent centrality intervals are shown. The vertical size of each box represents the quadrature sum of the systematic uncertainties described in \Sect{\ref{sec:systematics}}, and statistical uncertainties are shown with vertical lines. The horizontal positions of the points are the average \kt or \kys in each interval, and the horizontal lines indicate the standard deviation of \kt or \kys. The widths of the boxes differ among centrality intervals only for visual clarity.}
\label{fig:results_qosl_x}
\end{figure}

\begin{figure}[ht]
\centering
\subfloat[]{
\includegraphics[width=0.49\linewidth]{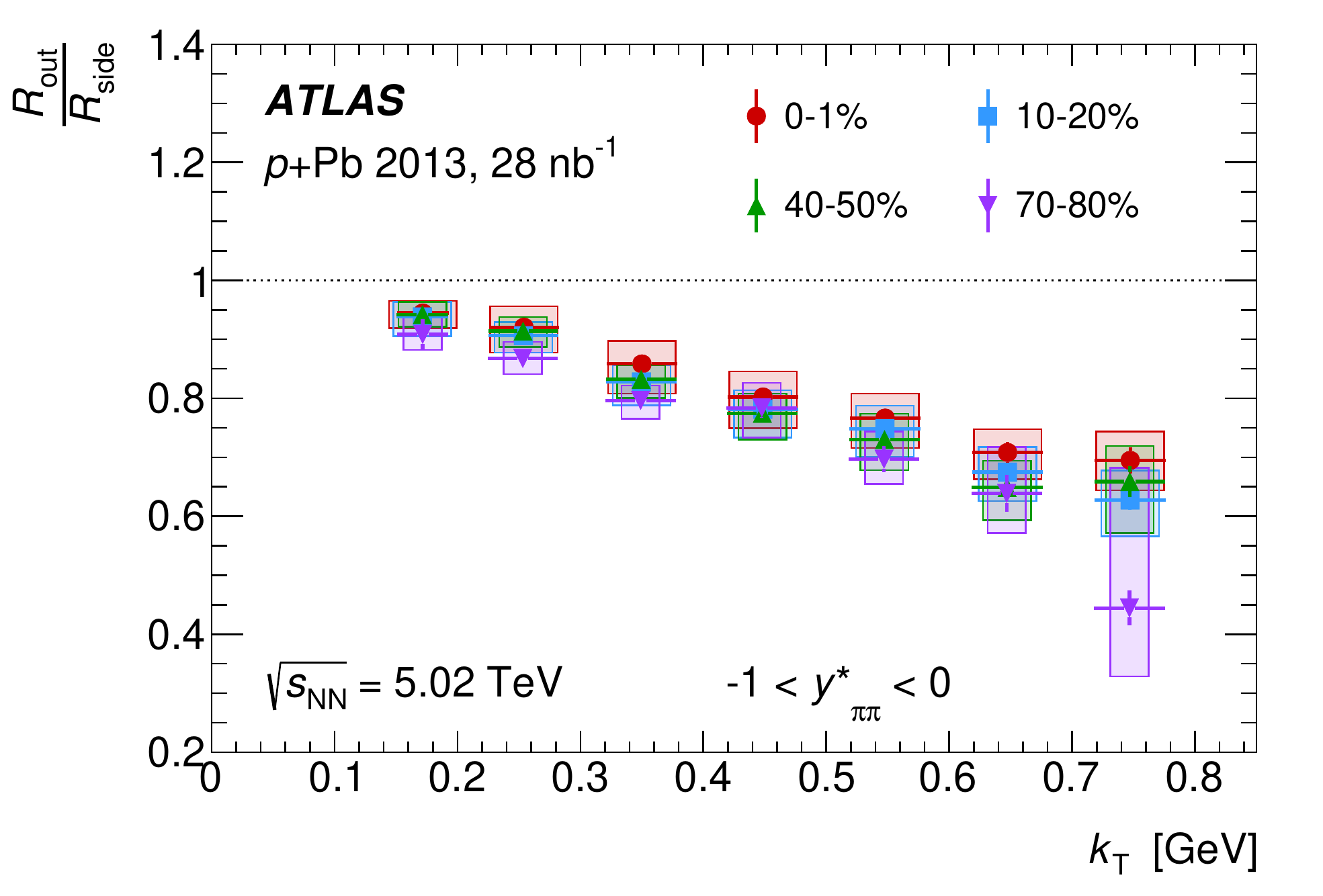}
}
\subfloat[]{
\includegraphics[width=0.49\linewidth]{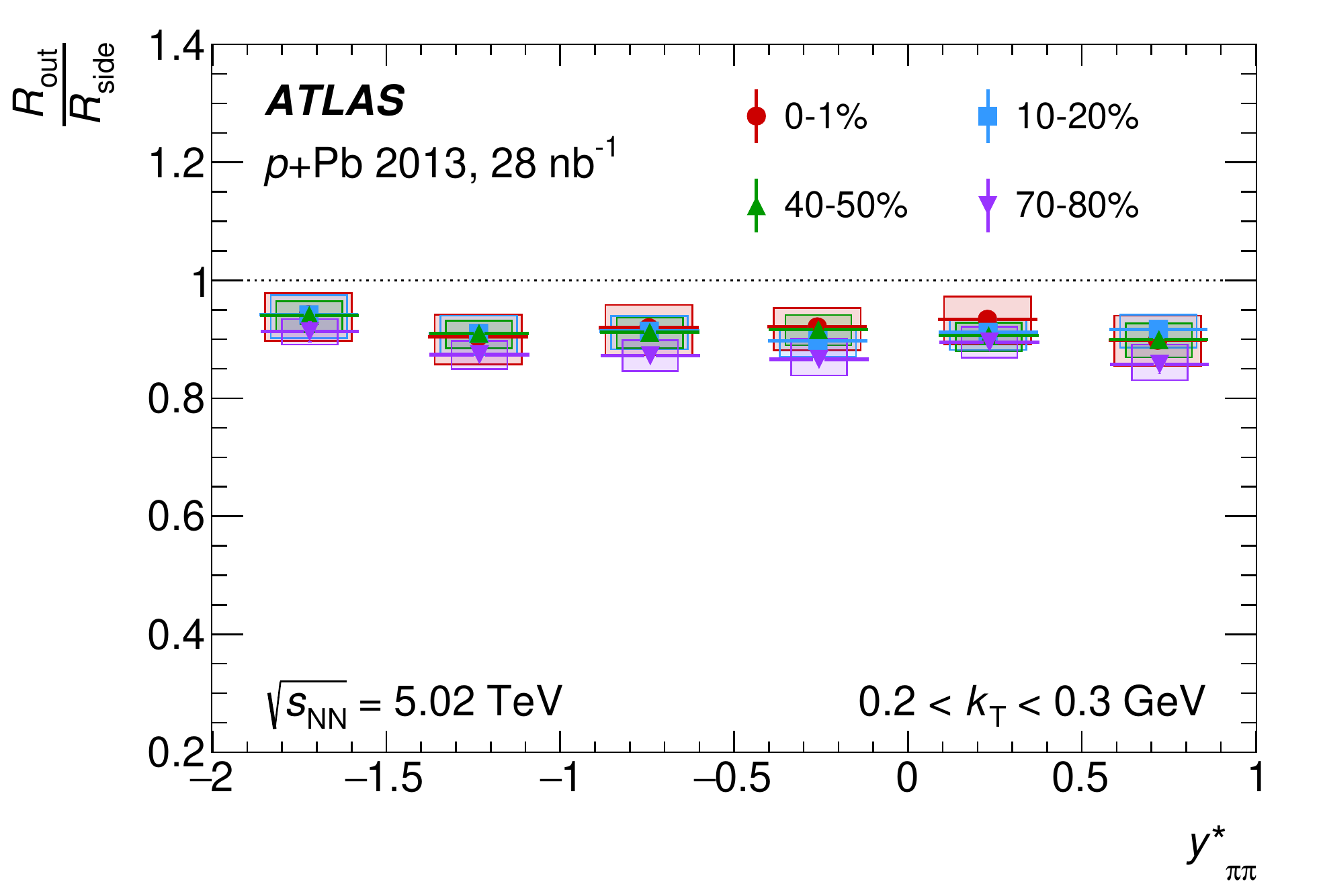}
}
\caption{The ratio of exponential radii $\Rout/\Rside$ as a function of pair transverse momentum, \kt, (a) and rapidity, \kys (b). Four non-adjacent centrality intervals are shown. The vertical size of each box represents the quadrature sum of the systematic uncertainties described in \Sect{\ref{sec:systematics}}, and statistical uncertainties are shown with vertical lines. The horizontal positions of the points are the average \kt or \kys in each interval, and the horizontal lines indicate the standard deviation of \kt or \kys. The widths of the boxes differ among centrality intervals only for visual clarity.}
\label{fig:results_RoutOverRside_kt}
\end{figure}

The Bose-Einstein amplitude in the 3D fits, \losl, is shown in \Fig{\ref{fig:results_qosl_x}} as a function of \kt and \kys. Like the invariant amplitude, at low \kt it is larger for central events than for peripheral ones. The three-dimensional amplitude does not decrease significantly with rising \kt as the invariant amplitude does, except in the most peripheral events. The 3D amplitude also exhibits no significant variation over rapidity.
\begin{figure}[ht]
\centering
\subfloat[]{
\includegraphics[width=0.49\linewidth]{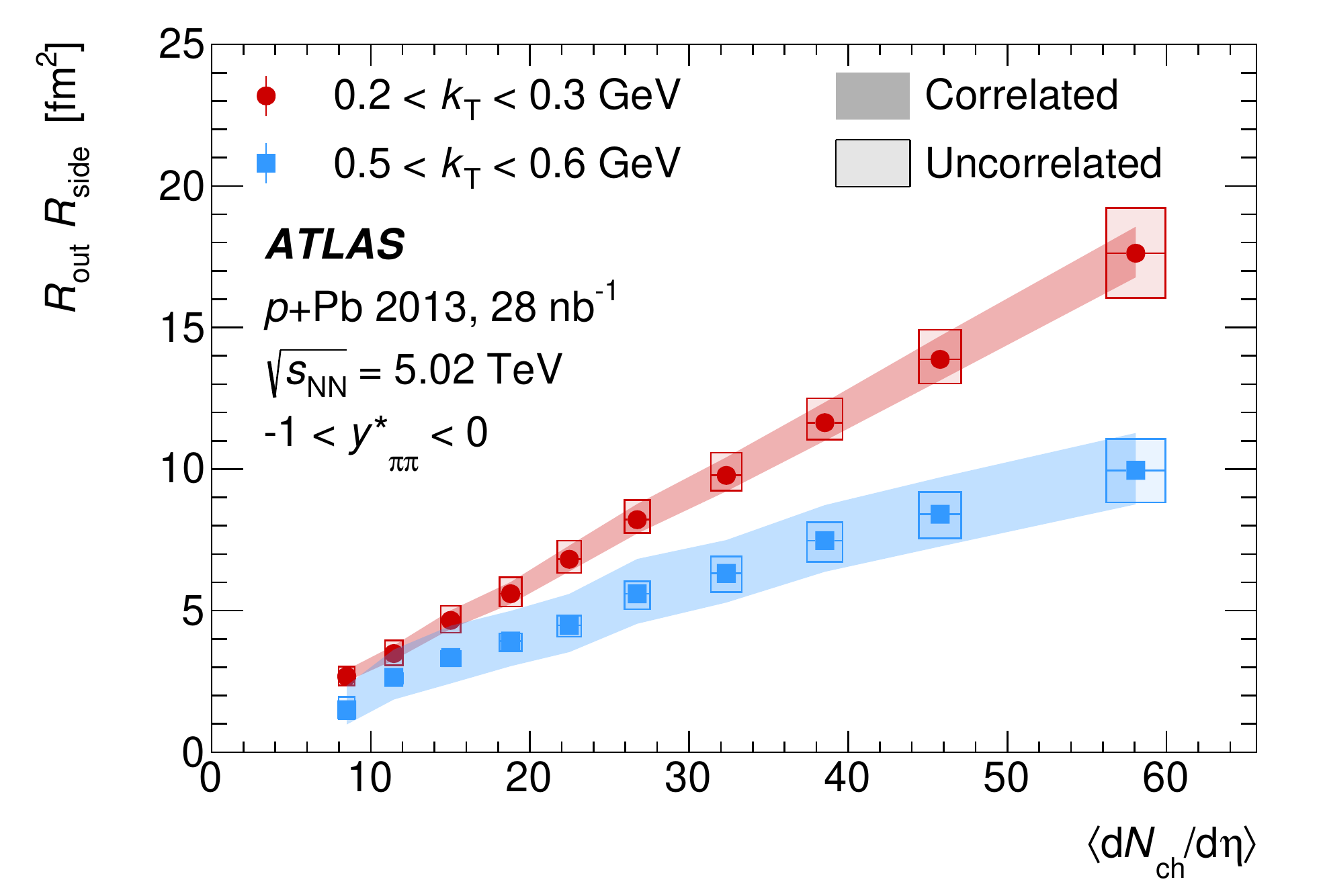}
}
\subfloat[]{
\includegraphics[width=0.49\linewidth]{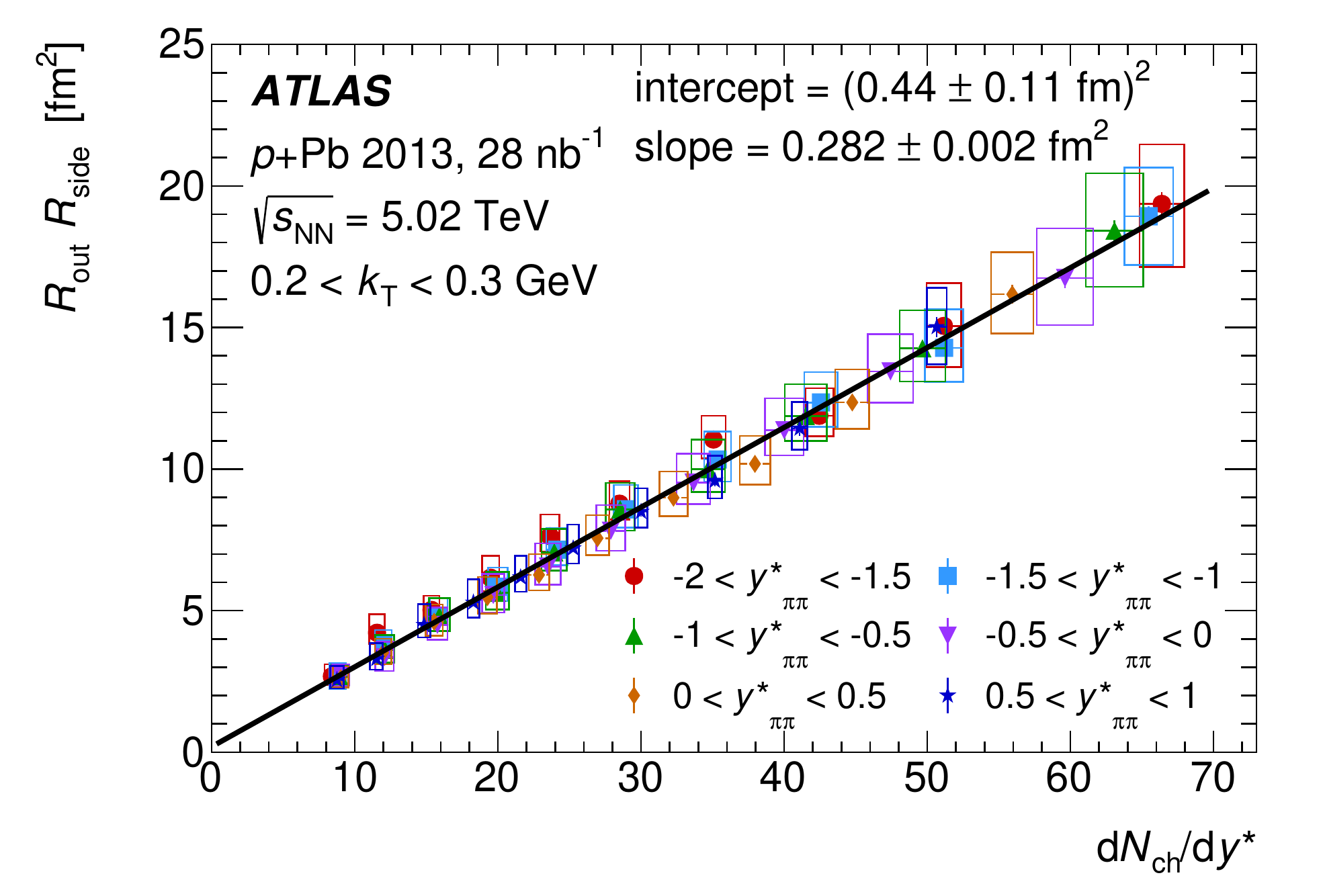}
}
\caption{The transverse area scale, $\Rout\Rside$, plotted against the average multiplicity, \avgdNdeta, (a) and the local density, \dNdy, as a function of rapidity (b). The systematic uncertainties from pion identification and the generator and collision system components of the jet background description are treated as correlated and shown as error bands. The systematic uncertainties from charge asymmetry, \Reff, rapidity variation of the jet fragmentation, and two-particle track reconstruction effects are treated as uncorrelated and indicated by the height of the boxes. The horizontal error bars indicate the systematic uncertainty from \avgdNdeta or \dNdy. The slope and intercept of the best fit to the right-hand plot are shown with combined statistical and systematic uncertainties.}
\label{fig:results_RoutRside_mult}
\end{figure}

\begin{figure}[h!]
\centering
\subfloat[]{
\includegraphics[width=0.49\linewidth]{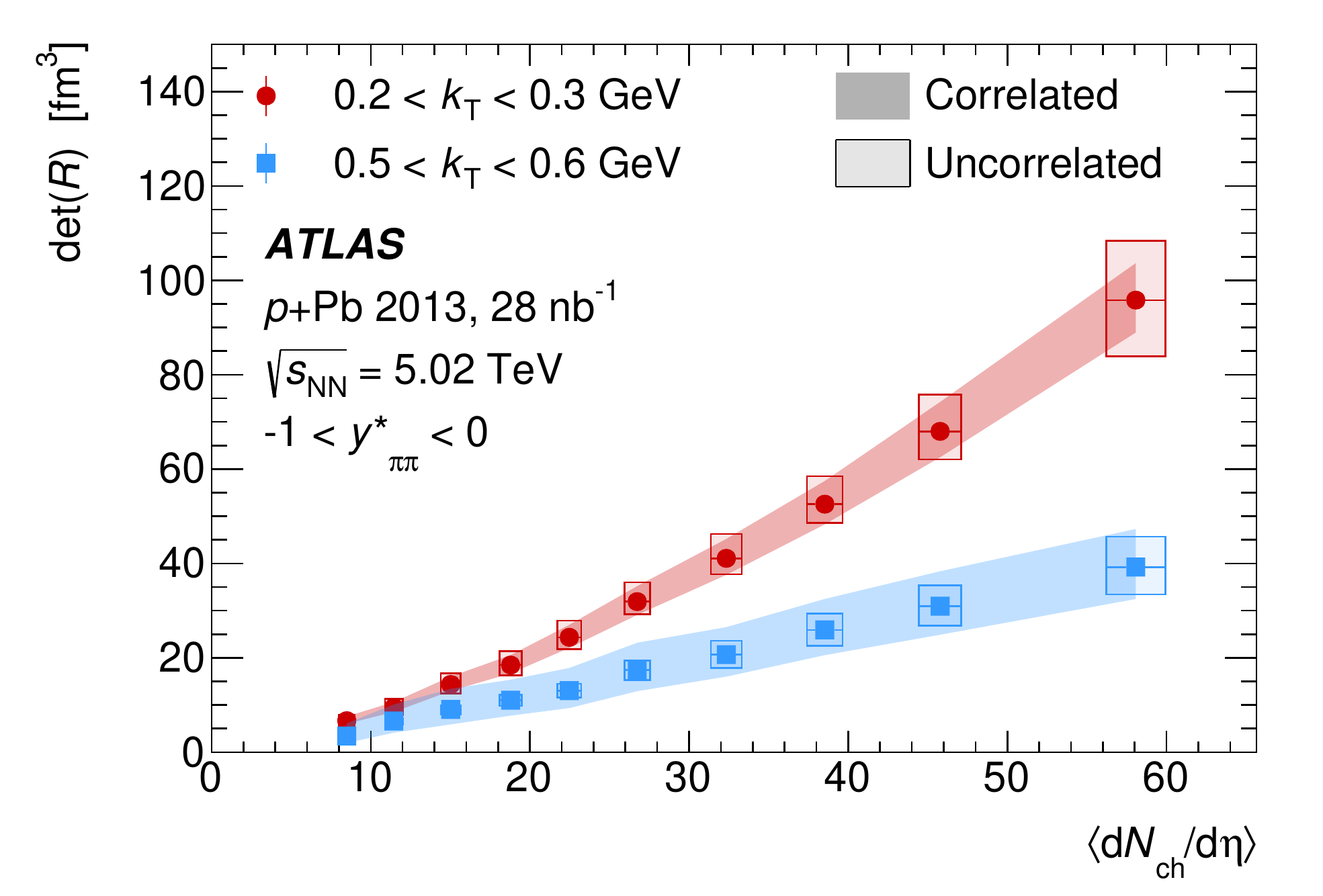}
}
\subfloat[]{
\includegraphics[width=0.49\linewidth]{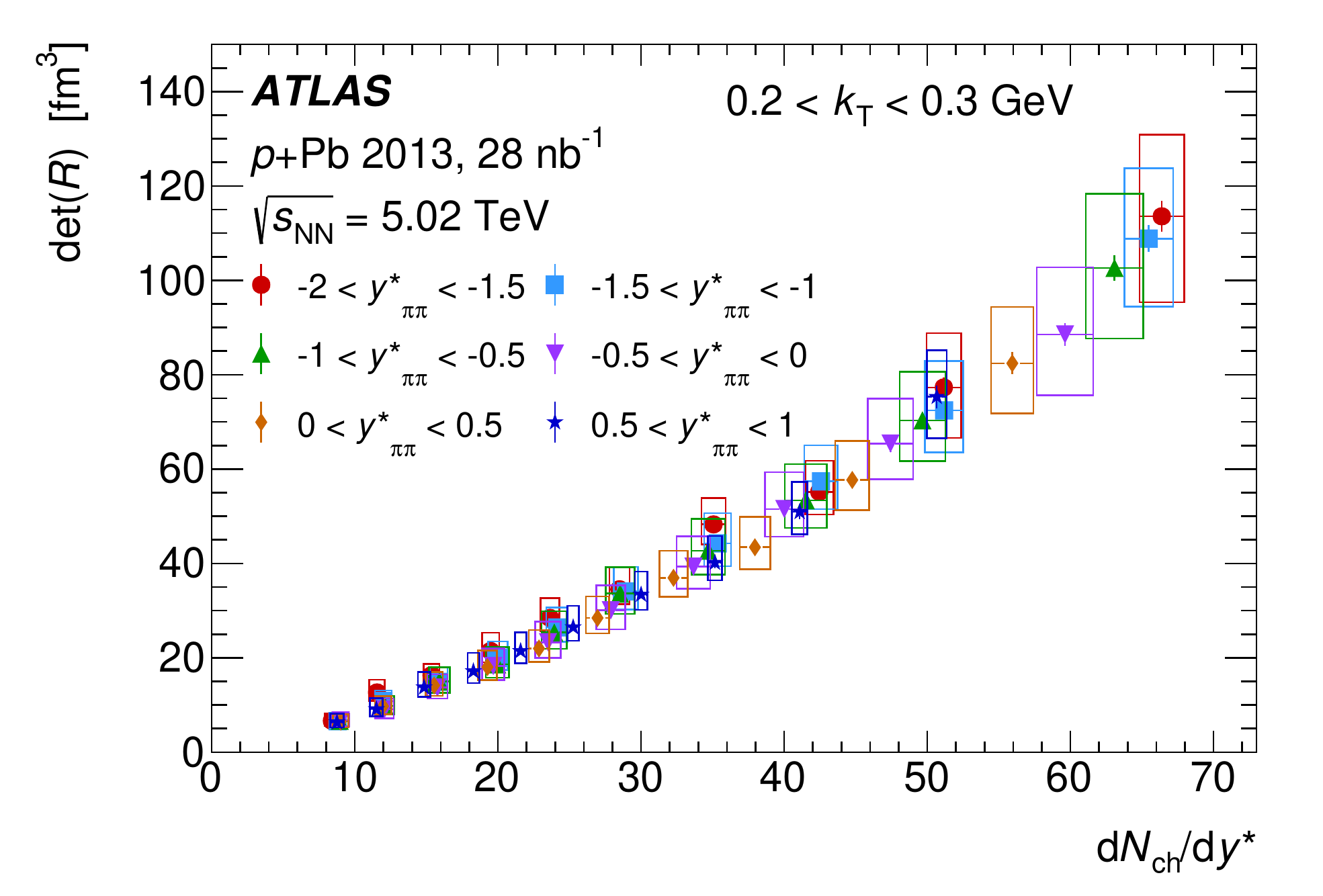}
}
\caption{The volume scale, \detR, plotted against the average multiplicity, \avgdNdeta, (a) and the local density, \dNdy, as a function of rapidity (b). In the left plot, the systematic uncertainties from pion identification and the generator and collision system components of the jet background description are treated as correlated and shown as error bands, while those from charge asymmetry, \Reff, rapidity variation of the jet fragmentation, and two-particle track reconstruction effects are treated as uncorrelated and indicated by the height of the boxes. The widths of the boxes indicate the systematic uncertainty in \avgdNdeta or \dNdy.}
\label{fig:results_RoutRsideRlong_dndeta}
\end{figure}

The ratio $\Rout/\Rside$ (\Fig{\ref{fig:results_RoutOverRside_kt}}) is often studied because in models with radial flow, $\Rout$ includes components of the source's lifetime but $\Rside$ does not (see, for instance, the discussion in \Ref{\cite{Lisa:2005dd}}).
A value of $\Rout/\Rside$ less than one is observed and it decreases with increasing \kt.
The ratio is observed to be the same in different centrality intervals within uncertainties.
As explained in \Ref{\cite{Pratt:2008qv}}, several improvements to naive hydrodynamic models---primarily pre-thermal acceleration, a stiffer equation of state, and shear viscosity---all result in more sudden emission.
This implies that a value of $\Rout/\Rside \lesssim 1$ does not necessarily rule out collective behavior.

The transverse area scale $\Rout\Rside$ is shown in \Fig{\ref{fig:results_RoutRside_mult}} as a function of both event and local density. At lower \kt, the transverse area scales linearly with multiplicity over all centralities and rapidities. This result is consistent with a picture in which the longitudinal dynamics can be separated from the transverse particle production, and low-\kt particles freeze out at a constant transverse area density.

\begin{figure}[tb]
\centering
\subfloat[]{
\includegraphics[width=0.49\linewidth]{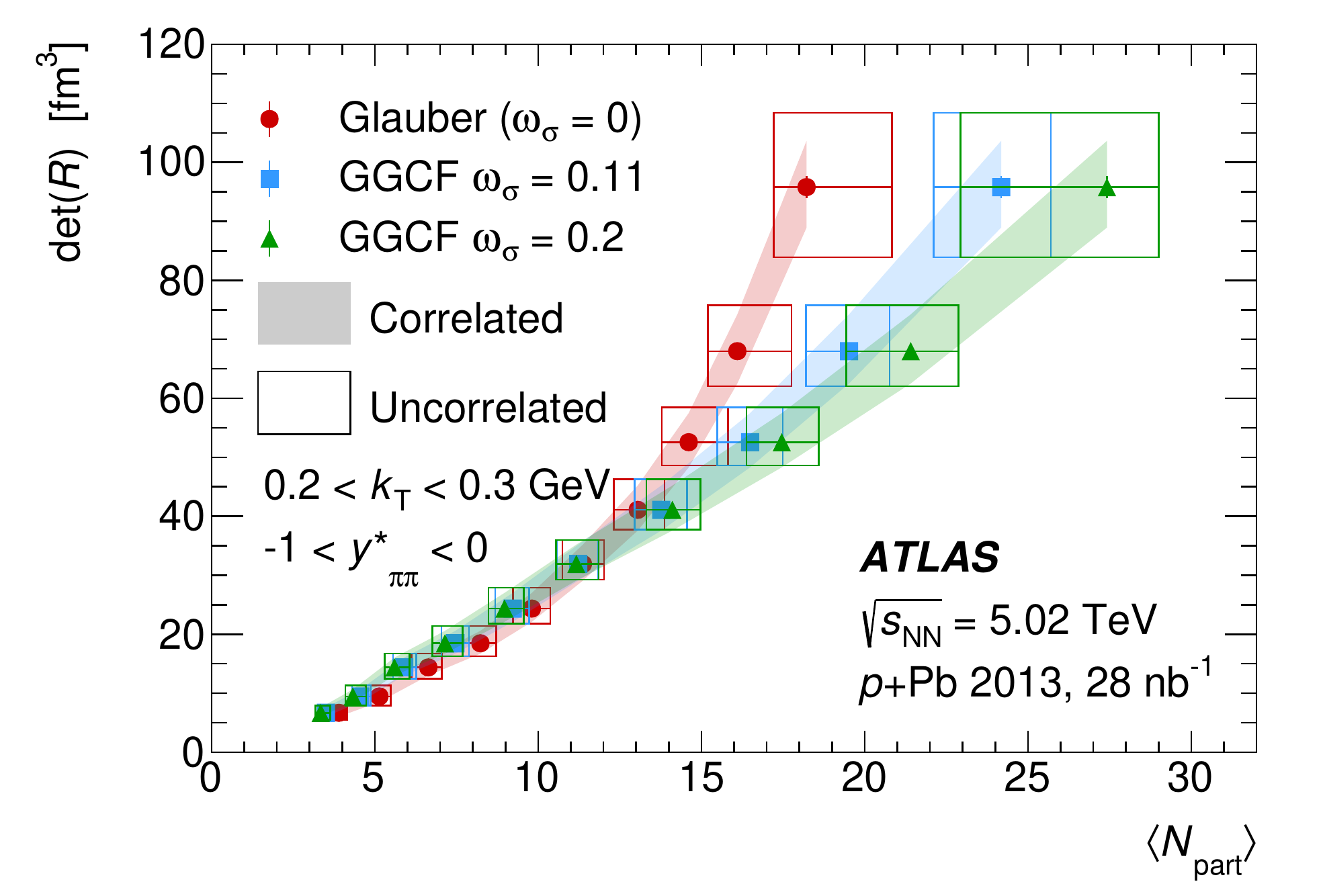}
}
\subfloat[]{
\includegraphics[width=0.49\linewidth]{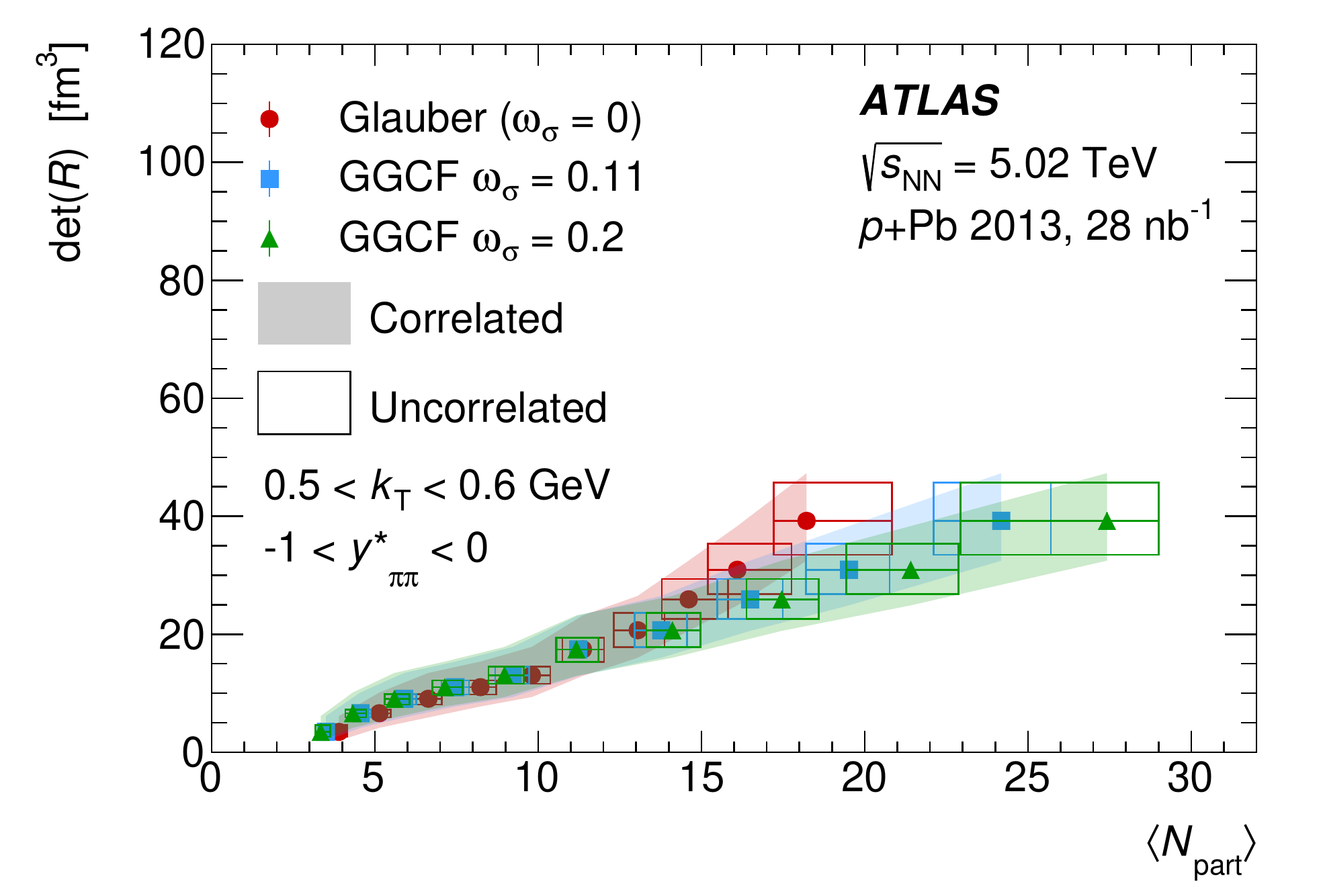}
}
\caption{The scaling of the volume element, \detR, with \avgNpart calculated with three initial geometry models: standard Glauber as well as Glauber-Gribov (GGCF) for two choices of the color fluctuation parameter, $\omega_\sigma$.
Each of the panels shows a different \kt interval.
The systematic uncertainties from the pion identification and the generator and collision system components of the background description are treated as correlated and shown as error bands.
The systematic uncertainties from charge asymmetry, \Reff, rapidity variation of the jet background, and two-particle reconstruction are treated as uncorrelated and indicated by the height of the boxes.
The horizontal error bars indicate the systematic uncertainties in \avgNpart.}
\label{fig:results_RoutRsideRlong_ggcf}
\end{figure}
\begin{figure}[h!]
\centering
\subfloat[]{
\includegraphics[width=0.49\linewidth]{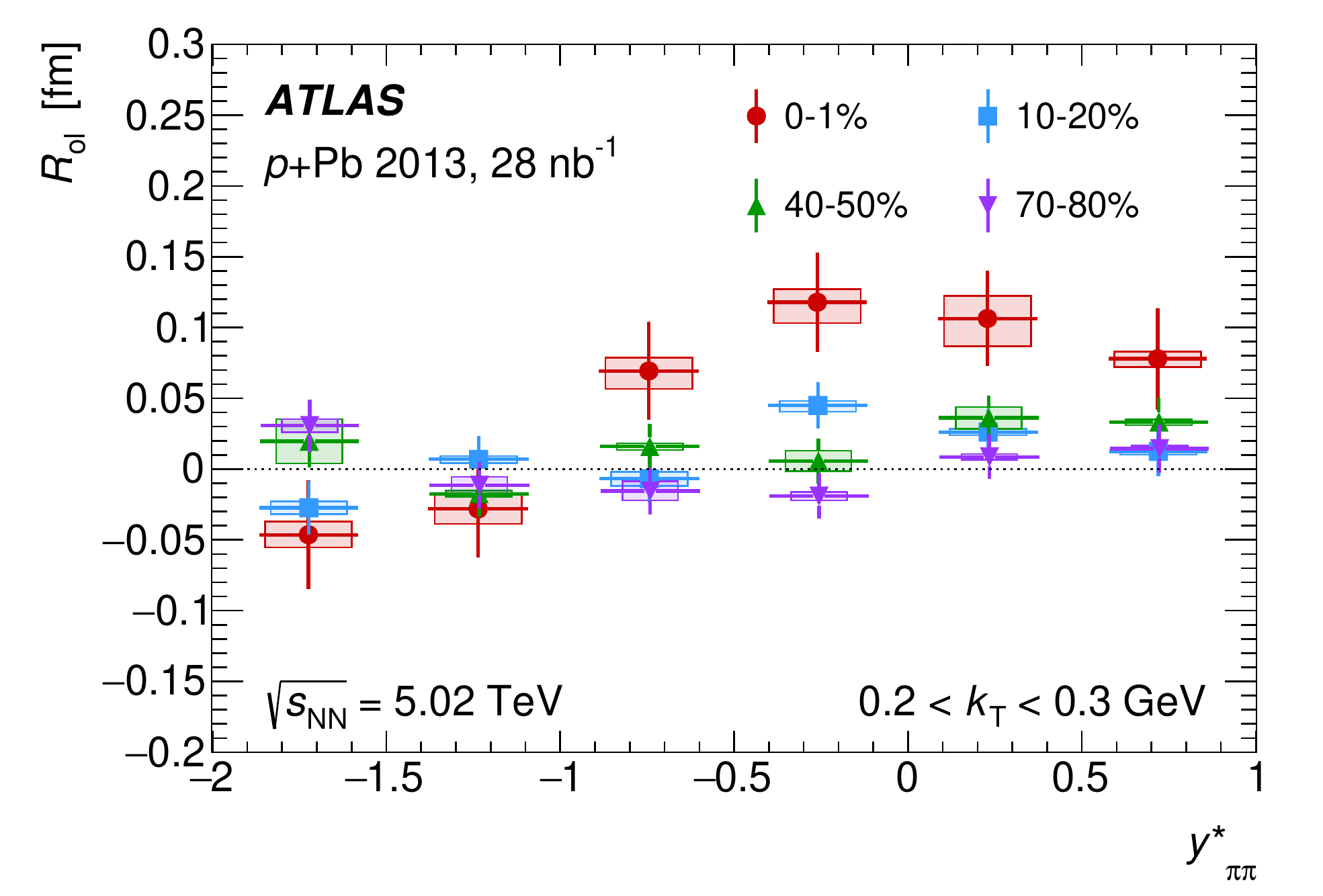}
}
\subfloat[]{
\includegraphics[width=0.49\linewidth]{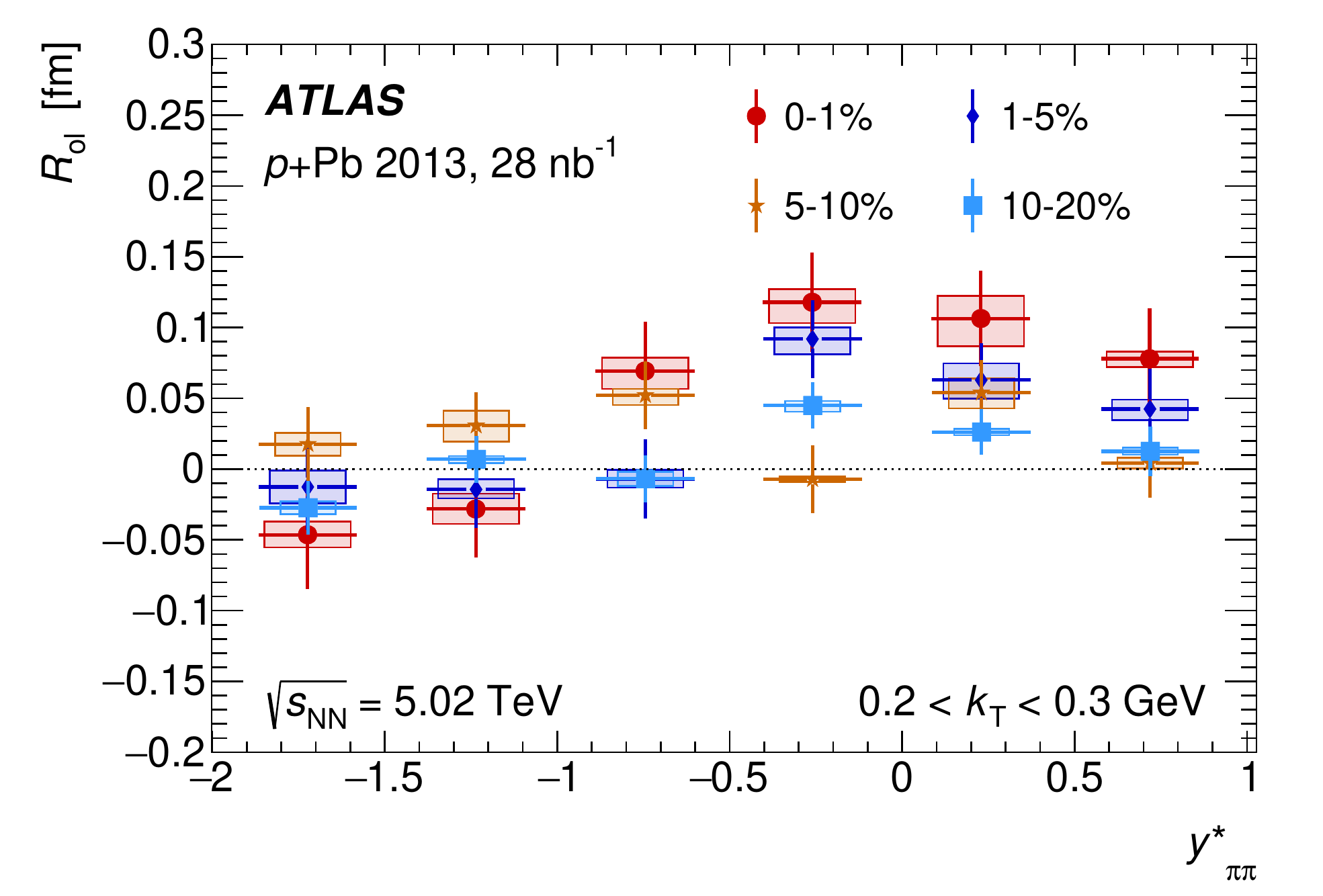}
}
\caption{The cross-term, \Rol, as a function of the pair's rapidity, \kys, in a wide range of centrality intervals (a) and in the four most central event classes (b). The vertical size of each box represents the quadrature sum of the systematic uncertainties described in \Sect{\ref{sec:systematics}}, and statistical uncertainties are shown with vertical lines. The horizontal positions of the points are the average \kys in each interval, and the horizontal lines indicate the standard deviation of \kys in the corresponding interval. The widths of the boxes differ only for visual clarity.}
\label{fig:results_Rol_kys}
\end{figure}

The determinant of the 3D radius matrix, \detR (\Eqn{\ref{eq:r_matrix}}), is shown in \Fig{\ref{fig:results_RoutRsideRlong_dndeta}} as a function of both the average and local density.
While the transverse area scales linearly with  multiplicity at low \kt, the volume scale grows linearly at higher \kt, implying a constant freeze-out volume density for particles with higher momentum.
\Fig{\ref{fig:results_RoutRsideRlong_ggcf}} compares the volume scaling with \avgNpart for the standard Glauber model as well as for two choices of the Glauber-Gribov color fluctuation (GGCF) model \cite{Alvioli:2013vk}.
The parameter $\omega_{\sigma}$ controls the size of the fluctuations in the nucleon-nucleon cross section within the Glauber-Gribov model.
With the Glauber model, the scaling of the volume element with \avgNpart has a significant upwards curvature.
Including Glauber-Gribov fluctuations in the \avgNpart calculation results in a more modest curvature in the scaling of \detR.
This result suggests that the fluctuations in the nucleon-nucleon cross section are a crucial component of the initial geometry description in \pPb systems.
The values and systematic uncertainties of \avgNpart in each model are listed in \Tab{\ref{table:npart}}.

\begin{figure}[t]
\centering
\subfloat[]{
\includegraphics[width=0.49\linewidth]{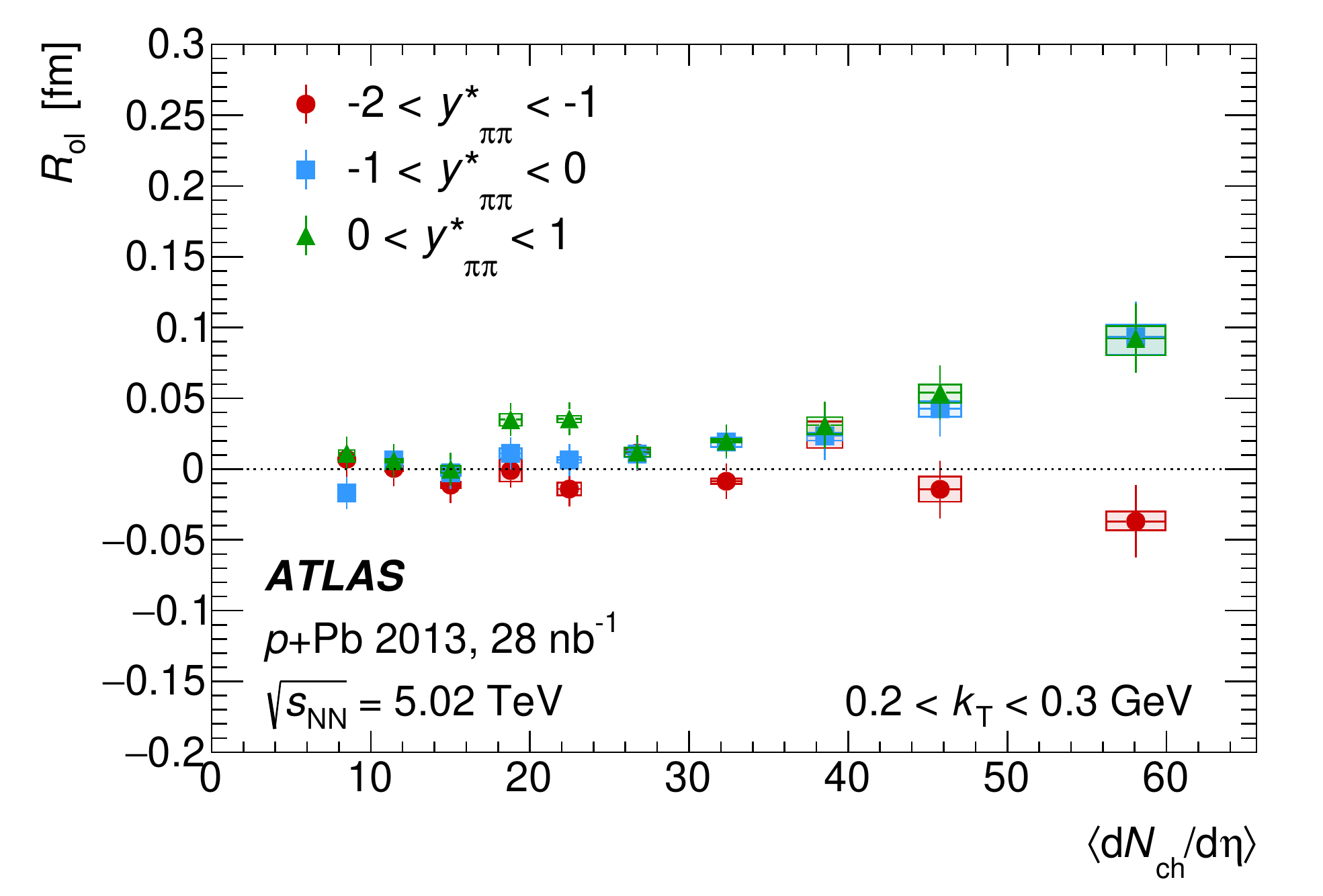}
}
\subfloat[]{
\includegraphics[width=0.49\linewidth]{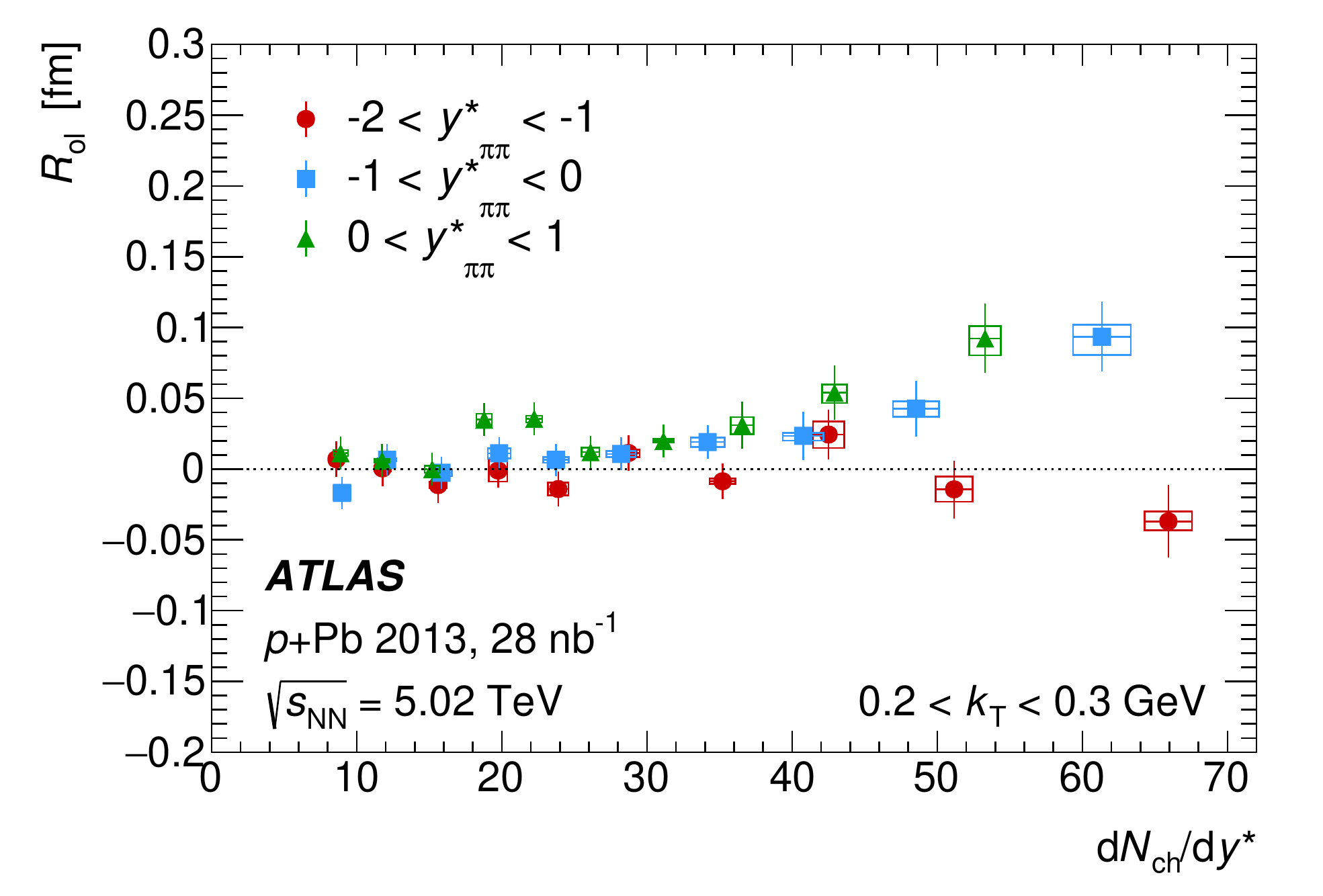}
}
\caption{The cross-term, \Rol, as a function of average event multiplicity, \avgdNdeta, (a) and the local density, \dNdy (b). The vertical size of each box represents the quadrature sum of the systematic uncertainties described in \Sect{\ref{sec:systematics}}, and statistical uncertainties are shown with vertical lines. The widths of the boxes indicate the systematic uncertainties in the corresponding quantities.}
\label{fig:results_Rol_mult}
\end{figure}

The cross-term, \Rol (\Eqn{\ref{eq:r_matrix}}), which couples to the lifetime of the source \cite{Chapman:1994yv}, is shown in \Figs{\ref{fig:results_Rol_kys}}{\ref{fig:results_Rol_mult}}.
A significant departure from zero is observed in this parameter in central events, but only for rapidities $\kys \gtrsim -1$.
For the 0--1\% centrality interval, in $ 0.2 < \kt < 0.4$ and $-1 < \kys < 1$, \Rol is measured to be nonzero with a significance of 7.1/7.3/5.1 $\sigma$ (statistical/systematic/combined).
The next most central interval, 1--5\%, has a nonzero \Rol with a significance of 5.2/5.8/3.9 $\sigma$ (statistical/systematic/combined).
This suggests that the particle production at middle and forward rapidities is sensitive to the local $z$-asymmetry of the system.

The argument from \Sect{\ref{subsec:parameterization}} for why the order of the particles in a pair can be chosen so that \qout is greater than zero relies on the assumption that both particles in the pair are the same species, or at least that they are characterised by the same momentum distributions.
In principle, final-state interactions between different particle species could break this symmetry of the correlation function and lead to a nonzero \Rol term. 
However, the systematic uncertainties shown in \Fig{\ref{fig:syst_qosl_Rol}} demonstrate that \Rol is not sensitive to particle identification, particularly at low \kt. At larger \kt the systematic effect from PID looks larger, but the variations are likely driven by statistical fluctuations.

\FloatBarrier

\section{Summary and conclusions}
\label{sec:conclusions}

This paper presents ATLAS measurements of two-identified-pion HBT correlations in \pPb collisions at the LHC at \pPbenergy{} using a total integrated luminosity of 28 $\mathrm{nb}^{-1}$.
Two-particle correlation functions were measured in one dimension as a function
of \qinv\ and in three dimensions using the out-side-long decomposition as a function of the pair's average transverse momentum and the pair's rapidity.
The measurements were performed for several intervals of \pPb\ centrality characterized by \sumETPb, the total transverse energy measured in the Pb-going forward calorimeter.
A data-driven technique was developed for constraining the contribution of jet fragmentation to the correlation function.
The correlation functions were fit to a Bowler-Sinyukov parameterization of the Bose-Einstein and Coulomb form with a contribution that describes the jet background.
The HBT correlation is described by an exponential form that provides a good description of the data.
For the out-side-long fits, the parameterization includes a non-diagonal coupling between \qout\ and \qlong\ in the correlation function. 

The radii extracted from the one-dimensional and three-dimensional fits show a significant variation with transverse momentum \kt that is strongest for the most central events and weakest or not present in the most peripheral centrality interval.
For the three-dimensional fits, the \kt dependence is found to be the largest for the out and long directions.
A small but significant dependence of the three-dimensional source radii on the pair's rapidity is observed in the more central collisions while in the most peripheral collisions the radii do not depend on rapidity.

The one-dimensional and three-dimensional source radii increase monotonically between peripheral and central collisions with a slope that decreases with rising \kt.
The dependence of the radii on centrality was studied as a function of both the cube root of the rapidity-averaged charged-particle multiplicity, $\avgdNdeta^{1/3}$, and the local charged-particle density, \dNdy.
When evaluated in intervals of both centrality and rapidity, the radii as a function of \dNdy fall on a single curve.
At low \kt the rapidity-averaged radii are observed to increase approximately linearly with $\avgdNdeta^{1/3}$.
A nonzero out-long cross-term \Rol is observed in central (0--5\%) collisions for rapidities greater than $-1$, with a combined significance of $5.1\sigma$ in the most central (0--1\%) events.

The transverse area, $\Rside \, \Rlong$, is observed to vary linearly with \dNdy\ at low \kt.
The volume scale represented by \detR increases faster than linearly with $\avgdNdeta^{1/3}$ or \dNdy\ at low \kt, but increases approximately linearly with $\avgdNdeta^{1/3}$ at higher momentum ($\kt > 0.5$~\GeV).
When plotted versus the mean number of nucleon participants \Npart\ obtained from three different geometric models, the volume shows a steady increase with \avgNpart\ for the GGCF-derived \avgNpart\ values, but a sudden increase with \avgNpart\ for $\avgNpart \gtrsim 12$ when \avgNpart\ is obtained from the Glauber model.
While the freeze-out volume scale \detR should only be strictly linear with the initial size, represented by \Npart, if the expansion is independent of centrality, an extreme deviation from a naive linear scaling is not expected.
This observation supports the conclusion drawn from previous studies that the Glauber-Gribov color fluctuation model provides a better description of \pPb\ collision geometry. 

The \Rout to \Rside ratio is found to be less than unity for all centrality and kinematic selections studied in this analysis, and it is observed to decrease approximately linearly with increasing \kt.
This result, combined with the \kt\ dependence of the radii, suggests a collective, explosive expansion of the source.
The nonzero out-long cross-term indicates that the freeze-out behavior of the source is sensitive to the local $z$-asymmetry of the particle production away from the Pb-going region. 

The results presented in this paper provide detailed measurements of the space-time extent of the particle source in \pPb systems.
In particular, the rapidity sensitivity of the results demonstrate the asymmetry of the \pPb system and show that \kt and local charged-particle density are sufficient to predict the radii.
These conclusions present a significant opportunity for theoretical models.

\FloatBarrier

\section*{Acknowledgments}
We thank CERN for the very successful operation of the LHC, as well as the
support staff from our institutions without whom ATLAS could not be
operated efficiently.

We acknowledge the support of ANPCyT, Argentina; YerPhI, Armenia; ARC, Australia; BMWFW and FWF, Austria; ANAS, Azerbaijan; SSTC, Belarus; CNPq and FAPESP, Brazil; NSERC, NRC and CFI, Canada; CERN; CONICYT, Chile; CAS, MOST and NSFC, China; COLCIENCIAS, Colombia; MSMT CR, MPO CR and VSC CR, Czech Republic; DNRF and DNSRC, Denmark; IN2P3-CNRS, CEA-DSM/IRFU, France; SRNSF, Georgia; BMBF, HGF, and MPG, Germany; GSRT, Greece; RGC, Hong Kong SAR, China; ISF, I-CORE and Benoziyo Center, Israel; INFN, Italy; MEXT and JSPS, Japan; CNRST, Morocco; NWO, Netherlands; RCN, Norway; MNiSW and NCN, Poland; FCT, Portugal; MNE/IFA, Romania; MES of Russia and NRC KI, Russian Federation; JINR; MESTD, Serbia; MSSR, Slovakia; ARRS and MIZ\v{S}, Slovenia; DST/NRF, South Africa; MINECO, Spain; SRC and Wallenberg Foundation, Sweden; SERI, SNSF and Cantons of Bern and Geneva, Switzerland; MOST, Taiwan; TAEK, Turkey; STFC, United Kingdom; DOE and NSF, United States of America. In addition, individual groups and members have received support from BCKDF, the Canada Council, CANARIE, CRC, Compute Canada, FQRNT, and the Ontario Innovation Trust, Canada; EPLANET, ERC, ERDF, FP7, Horizon 2020 and Marie Sk{\l}odowska-Curie Actions, European Union; Investissements d'Avenir Labex and Idex, ANR, R{\'e}gion Auvergne and Fondation Partager le Savoir, France; DFG and AvH Foundation, Germany; Herakleitos, Thales and Aristeia programmes co-financed by EU-ESF and the Greek NSRF; BSF, GIF and Minerva, Israel; BRF, Norway; CERCA Programme Generalitat de Catalunya, Generalitat Valenciana, Spain; the Royal Society and Leverhulme Trust, United Kingdom.

The crucial computing support from all WLCG partners is acknowledged gratefully, in particular from CERN, the ATLAS Tier-1 facilities at TRIUMF (Canada), NDGF (Denmark, Norway, Sweden), CC-IN2P3 (France), KIT/GridKA (Germany), INFN-CNAF (Italy), NL-T1 (Netherlands), PIC (Spain), ASGC (Taiwan), RAL (UK) and BNL (USA), the Tier-2 facilities worldwide and large non-WLCG resource providers. Major contributors of computing resources are listed in Ref.~\cite{ATL-GEN-PUB-2016-002}.

\FloatBarrier

\printbibliography

\newpage
\begin{flushleft}
{\Large The ATLAS Collaboration}

\bigskip

M.~Aaboud$^\textrm{\scriptsize 136d}$,
G.~Aad$^\textrm{\scriptsize 87}$,
B.~Abbott$^\textrm{\scriptsize 114}$,
J.~Abdallah$^\textrm{\scriptsize 65}$,
O.~Abdinov$^\textrm{\scriptsize 12}$,
B.~Abeloos$^\textrm{\scriptsize 118}$,
R.~Aben$^\textrm{\scriptsize 108}$,
O.S.~AbouZeid$^\textrm{\scriptsize 138}$,
N.L.~Abraham$^\textrm{\scriptsize 152}$,
H.~Abramowicz$^\textrm{\scriptsize 156}$
% ...
\bigskip
\\
% ...
$^{12}$ Institute of Physics, Azerbaijan Academy of Sciences, Baku, Azerbaijan\\
% ...
$^{65}$ University of Iowa, Iowa City IA, United States of America\\
% ...
$^{87}$ CPPM, Aix-Marseille Universit{\'e} and CNRS/IN2P3, Marseille, France\\
% ...
$^{108}$ Nikhef National Institute for Subatomic Physics and University of Amsterdam, Amsterdam, Netherlands\\
% ...
$^{114}$ Homer L. Dodge Department of Physics and Astronomy, University of Oklahoma, Norman OK, United States of America\\
% ...
$^{118}$ LAL, Univ. Paris-Sud, CNRS/IN2P3, Universit{\'e} Paris-Saclay, Orsay, France\\
% ...
$^{136}$ $^{(a)}$ Facult{\'e} des Sciences Ain Chock, R{\'e}seau Universitaire de Physique des Hautes Energies - Universit{\'e} Hassan II, Casablanca; $^{(b)}$ Centre National de l'Energie des Sciences Techniques Nucleaires, Rabat; $^{(c)}$ Facult{\'e} des Sciences Semlalia, Universit{\'e} Cadi Ayyad, LPHEA-Marrakech; $^{(d)}$ Facult{\'e} des Sciences, Universit{\'e} Mohamed Premier and LPTPM, Oujda; $^{(e)}$ Facult{\'e} des sciences, Universit{\'e} Mohammed V, Rabat, Morocco\\
% ...
$^{138}$ Santa Cruz Institute for Particle Physics, University of California Santa Cruz, Santa Cruz CA, United States of America\\
% ...
$^{152}$ Department of Physics and Astronomy, University of Sussex, Brighton, United Kingdom\\
% ...
$^{156}$ Raymond and Beverly Sackler School of Physics and Astronomy, Tel Aviv University, Tel Aviv, Israel\\
% ...
$^{a}$ Also at Department of Physics, King's College London, London, United Kingdom\\
$^{b}$ Also at Institute of Physics, Azerbaijan Academy of Sciences, Baku, Azerbaijan\\
$^{c}$ Also at Novosibirsk State University, Novosibirsk, Russia\\
$^{d}$ Also at TRIUMF, Vancouver BC, Canada\\
$^{e}$ Also at Department of Physics {\&} Astronomy, University of Louisville, Louisville, KY, United States of America\\
% ...
\end{flushleft}


\end{document}